\documentclass[prx,amsmath,amssymb,twocolumn,showpacs,nofootinbib,superscriptaddress]{revtex4}
\usepackage{graphicx}
\usepackage{subfigure}
\usepackage{adjustbox}
\usepackage{bm}
\usepackage{color}
\usepackage{braket}
\usepackage{standalone}
\usepackage{multirow}
\usepackage{tikz}
\usepackage{mathrsfs}
\usepackage[colorlinks,bookmarks=true,citecolor=blue,linkcolor=red,urlcolor=blue]{hyperref}

\definecolor{emil}{rgb}{1,0,0}

\definecolor{flore}{rgb}{0,0,1}

\DeclareMathAlphabet{\mathpzc}{OT1}{pzc}{m}{it}

\newcommand{\bfk}{{\bf k}}
\newcommand{\bfq}{{\bf q}}

\begin{document}

\title{Anatomy of Topological Surface States:\\ Exact Solutions from Destructive Interference on Frustrated Lattices}

\author{Flore K. Kunst}
\email[]{flore.kunst@fysik.su.se}
\affiliation{Department of Physics, Stockholm University, AlbaNova University Center, 106 91 Stockholm, Sweden}
\affiliation{Dahlem Center for Complex Quantum Systems and Insitut f\"ur Theoretische Physik, Freie Universit\"at Berlin, Arnimallee 14, 14195 Berlin, Germany}
\author{Maximilian Trescher}
\affiliation{Dahlem Center for Complex Quantum Systems and Insitut f\"ur Theoretische Physik, Freie Universit\"at Berlin, Arnimallee 14, 14195 Berlin, Germany}
\author{Emil J. Bergholtz}\email[]{emil.bergholtz@fysik.su.se}
\affiliation{Department of Physics, Stockholm University, AlbaNova University Center, 106 91 Stockholm, Sweden}
\affiliation{Dahlem Center for Complex Quantum Systems and Insitut f\"ur Theoretische Physik, Freie Universit\"at Berlin, Arnimallee 14, 14195 Berlin, Germany}
\date{\today}
%

\begin{abstract}
The hallmark of topological phases is their robust boundary signature whose intriguing properties---such as the one-way transport on the chiral edge of a Chern insulator and the sudden disappearance of surface states forming open Fermi arcs on the surfaces of Weyl semimetals---are impossible to realize on the surface alone. Yet, despite the glaring simplicity of noninteracting topological bulk Hamiltonians and their concomitant energy spectrum, the detailed study of the corresponding surface states has essentially been restricted to numerical simulation. In this work, however, we show that exact analytical solutions of both topological and trivial surface states can be obtained for generic tight-binding models on a large class of geometrically frustrated lattices in any dimension without the need for fine-tuning of hopping amplitudes. Our solutions derive from local constraints tantamount to destructive interference between neighboring layer lattices perpendicular to the surface and provide microscopic insights into the structure of the surface states that enable analytical calculation of many desired properties including correlation functions, surface dispersion, Berry curvature, and the system size dependent gap closing, which necessarily occurs when the spatial localization switches surface. This further provides a deepened understanding of the bulk-boundary correspondence. We illustrate our general findings on a large number of examples in two and three spatial dimensions. Notably, we derive exact chiral Chern insulator edge states on the spin-orbit-coupled kagome lattice, and Fermi arcs relevant for recently synthesized slabs of pyrochlore-based Eu$_2$Ir$_2$O$_7$ and Nd$_2$Ir$_2$O$_7$, which realize an all-in-all-out spin configuration, as well as for spin-ice-like two-in-two-out and one-in-three-out configurations, which are both relevant for Pr$_2$Ir$_2$O$_7$. Remarkably, each of the pyrochlore examples exhibit clearly resolved Fermi arcs although only the one-in-three-out configuration features bulk Weyl nodes in realistic parameter regimes. Our approach generalizes to symmetry protected phases, e.g. quantum spin Hall systems and Dirac semimetals with time-reversal symmetry, and can furthermore signal the absence of topological surface states, which we illustrate for a class of models akin to the trivial surface of Hourglass materials KHg$X$ where the exact solutions apply but, independently of Hamiltonian details, yield eigenstates delocalized over the entire sample. 
\end{abstract} 

\pacs{71.10.Fd, 73.21.Ac, 73.20.At, 03.65.Vf}

\maketitle

\section{Introduction}
The experimental discovery of the quantum Hall effect in 1980 [\onlinecite{klitzingdordapepper}] decisively put topological phases in the limelight, and especially during the past decade the interplay between theoretical ideas and experimental advances has led to spectacular developments with intriguing prospects for future technological applications \cite{hasankane,qizhang,Weylreview,topocomp}. Most early work focused on topological insulators [\onlinecite{hasankane,qizhang}], the most basic of which are simple two-dimensional lattice generalizations of the quantum Hall states, namely Chern insulators [\onlinecite{haldane,tknn,hofstadter,changzhangfengshenzhang,jotzumesserdesbuquoislebratuehlingergreifesslinger}], while quantum spin Hall insulators stem from two time-reversed copies thereof [\onlinecite{kanemele, kanemele2}]. Weyl semimetals, experimentally realized in 2015 [\onlinecite{xubelopolskialidoustetal, lvwengwantmiaoetal,luwangyeranfujoannopoulossoljacic,hasanreview}], are paradigmatic examples of a gapless topological phase existing in three dimensions [\onlinecite{volovik,murakami, wanturnerbishwanathsavrasov, burkovbalents}], whose time-reversal invariant cousins, the Dirac semimetals, were unraveled in 2014 [\onlinecite{liuzhouwangwengprabhakaran, liujianzhouwangzhangweng}].

What makes these topological phases so intriguing is their robust and novel boundary states. Despite their central importance, and the simplicity of their bulk description, explicit solutions for the boundary states of topological phases are only known in a very limited number of special cases \cite{kitaev,liuqizhang,maokuramotoimurayamakage, shenshanlu, koenigbuhmannmolenkamphughesliuqizhang, mongshivamoggi, zhouluchushenniu,ojanen, aklt}. While powerful transfer matrix methods, which in some special cases allow analytical progress, have been developed \cite{transfer1,transfer2,transfer3}, there is a glaring absence of generic analytical solutions that do not require fine-tuning, that are valid in any dimension, in the entire surface Brillouin zone, at finite size, and without the need for approximations. 

\begin{figure*}[t]
  \centering
  \adjustbox{trim={0\width} {0.3\height} {0\width} {0\height},clip}
  {\includegraphics[width=\textwidth]{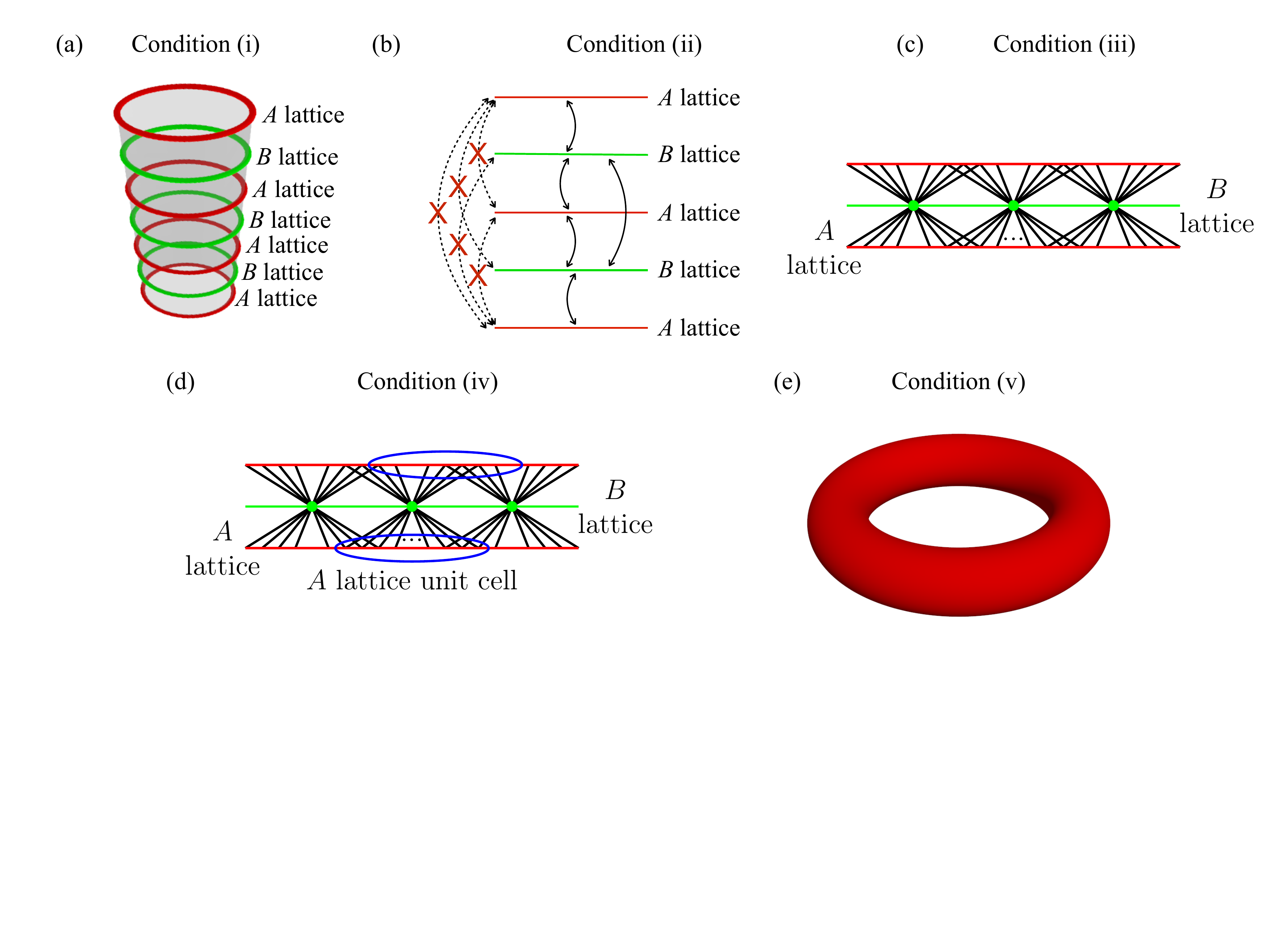}}
\caption{Schematic figures to illustrate conditions (i)-(v). (a) The periodic, $(d-1)$-dimensional $A$ and $B$ lattices are shown in red and green, respectively. They are stacked such that condition (i) is fulfilled. (b) Schematic figure of which hoppings between the $A$ and $B$ lattices are allowed to fulfill condition (ii). Allowed hoppings are shown with arrows, and forbidden hoppings are indicated by crossed-out, dashed arrows. (c) The $A$ and $B$ lattices are shown in red and green, respectively, and the black lines indicate the hoppings between them. The lattices are coupled in a geometrically frustrated fashion, condition (iii). (d) Schematic figure of the frustrated lattice models with the minimal unit cell of the $A$ lattice indicated by a blue ellipse. The unit cell is chosen to be the same for each $A$ lattice. The hopping constraint to the intermediate site involves a nontrivial connection between unit cells fulfilling condition (iv). (e) Condition (v) states that the bulk Hamiltonian must support a topological phase, here symbolized by a torus.}
\label{figureallconditions}
\end{figure*}

In this work, we devise a general strategy for finding exact surface state solutions for trivial as well as for topological phases in any dimension, notably including Chern insulators and Weyl semimetals as well as their time-reversal invariant counter parts in quantum spin Hall insulators and Dirac semimetals. Rather than stemming from fine-tuning of hopping amplitudes our method is rooted in the underlying lattice structure. The lattices we consider can be seen as composed by $(d-1)$-dimensional layers of different variety, referred to as $A$ and $B$ lattices, that are stacked on top of each other in an alternating fashion such that the full $d$-dimensional lattice is geometrically frustrated (Fig.~\ref{figureallconditions}). Prominent examples of this type are kagome lattices in $d=2$ and pyrochlore in $d=3$. 

Frustrated lattices are usually studied in the context of magnetism and lead to rich physics while being notoriously difficult to understand even at a qualitative level \cite{balents10}. In glaring contrast, we find that frustration greatly simplifies the study of surface states of both trivial and topological variety. Under very general conditions we find exact eigenstates of the form 
\begin{align}
\ket{\Psi ({\bf k})} & = \mathcal{N} ({\bf k}) \bigoplus_{m = 1}^N \left(r ({\bf k})\right)^m \ket{\Phi ({\bf k})}_m, \label{exactstate}
\end{align}
where ${\bf k}$ is the  $(d-1)$-dimensional quasi-momentum parallel to the surface, $\ket{\Phi ({\bf k})}_m$ is a Bloch state of the $m$th $A$ lattice layer in a system composed of $N$ such layers and $N-1$ intermediate $B$ lattice layers, and $r ({\bf k})$ is a simple function determined by the local connectivity between neighboring layers and the Bloch states of the individual $A$ layers. A salient feature of (\ref{exactstate}) is the vanishing amplitudes on the $B$ lattice layers, which is directly related to how the exact solutions are found: assuming vanishing amplitudes on the $B$ lattices puts constraints on $r ({\bf k})$ and provides a bootstrapping procedure uniquely leading to (\ref{exactstate}). The existence of these solutions hinges only on the counting of local constraints in combination with locality and translation invariance, and as such is insensitive to Hamiltonian details. In this context we stress that the local constraints are not a feature of the Hamiltonian but rather an emergent exact property of the eigenstates in Eq. (\ref{exactstate}), which is, however, not fulfilled for any other eigenstate. It is also noticeable that the exact solutions are for the full tight-binding model and thus extend in the full $(d-1)$-dimensional surface Brillouin zone and thereby also describe the attachment to bulk bands as the states switch surface. Moreover, the solutions remain exact at any finite size, i.e., for any number of layers, $N$. 

Our approach is akin to the construction of flat band models arising due to local constraints on 'line graphs' such as kagome and pyrochlore lattices (see e.g. Ref.~\onlinecite{bergmanwubalents}). In contrast to our setup, these models require precise fine-tuning of the hopping amplitudes, typically allowing real and strictly nearest-neighbor hopping only. A second key difference is that the flat bands studied earlier are $d$-dimensional bulk bands while our solutions provide a $(d-1)$-dimensional manifold corresponding to the surface Brillouin zone. A similarity is, however, that band touchings necessarily occur in both setups.
 
Expanding on the seminal work by Mielke \cite{mielke}, a large body of work, including effects of interactions and disorder on line graphs, has accumulated during the past 25 years. In this context, valuable insights have been obtained for antiferromagnetic Heisenberg models on frustrated lattices with a flat band corresponding to a localized magnons \cite{schulenburg,schnack,zengelser, asakawasuzuki} as well as on flat-band Hubbard models \cite{derzhkorichtermaksymenko, gulacsikampfvollhardt}. Alongside the extensive literature on theory (see also Refs.~\onlinecite{mielke, tasaki, mielketasaki, flachleykambodyfeltmatthies, bodyfeltleykamdanieliyu, huberaltman, goldmanurbanbercioux}), intriguing recent experiments \cite{babouxjacqimbiodigalopin, mukherjeespracklenchoudhurygoldman} have underscored the value of these works. 

It is conceivable that a similar progress on topological (and trivial) surface states can be spurred by the present work. Indeed, earlier work by two of us exploring Eq. (\ref{exactstate}) in the special case of $\{111\}$-oriented slabs of the pyrochlore lattice \cite{trescherbergholtz,bergholtzliutreschermoessnerudugawa} has already borne fruit: for thin slabs this provided a natural platform for nearly flat bands with higher Chern numbers \cite{trescherbergholtz} and led to the subsequent discovery of an entire zoo of novel fractional Chern insulators qualitatively different from their quantum Hall relatives \cite{ChernN,ChernN2,bergholtzliutreschermoessnerudugawa}. For thicker slabs, we discovered that Fermi arcs can in fact persist without Weyl nodes in the bulk---and that when Weyl nodes do occur their dispersion is generally both anisotropic and tilted \cite{bergholtzliutreschermoessnerudugawa, titledweylcones}. In particular, the tilting can easily be so strong that the Weyl cones become ``over-tilted" forming a compensated metal where the Weyl point is a singular point connecting two Fermi pockets \cite{bergholtzliutreschermoessnerudugawa}. These systems were later popularly coined type-II Weyl semimetals \cite{saluyanovgreschwangwutroyerdaibernevig} and subsequently experimentally identified in a growing list of intriguing materials \cite{weyl2exp,weyl2exp2,weyl2exp3,weyl2exp4,weyl2exp5,weyl2expWTe2_1,weyl2expWTe2_2,weyl2expMoxW1?xTe2}. The phenomenon of Fermi arcs without the presence of Weyl nodes has been corroborated by recent experimental findings [\onlinecite{brunotamaiwucucchi, xuautesmattlvyao}].

Building on our previous work, we here explore the connection between frustration and surface topology in much more detail and generality whereby we derive a number of generic results regarding correlation functions, surface dispersion, Berry curvature, energy gaps, and the bulk-boundary correspondence. While we mostly focus on two- and three-dimensional examples our results apply {\it mutatis mutandis} to any dimension. We also refine the earlier analysis of pyrochlore slabs to relate more directly to experiments on pyrochlore iridates. In particular, this makes contact to beautiful recent experimental progress in growing (thin) single-crystal slabs of the pyrochlore iridates Eu$_2$Ir$_2$O$_7$ \cite{fujita2015} and Nd$_2$Ir$_2$O$_7$ \cite{gallagher2016}---two materials that both exhibit an all-in-all-out spin ordering and that we conclude are likely to have Fermi-arc-like surface states without possessing Weyl nodes in the bulk. For the yet to be grown slabs of Pr$_2$Ir$_2$O$_7$, we, however, find that Weyl nodes exist depending on the particular spin ordering \cite{goswami2016}, which is either spin-ice-like two-in-two-out and one-in-three-out, while Fermi arcs exist in either configuration.

This work is structured as follows. In Sec.~\ref{sectiontwo}, we introduce a generic recipe for constructing our models and present the exact solutions and consequences thereof in general terms. In Sec. ~\ref{sectionthree}, we illustrate the effectiveness of our recipe with a number of examples. In Sec.~\ref{sectionthreezero}, we introduce a one-dimensional chain, and analyze the exact expressions for its end modes. In Sec.~\ref{sectionthreeA}, we focus on two dimensions and derive exact edge state solutions on the kagome lattice including the chiral edge states occurring when the system is a Chern insulator. We also discuss the connection between the local lattice structure and topology. In Secs.~\ref{sectionfour}-\ref{sectionfive}, we investigate the surface states of three-dimensional lattice models, most saliently obtaining exact solutions for the Fermi arcs of Weyl semimetals. Throughout this exposition, we comment on the relevance of our solutions for naturally existing, synthesized and artificial materials. We conclude with a discussion in Sec.~\ref{conclusion}.

\section{Setup and general considerations} \label{sectiontwo}
In this section, we introduce five conditions---illustrated in Fig. \ref{figureallconditions} and detailed below---, which, when they are fulfilled, allow us to find a $(d-1)$-dimensional manifold of exact wave functions and energies corresponding to the topological surface theory of a given $d$-dimensional topological phase. After detailing the general setup, we describe a number of results that can be derived directly from the exact surface state solutions. 

\subsection{Lattice structure}
We study $d$-dimensional models with periodic boundary conditions in $(d-1)$ dimensions while they are left open in the remaining dimension giving the possibility of surface state solutions. More precisely, we consider models that can be decomposed in terms of alternating layers of two different $(d-1)$-dimensional periodic lattices; a lattice with $n$ degrees of freedom ($A$ lattice) and a lattice with $n'$ degrees of freedom ($B$ lattice) such that the surfaces of the material are formed by $A$ lattices as shown schematically in Fig.~\ref{figureallconditions}(a), which we refer to as condition (i). In this work, for the sake of transparency, we mostly consider examples in which the degrees of freedom equal the number of sites in the unit $A$ lattice cells, there is one available state per site. Note, however, that our results can readily be generalized to include more degrees of freedom which is necessary for instance for time-reversal symmetric models including onsite spin degrees of freedom.

The key assumption is that the $A$ lattices are only connected to each other via the intermediate $B$ lattices, and hence that direct hopping between different $A$ lattices is prohibited as shown in Fig.~\ref{figureallconditions}(b). We refer to this as condition (ii). This is a realistic scenario, because it is unlikely that the orbitals of electrons sitting on different $A$ lattices will overlap. Upon solving the Schr\"odinger equation, we find precisely $n$ exact solutions to the wave function, which have total-zero weight on the intermediate-lattice sites if the layers are connected such that the full model is geometrically frustrated, condition (iii), as shown schematically in Fig.~\ref{figureallconditions}(c). This is due to geometric frustration, which allows the hoppings from the $A$ lattices to the intermediate $B$ lattice to interfere out. We refer to this interference as the local hopping constraint, and wave functions obeying this constraint can always be found when the lattice satisfies conditions (ii) and (iii). This shows that the connection of the $A$ lattices via the intermediate $B$ lattices is essential for our problem. Hopping within the $A$ and within the $B$ lattices is allowed and will be elaborated upon in the next section.

We distinguish two types of connectivity in this stacking construction. In the first, the intermediate sites on the $B$ lattice are connected to sites in the minimal unit cells on both neighboring $A$ lattices in a symmetric way. In the second case, the intermediate sites on the $B$ lattice are connected differently to the sites in the minimal unit cells of the $A$ lattice below than to the sites in the unit cells in the $A$ lattice above, as shown in the bottom panel of Fig.~\ref{figureallconditions}(d). In either case, there is a natural constraint---zero total hopping amplitude to the $B$ lattice sites---that leads to a bootstrapping procedure and exact eigenstates that can be expressed entirely in terms of the Bloch eigenstates of the $A$ lattice layers. In the symmetric case, the solutions are rather mundane with $|r(\mathbf k)|=1$, while the latter case, where the local constraint necessarily connects multiple minimal unit cells, gives more interesting solutions for $r(\mathbf k)$ including those that correspond to topological surface states. The latter situation is referred to as condition (iv). Note that in this discussion we have assumed that the coupling strength between the intermediate $B$ lattice and the degrees of freedom in the unit cell of the $A$ lattice above and equals the coupling strength between the intermediate $B$ lattice and the degrees of freedom in the unit cell of the $A$ lattice below. We refer to this as the isotropic case. If this coupling were anisotropic, we can find nontrivial solutions for $|r(\mathbf k)|$, which depend on both the crystal momentum $\bfk$ and the strength of the various perpendicular hopping parameters, regardless of whether the local constraint involves multiple unit cells. If we now include pertinent terms in the Hamiltonian such that the system indeed supports a topological phase, as shown in Fig.~\ref{figureallconditions}(e), we find that our wave function solutions describe a topological boundary state. This final condition is referred to as condition (v).

We can thus list five conditions that need to be fulfilled to find topological boundary states.
\begin{enumerate}
\item[(i.)] The lattices are formed by alternating $(d-1)$-dimensional lattices, referred to as $A$ and $B$ lattices, which have periodic boundary conditions. There are open boundary conditions in the direction of stacking and the outermost layers are $A$ lattices [Fig.~\ref{figureallconditions}(a)];
\item[(ii.)] The $A$ lattices are only coupled to each other via intermediate $B$ lattices and cannot directly communicate [Fig.~\ref{figureallconditions}(b)];
\item[(iii.)] The $A$ and $B$ lattices are connected in a geometrically frustrated fashion meaning that there are several inequivalent ways of hopping from the neighboring $A$ lattices to the single orbital in the $B$ lattice unit cell
leading, together with (ii), to the emergence of a local constraint obeyed by the exact solutions in Eq. (\ref{exactstate})
[Fig.~\ref{figureallconditions}(c)]. (In the presence of pertinent symmetries, this can be generalized to several orbitals in the $B$ lattice unit cell.);
\item[(iv.)] There is no way of choosing a minimal unit cell such that the local constraint obeyed by the exact solutions in Eq. (\ref{exactstate}) takes place within a single unit cell on both of the involved $A$ lattices [Fig.~\ref{figureallconditions}(d)]. Alternatively, this condition can be satisfied if the coupling between the $A$ and $B$ lattices is anisotropic;
\item[(v.)] The bulk Hamiltonian supports the pertinent topological phase [Fig.~\ref{figureallconditions}(e)]. 
\end{enumerate}
{\it
Exact wave-function solutions corresponding to $n$ $(d-1)$-dimensional bands can be found whenever conditions (i)-(iii) are fulfilled. Fulfilling condition (iv), the exact solution generically yields exponentially localized boundary states, and whenever the bulk supports a given topological phase, condition (v), the exact solution corresponds to its surface theory.}

\subsection{Generic tight-binding models}
We consider tight-binding models describing noninteracting identical particles on the lattices described above. For the sake of clarity we set $n' = 1$, i.e. we consider intermediate $B$ lattices with a single degree of freedom per unit cell. The Hamiltonian describing a system with $N$ stacked $A$ lattices is written directly in momentum space and reads $H^N ({\bf k}) = \Psi^\dagger ({\bf k}) \mathcal{H}^N_{\bf k} \Psi ({\bf k})$ with $\Psi ({\bf k})= \bigoplus_{s=1}^{l} \psi_{s} (\bfk), \, l \equiv (n + 1) N - 1$, the annihilation operator of an electron in the full lattice, and $\mathcal{H}^N_{\bf k}$ an $((n + 1) N - 1) \times ((n + 1) N - 1)$-dimensional matrix given by
\begin{equation}
\mathcal{H}^N_{\bf k} = \begin{pmatrix}
\mathcal{H}_{\bf k} & \mathcal{H}^A_{\perp} & 0 & 0 & 0 & 0 & 0 \\
\mathcal{H}_{\perp}^{\dagger, A} & h_{{\bf k}} & \mathcal{H}_{\perp}^{\dagger, B} & 0 & 0 & 0  & 0\\
0 & \mathcal{H}_{\perp}^B & \mathcal{H}_{\bf k} & \mathcal{H}^A_{\perp} & 0 & 0 & 0 \\
0 & 0 & \mathcal{H}^{\dagger, A}_{\perp} & h_{{\bf k}} & \mathcal{H}^{\dagger, B}_{\perp} & 0 & 0 \\
0 & 0 & 0 & \mathcal{H}^{B}_{\perp} & \mathcal{H}_{\bf k} & \ldots & 0 \\
0 & 0 & 0 & 0 & \vdots & \ddots & \mathcal{H}^{\dagger, B}_{\perp} \\
0 & 0 & 0 & 0 & 0 & \mathcal{H}_{\perp}^B & \mathcal{H}_{\bf k}
\end{pmatrix}, \label{generalmultilayerhamiltonian}
\end{equation}
where $\mathcal{H}_{\bf k}$ is the $(n \times n)$-dimensional Hamiltonian for the $A$ lattice, $h_{{\bf k}}$ is the $(1 \times 1)$-dimensional Hamiltonian for the intermediate $B$ lattice, and $\mathcal{H}_\perp^\alpha$ is an $(n \times 1)$ matrix connecting the $A$ lattice to the intermediate $B$ lattice. In general, this connecting Hamiltonian can be written as
\begin{equation}
\mathcal{H}_\perp^\alpha = \bigoplus_{s=1}^n t_{\perp,\alpha, s} \, f_{\alpha, s}({\bf k}), \label{generalmultilayerhamiltonianperp}
\end{equation}
where $t_{\perp,\alpha,s}$ is the hopping amplitude from site $s$ in the unit cell of the $A$ lattice to the intermediate $B$ lattice, and $f_{\alpha, s}({\bf k})$ is a ${\bf k}$-dependent phase derived from the local lattice structure. Note that all hopping amplitudes are allowed to be complex, i.e. allowing for spin-orbit coupling as well as commensurate magnetic fields incorporated via Peierls substitution.

\subsection{Exact eigenstates}

Using the Hamiltonian in Eq.~(\ref{generalmultilayerhamiltonian}), we find a subset of solutions to the Schr\"odinger equation, $\mathcal{H}^N_{\bf k} \ket{\Psi_i ({\bf k})} = E_i^N ({\bf k}) \ket{\Psi_i ({\bf k})}, \, i = 1, 2 ..,n$, corresponding to the number of degrees of freedom in the $A$ lattice, given by
\begin{align}
E_i^N ({\bf k}) &= E_i ({\bf k}), \\
\ket{\Psi_i ({\bf k})} & = \mathcal{N}_i ({\bf k}) \bigoplus_{m = 1}^N \left(r_i ({\bf k})\right)^m \ket{\Phi_i ({\bf k})}_m, \label{exactsolutionedgestate}
\end{align}
where $E_i ({\bf k})$ are the eigenvalues of the $A$ lattice Hamiltonian $\mathcal{H}_\bfk$ and $\Phi_i ({\bf k})$ the eigenstates with components $\phi_{i,s} (\bfk), \, s = 1, .. , n$, thereof, $\bfk$ is the $(d-1)$-dimensional momentum,
\begin{equation}
\mathcal{N}_i ({\bf k})= \frac{1}{|r_i({\bf k})|}\sqrt{ \frac{|r_i({\bf k})|^2-1}{(|r_i({\bf k})|^2)^N - 1}} \ ,
\end{equation}
ensures normalization,\footnote{As $|r({\bf k})|\rightarrow 1$, the normalization factor is smooth approaching $\mathcal{N}({\bf k})=1/\sqrt{N}$.} $m$ labels the $A$ lattice layer, and $r_i ({\bf k})$ is a prefactor given by
\begin{equation}
    r_i({\bf k}) = - \frac{\psi_{i,m,s} ({\bf k})}{\psi_{i,m+1,s} ({\bf k})}, \quad \forall \, s \in \{1 \cdots n\}, m \in \{1 \cdots N-1\},\label{equationgeneralrofkintermsoffullwavefct}
\end{equation}
where $\psi_{i,m,s}$ are the components of $\Psi_i({\bf k})$ in the $m$th $A$ lattice on sublattice site $s$. Using that the weight of the wave function on the intermediate site of the $B$ lattice is zero, $r_i({\bf k})$ can also be expressed in terms of the components of $\Phi_i({\bf k})$:
\begin{equation}
r_i({\bf k}) = - \frac{\mathcal{H}_\perp^{\dagger, A} \Phi_{i} (\bfk)}{\mathcal{H}_\perp^{\dagger, B} \Phi_{i} (\bfk)} = - \frac{\sum_{s=1}^{n} t_{\perp,A,s} f^\dagger_{A, s}({\bf k}) \phi_{i,s}({\bf k})}{\sum_{s=1}^{n} t_{\perp,B,s} f^\dagger_{B, s}({\bf k}) \phi_{i, s}({\bf k})}\ , \label{equationrofkintermsofbasishamiltonianeigenstatesandphase}
\end{equation}
where $f_{\alpha, s}({\bf k})$ is given in Eq.~(\ref{generalmultilayerhamiltonianperp}). From this equation, we can formalize condition (iv): this condition is fulfilled when $|r_i({\bf k})| \neq 1$, i.e. $t_{\perp,A,s} f_{A,s} \neq t_{\perp,B,s}f_{B,s}$, and broken when $|r_i({\bf k})| = 1$, i.e. $t_{\perp,A,s} f_{A,s} = t_{\perp,B,s}f_{B,s}$. Note that explicitly calculating $|r_i({\bf k})|$, one can still find $|r_i({\bf k})| = 1$ when $t_{\perp,A,s} f_{A,s} \neq t_{\perp,B,s}f_{B,s}$. This is due to the explicit form of $\phi_{i, s}({\bf k})$ and closely related to topology in the model, condition (v), which is further discussed towards the end of this section. We want to emphasize that the exact solution is completely independent of the Hamiltonian $h_\bfk$ on the intermediate $B$ lattice. It should, however, be mentioned that the remaining $((n + 1) N - (n + 1))$-solutions to the Schr\"odinger equation, which are only numerically accessible, do depend on this Hamiltonian and are subject to deformation by changing the perpendicular hopping strength.

Inspecting the solution in Eq.~(\ref{exactsolutionedgestate}), we notice three properties. First, the solution has zero weight on the intermediate sites, which means that the Hamiltonian $h_{\bf k}$ for the intermediate $B$ lattice is completely irrelevant to the solution and as such can generally include arbitrary terms. Second, only the connectivity of the $A$ lattices via the intermediate $B$ lattices encoded by $t_{\perp,\alpha,s} f_{\alpha, s}({\bf k})$ is relevant. The coupling between $A$ lattices may differ in strength effectively yielding strongly or weakly coupled layers. Third, we can now understand why satisfying condition (iv) leads to boundary states. If $|r_i({\bf k})|=1$, the weight of the wave function on layer $m$,
\begin{equation}
P_{i,m}({\bf k}) = \left|\mathcal{N}_i(\bfk)\right|^2 |r_i (\bfk)|^{2m}, \label{equationweightofwavefctoneachlayer}
\end{equation}
is the same for all $m$, and the wave function is equally localized on each $A$ lattice. However, if $|r_i({\bf k})|\neq1$, the eigenstate will localize to one of the boundaries. If $|r_i ({\bf k})|<1$, $P_{i,m}({\bf k})$ decreases with increasing $m$ and the wave function is strongly localized on the first layer $m=1$ corresponding to the surface on one side. When $|r_i ({\bf k})|>1$, $P_{i,m}({\bf k})$ increases with increasing $m$ and the state is localized on the last layer $m = N$ corresponding to the surface on the other side. Therefore, when $|r_i ({\bf k})|\neq1$, we have exponentially localized boundary states with a localization length, and the solution in Eq.~(\ref{exactsolutionedgestate}) thus corresponds to the solution for the boundary state. Now, if $|r_i ({\bf k})|$ also has a $\bfk$-dependent structure the boundary state can switch surfaces, which for a three-dimensional material means a constant energy contour represents a Fermi arc. When discussing the exponential surface localization we will make use of the localization length $\xi({\bf k}) = ({\rm ln}|r({\bf k})|)^{-1}$.

Satisfying condition (iv) means we have found a suitable geometry for the lattices to find boundary states. However, when one plugs the solution to the eigenfunctions into Eq.~(\ref{equationrofkintermsofbasishamiltonianeigenstatesandphase}), we may still find $|r_i({\bf k})| = 1$ for systems deep in the topologically trivial regime. In the trivial regime it is also possible to find unprotected, weakly-localized boundary states for which $|r_i({\bf k})|$ has a nontrivial structure. Therefore, we need to impose a fifth condition, condition (v), that introduces nontrivial topology in the models such that we for instance obtain a Chern insulator and Weyl semimetal, which are examples of two- and three-dimensional models, respectively. One needs to minimally break time-reversal symmetry to find a Chern insulator, such that in two dimensions, the Hamiltonian should include at least one such term. To obtain a Weyl semimetal, one could either break inversion or time-reversal symmetry. In the cases studied in this work, we break the latter symmetry by turning the $A$ lattice into a Chern insulator. In section~\ref{sectiontwof}, we present an argument to understand why this leads to an eigenstate that switches surfaces.

It is worth emphasizing that the exact wave function solution directly enables the computation of correlation functions within the surface bands which are otherwise only numerically or approximately tractable. For example, the expectation value of any operator $\mathcal{A}$, which acts the same within each layer, reads $\left<\mathcal{A}\right> = |\mathcal{N}(\bfk)|^2 \sum_{m = 1}^{N} |r_i (\bfk)|^{2m} \bra{\Phi_i ({\bf k})} \mathcal{A} \ket{\Phi_i ({\bf k})}$. This expression can readily be extended to more complicated, layer dependent, operators.

Diagonalizing the Hamiltonian in Eq.~(\ref{generalmultilayerhamiltonian}) leads to the band spectrum of the full system with $((n+1)N - 1)$ bands, which are divided into $(n + 1)$-bulk parts separated by band gaps. When conditions (i) - (v) are met, boundary states are present, which can be identified in the bulk spectrum as bands crossing a gap and connecting two bulk parts. 

It should be noted that, while the exact solutions remain unchanged, all other eigenstates change while deforming the coupling to the $B$ lattices. For instance, as we will demonstrate in Section \ref{sectionthree}, increasing the inter-layer coupling strength drives a transition between the quasi-two-dimensional layered Chern insulators phase and a truly three-dimensional Weyl semimetal regime. 

\subsection{Attachment of bulk bands and surface switching}

It is a generic property of our exact solutions that in part of the surface Brillouin zone they attach to bulk bands in the limit of many stacked layers. This happens precisely at those points ${\bf q}$  in the $(d-1)$-dimensional (surface) Brillouin zone where the boundary state connects to the bulk where $|r({\bf q})| = 1$, i.e. where the penetration depth $\xi$ diverges. Below we will provide a variational argument that shows that the gap---either from below or from above---vanishes as $N^{-2}$ for large $N$. In contrast, whenever the penetration is finite there is a gap to neighboring bands also in the limit $N\rightarrow \infty$. 

There are two types of penetration depth divergences and concomitant bulk band attachments. First, at ${\bf q}=0$ it immediately follows that $|r({\bf q=0})| = 1$ independently of model details. 
Second, many models feature $(d-2)$-dimensional families of such ${\bf q}$ points, typically along high symmetry paths cutting through the surface Brillouin zone. The existence of these motifs depend on details of the lattice geometry and the tight-binding Hamiltonian.  
In particular, the boundary states may switch surfaces at ${\bf q}$, i.e. ${\rm sign}[{\rm ln}|r_i({\bf q - \epsilon k})|] \neq {\rm sign}[{\rm ln}|r_i({\bf q + \epsilon k})|]$ as $\epsilon \rightarrow 0$. Notably, $(d-2)$-dimensional families of such ${\bf q}$ points necessarily exist for topologically nontrivial models although they can also occur in topologically trivial models. 

It is intuitively plausible that the energy gap between the boundary states and the bulk bands at the points ${\bf q}$ should disappear. We can explicitly demonstrate that this is indeed the case by introducing a class of ansatz wave functions describing bulk state with nearly the same energy as the exact solution $|r({\bf q})| = 1$. In this case, the exact solution to the low-energy boundary state can be written as
\begin{equation}
\ket{\psi ({\bf q})} = \frac{1}{\sqrt{N}} \bigoplus_{m = 1}^N {\rm e}^{i \alpha(\bfq) m} \ket{\phi ({\bf q})}_m. \label{equationexactsolutionforrofkequal1}
\end{equation}
We can now make an ansatz for a class of states expected to be close in energy,
\begin{equation}
\ket{\chi^0 ({\bf q})} =\frac{1}{\sqrt{N}} \bigoplus_{m = 1}^N {\rm e}^{i \tilde{\alpha}(\bfq) m} \ket{\phi ({\bf q})}_m, \label{equationfirstansatznoteigenfunction}
\end{equation}
which has to be orthogonal to the exact solution in Eq.~(\ref{equationexactsolutionforrofkequal1}), such that we find
\begin{equation}
\alpha(\bfq) - \tilde{\alpha}(\bfq) = \frac{2 \pi}{N} p, \quad p = 1, 2, ..., N-1.
\end{equation}
This leads to
\begin{equation}
{\rm e}^{i \tilde{\alpha}(\bfq)m} = {\rm e}^{i (\alpha (\bfq) - 2 \pi p/N) m}.
\end{equation}
However, the trial wave function in Eq.~(\ref{equationfirstansatznoteigenfunction}) is not an eigenfunction of the Hamiltonian because it has zero weight on the intermediate site of the $B$ lattice, which it should not have. 

Therefore, the trial wave function can be made an eigenstate by mixing in the state for the intermediate site $\ket{\tilde{\chi} ({\bf q})}$, which results in altering the entries given in $\ket{\phi ({\bf q})}$, such that we can create an approximate eigenstate, which becomes exact in the limit of large N:
\begin{equation}
\ket{\chi ({\bf q})} =\frac{1}{\sqrt{N}} \bigoplus_{m = 1}^N {\rm e}^{i \tilde{\alpha}(\bfq) m} \ket{\phi ({\bf q})}_m + \frac{1}{\sqrt{N}} \bigoplus_{m=1}^N a_m \ket{\tilde{\chi} ({\bf q})}_m,
\end{equation}
where
\begin{equation}
a_m \sim - \frac{t_\perp \, p}{N}.
\end{equation}
Therefore, we find that the energy difference between the two states behaves as
\begin{equation}
\Delta \sim - \frac{t_\perp^2 p^2 + \mathcal{O} (t_\perp^4)}{N^2}
\end{equation}
for large $N$. Therefore, the energy difference between the exact solution and our variational bulk state disappears as $N^{-2}$ at $\bfk = {\bf q}$ for large $N$. In fact, there will be many such states as signaled by the family of states constructed (varying $p$). Note, however, that depending on details, in particular the strength of the inter-layer tunneling $t_\perp$, the variational state may be lower or higher in energy. For weak $t_\perp$, the variational state is always lower in energy than the exact solution. However, for stronger $t_\perp$, this can change in the Brillouin zone, as is strikingly manifested in the case of Weyl points at which these energies are equal and a sign change of $\Delta$ takes place.

\subsection{Berry curvature and surface state topology} \label{sectiontwof}
The appearance of boundary states is closely related to nontrivial topology in the bulk of a material. For Chern insulators and Weyl semimetals, this bulk topology is manifested by a nonzero Chern number. The Chern number for an isolated band $\ket{\rho_s({\bf k})}$ of a two-dimensional periodic lattice is computed by integrating the Berry curvature over the Brillouin zone:
\begin{equation}
C_s = \frac{1}{2 \pi} \int_{BZ} \mathcal{F}_{xy, s} ({\bf k}) \, {\rm d}^2 k, \label{equationchernnumber}
\end{equation}
where $\mathcal{F}_{ij, s} ({\bf k})$ is the Berry curvature given by $\mathcal{F}_{ij,s} ({\bf k}) = \partial_{k_i} A_{j,s} ({\bf k}) - \partial_{k_j} A_{i,s} ({\bf k})$ with $A_{j,s} ({\bf k}) = - i \bra{\rho_s({\bf k})} \partial_{k_j} \ket{\rho_s({\bf k})}$ the Berry connection. The total Chern number of any system has to be zero, i.e. $\sum_{s} C_s = 0$. While correlation functions are easily calculated, the derivatives entering the Berry curvature complicate analytical calculations thereof. Alternatively, this problem can be seen from the fact that a (derivative-free) expression of the Berry curvature involves all energy eigenstates of the model---not just the solvable surface bands. Nevertheless, a number of instructive results can be derived. 

For (quasi-)three-dimensional models we use the solution to the surface state in Eq.~(\ref{exactsolutionedgestate}) such that one can write the Berry curvature of a system with $N$ two-dimensional $A$ lattices as
\begin{align}
    &\mathcal{F}_{N,xy,i} = \mathcal{F}_{1,xy,i}  -i \, F\left(N, r_i({\bf k}) \right) \nonumber \\
& \times \left\{\left[\partial_{k_x} r^*_i({\bf k}) \right] \left[\partial_{k_y} r_i({\bf k}) \right] - \left[\partial_{k_y} r_i^*({\bf k}) \right] \left[\partial_{k_x} r_i({\bf k}) \right] \right\}, \label{equationberrycurvature}
\end{align}
with
\begin{align*}
& \mathcal{F}_{1,xy,i} = -i \left[\left(\partial_{k_x} \bra{\Phi_i({\bf k})}\right) \partial_{k_y} \ket{\Phi_i({\bf k})} \right. \nonumber \\
& \left. - \left(\partial_{k_y} \bra{\Phi_i({\bf k})}\right) \partial_{k_x} \ket{\Phi_i({\bf k})} \right], 
\end{align*}
being the Berry curvature of a single $A$ layer and
\begin{align}
F\left(N, r_i({\bf k}) \right) &\equiv \left(1-\left|r_i({\bf k})\right|^2\right)^{-2} \label{equationprefactorberry} \\
& - N^2 \left|r_i({\bf k})\right|^{2N-2} \left(1-\left|r_i({\bf k})\right|^{2N}\right)^{-2}. \nonumber
\end{align}
We emphasize that $F\left(N, r_i({\bf k}) \right)$ contains the full $N$-dependence of the Berry curvature in Eq.~(\ref{equationberrycurvature}). In the limit $|r_i(\bfk)| \rightarrow 1$, Eq.~(\ref{equationprefactorberry}) reduces to $(N^2 - 1)/12$, thus for large $N$ the Berry curvature exhibits a peak scaling with $N^2$ at those parts in the Brillouin zone where the exact solution is not a surface state but is completely delocalized over all layers. In contrast, whenever $|r_i(\bfk)| \neq 1$, the Berry curvature saturates as a function of $N$ consistent with the exponential localization of the wave functions to the surface layers.  

The Chern number $C_{N,i}$ of the solvable bands in a system with $N$ $A$ lattices can be found upon integrating the Berry curvature in Eq.~(\ref{equationberrycurvature}) over the Brillouin zone as shown in Eq.~(\ref{equationchernnumber}). When $|r_i({\bf k})| = 1$ everywhere, the derivative over the second part of Eq.~(\ref{equationberrycurvature}) yields zero such that the total Chern number equals that of one layer $C_{1, i}$. However, in the generic situation when $|r_i({\bf k})| \neq 1$, we observe in our examples that the total Chern number grows with $N$ as
\begin{equation}
C_{N,i, |r| \neq 1} = N \, C_{1,i}, \label{equationchernnumberscaling}
\end{equation}
which means that the Chern number that is associated with each $A$ lattice is absorbed into the surface state.

Furthermore, it can be shown that a finite Chern number on the $A$ lattice implies Fermi arcs by considering a generic two-band model on the $A$ lattice whose Hamiltonian reads
\begin{equation}
\mathcal{H}_\bfk = {\rm d}(\bfk) \cdot \boldsymbol\sigma + d_0(\bf k) \sigma_0,
\end{equation}
where $\boldsymbol\sigma$ are the Pauli matrices, $\sigma_0 = \mathbb{I}_{2 \times 2}$, and 
\begin{equation}
\mathcal{P}_\pm(\bfk) = \frac 1 2\left( \sigma_0 \pm\hat{\bf d}(\bfk)\cdot \sigma\right),
\end{equation}
projects on to the upper (+) and lower (-) bands, respectively, with $\hat{\bf d}(\bfk) = {\bf d}(\bfk)/|{\bf d}(\bfk)|$. Using the projector, the Chern number can be written in terms of $\hat{\bf d}(\bfk)$ as
\begin{equation}
C = \frac{1}{4 \pi} \int {\rm d}k_x \int {\rm d}k_y \, \hat{\bf d} (\bfk)\cdot \left(\frac{\partial \hat{\bf d}(\bfk)}{\partial k_x} \times \frac{\partial \hat{\bf d}(\bfk)}{\partial k_y} \right).
\end{equation}
The Chern number can thus be interpreted as the number of times $\hat{\bf d}(\bfk)$ wraps the unit sphere. Provided the general structure of $r_i(\bfk)$ one can show that, for a very generic class of models, this implies that $\hat{\bf d}(\bfk)$ renders both the numerator and denominator of $r_i(\bfk)$ given in Eq.~(\ref{equationrofkintermsofbasishamiltonianeigenstatesandphase}) to vanish at {\it different} $\bfk$ whenever the single layer Chern number is finite. Thus $r_i(\bfk)$ has zeros and infinities when the $A$ lattice is a Chern insulator implying surface switching and the existence of Fermi arcs in the sense that the surface state is entirely localized at the top and bottom layer at different points in the surface Brillouin zone. An explicit example of this is provided in section~\ref{sectionfour}.

\section{Examples}\label{sectionthree}

\subsection{One dimension: end modes on a chain} \label{sectionthreezero}

\begin{figure}[h]
  \centering
  \adjustbox{trim={0.\width} {.2\height} {0.\width} {.5\height},clip}
  {\includegraphics[width=0.5\textwidth]{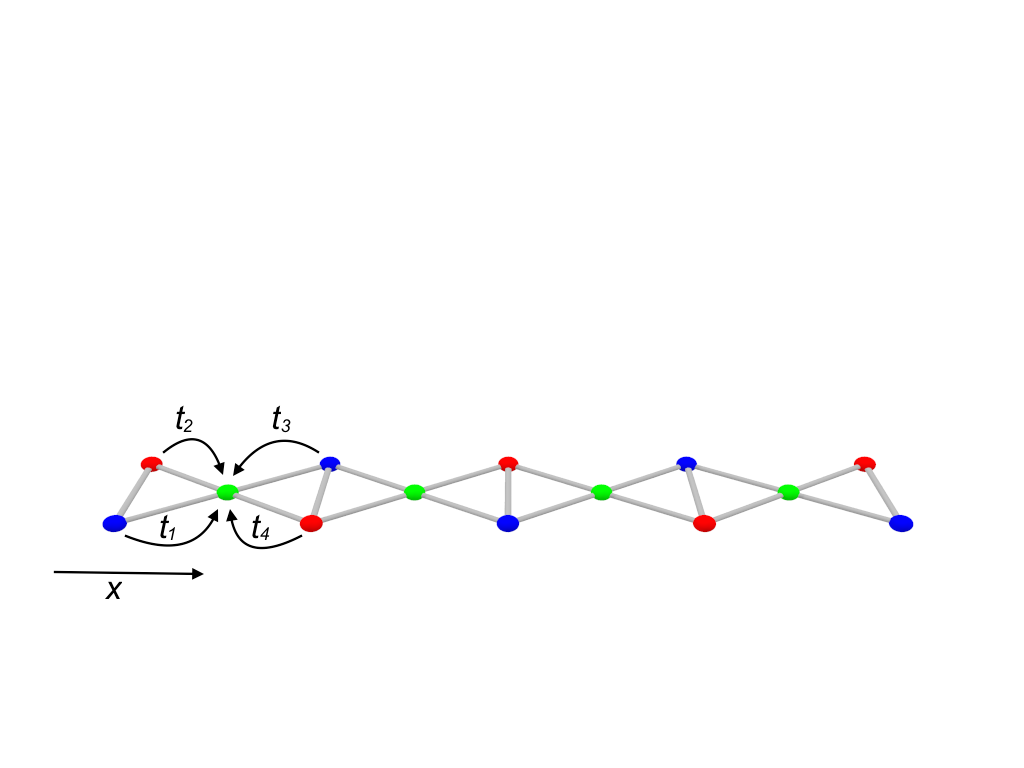}}
\caption{The chain model featuring a two-band model (red and blue sites) stacked on top of each other with a single (green) site in between.}
\label{figureonedchain}
\end{figure}

As a simple warmup, we start by studying a one-dimensional system, which has two degrees of freedom in the $A$ lattice and one degree of freedom in the intermediate $B$ lattice shown in Fig.~\ref{figureonedchain}. This chain readily satisfies conditions (i)-(iii), and the absence of a surface momentum parameter means that condition (iv) can only be satisfied if the $A$ and intermediate $B$ lattices are coupled in an anisotropic fashion.

In its most generic form, the Hamiltonian for each $A$ lattice reads $H = \Phi^\dagger \mathcal{H}^{\rm 1D} \Phi$, where $\Phi$ is the annihilation operator of an electron in the $A$ lattice and $\mathcal{H}^{\rm 1D}$ can be written in the Dirac form
\begin{equation}
\mathcal{H}^{\rm 1D} = {\bf d} \cdot \boldsymbol\sigma + d_0 \sigma_0,\label{standardformulahamiltonian1d}
\end{equation}
with the energy eigenvalues
\begin{equation}
E_\pm = \pm|{\bf d}| + d_0,
\label{standardformulaenergy1d}
\end{equation}
and eigenstates $\ket{\Phi_\pm}$. Due to the lack of a momentum parameter, the eigenvalues and the amplitudes of the eigenstates are constants. The Hamiltonian for the one-dimensional chain is given in Eqs.~(\ref{generalmultilayerhamiltonian}) and (\ref{generalmultilayerhamiltonianperp}) with the phases $f_{\alpha,s} = 1\, \forall \, \alpha, \, s$, and we set the intermediate $B$ lattice Hamiltonian $h = 0$ and $t_{\perp, A, s} = t_s$, and $t_{\perp, B, s} = t_{s+2}$ with $t_{\perp, \alpha, s}  \in \mathbb{R} \, \forall \, \alpha, \, s$. Using Eq.~(\ref{equationrofkintermsofbasishamiltonianeigenstatesandphase}), we find 
\begin{equation} r_\pm (t_s) = - \frac{t_1 \phi_{\pm, 1} + t_2 \phi_{\pm, 2}}{t_3 \phi_{\pm, 1} + t_4 \phi_{\pm, 2}},\end{equation} which is a function of $t_s$. When $t_1 = t_3$ and $t_2 = t_4$, $r_\pm = -1$ and according to Eq.~(\ref{equationweightofwavefctoneachlayer}) the wave function in Eq.~(\ref{exactsolutionedgestate}) has equal weight on each $A$ lattice $m$. However, when $t_1 \neq t_3$ and/or $t_2 \neq t_4$, we find $|r_\pm| \neq 1$ and there are end modes on the chain. These modes either reside at the same end, e.g. at $t_1 = 10 \, t_3$ and $t_2 = 10 \, t_4$ yields $r_\pm =- 10$ hence both end modes are exponentially localized around the $A$ lattice $m = N$---or they live at opposite ends, e.g. when the Hamiltonian for the $A$ lattice in Eq.~(\ref{standardformulahamiltonian1d}) reads ${\bf d} = (V, \, 0, \, 0)$ and we require the perpendicular hopping parameters to satisfy $t_1 + t_2 < t_3 + t_4$ and $t_1 - t_2 > t_3 - t_4$ yielding $|r_+| < 1$ and $|r_-| > 1$ such that the end modes $\Psi_+$ and $\Psi_-$ appear at $m = 1$ and $m=N$, respectively. The exactly obtained end modes of the chain thus switch ends as a function of the perpendicular hopping parameters $t_s$.


\begin{figure*}[t]
  \centering
  \subfigure[]{
  \adjustbox{trim={0.05\width} {.235\height} {0.05\width} {.3\height},clip}
  {\includegraphics[scale=0.25]{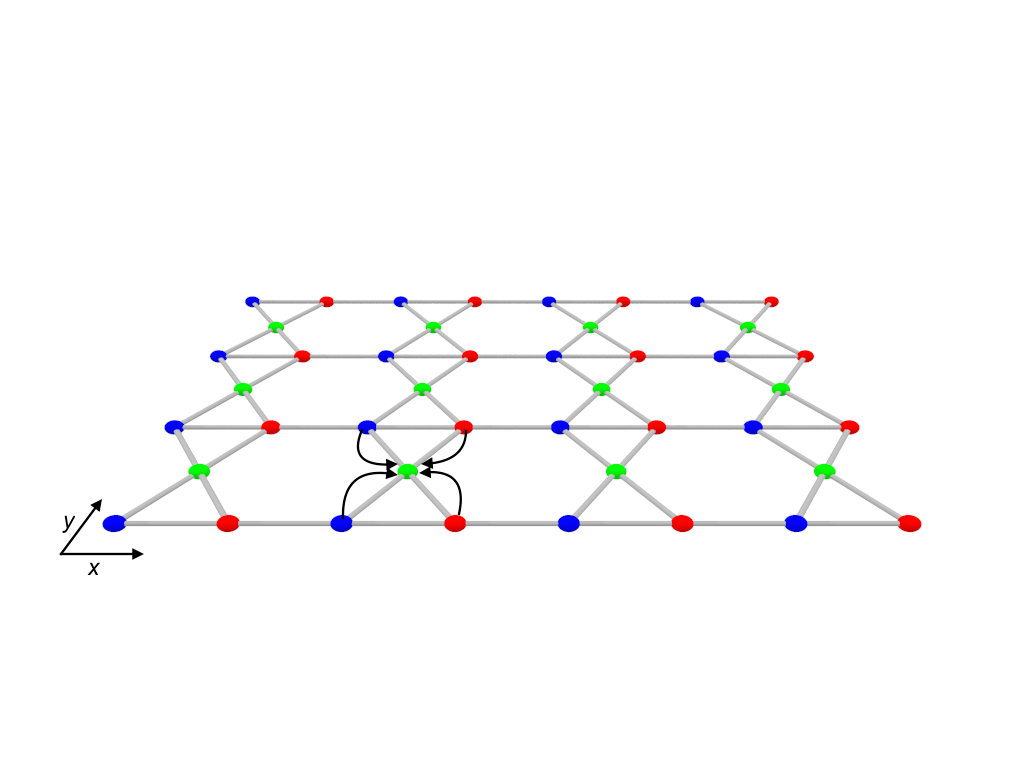}}
  \label{figurechainmodelsa}}\quad
  \subfigure[]{
  \adjustbox{trim={0.05\width} {.235\height} {0.05\width} {.3\height},clip}
  {\includegraphics[scale=0.25]{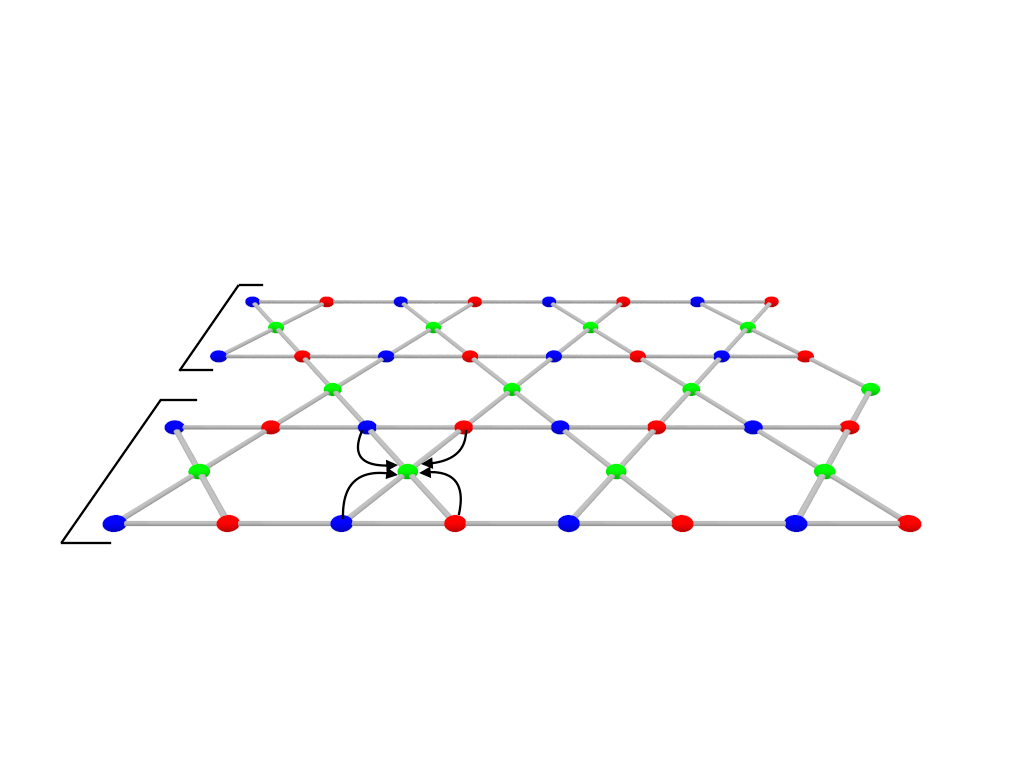}}
  \label{figurechainmodelsb}}\quad
  \subfigure[]{
  \adjustbox{trim={0.05\width} {.235\height} {0.05\width} {.3\height},clip}
  {\includegraphics[scale=0.25]{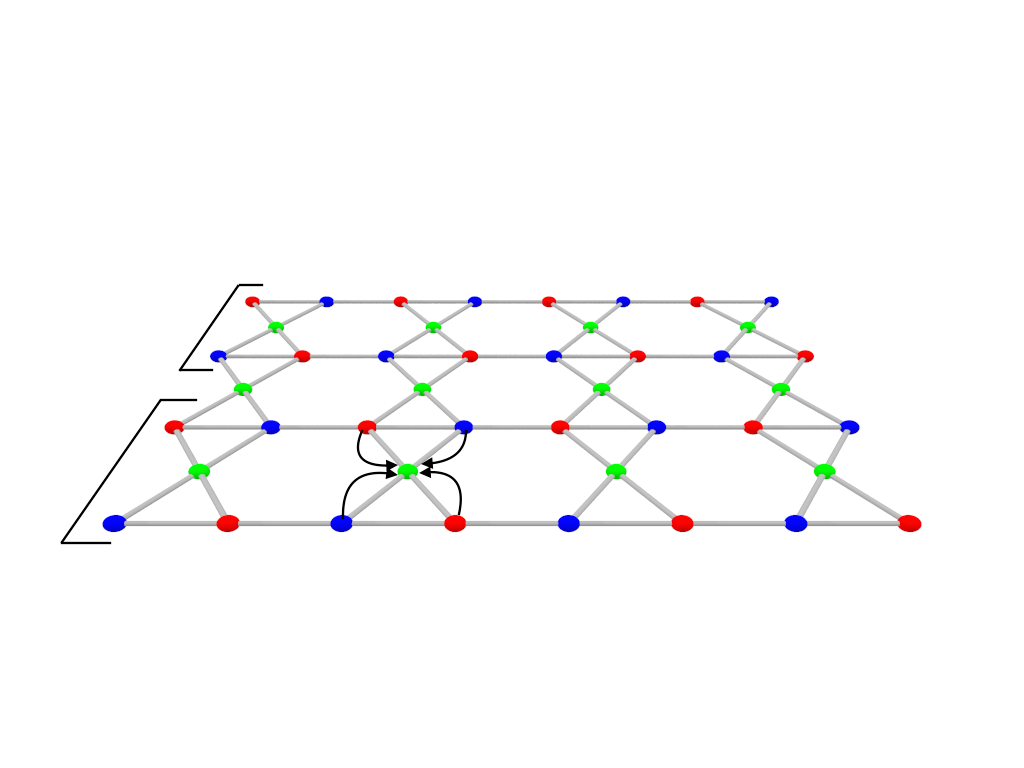}}
  \label{figurechainmodelsc}}\quad
  \subfigure[]{
  \adjustbox{trim={0.05\width} {.235\height} {0.05\width} {.3\height},clip}
  {\includegraphics[scale=0.25]{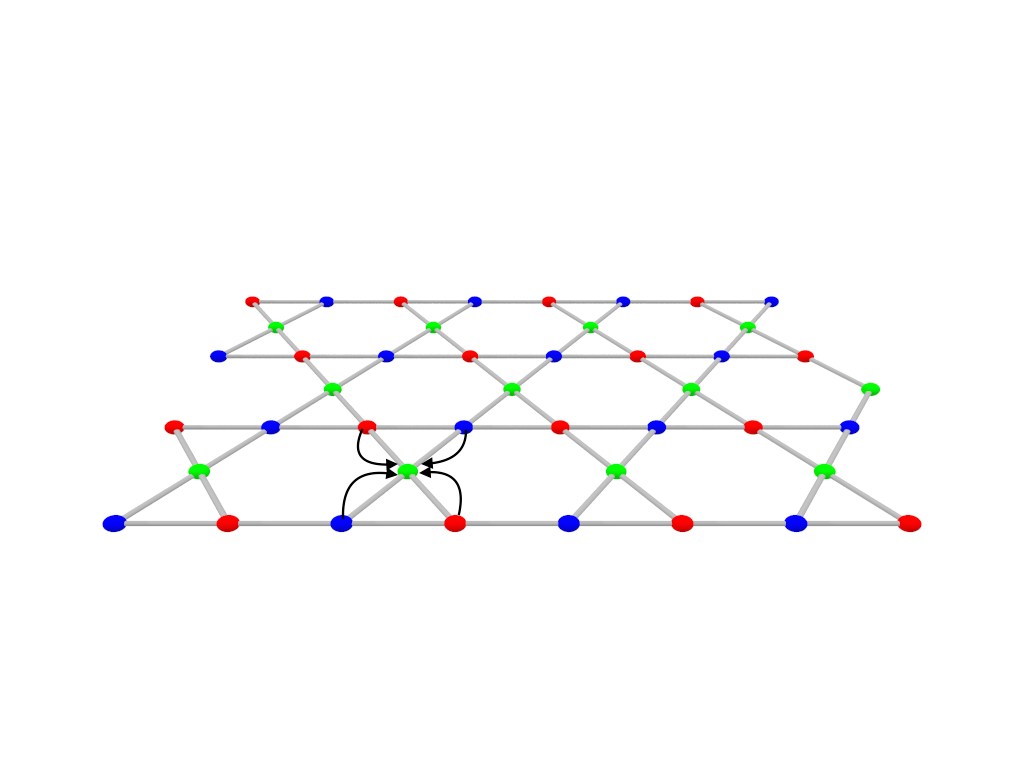}}
  \label{figurechainmodelsd}}
  \caption{Four ways to stack two different chains, a chain with a two-site unit cell in red and blue and a chain with a one-site unit cell in green. If the models were fully periodic, the lattices in (a) and (d) would have three sites in the unit cell, whereas the lattices in (b) and (c) would contain six sites in the unit cell.
  In (a) and (d), the two-site chains correspond to the $A$ lattices. In (b) and (c), the $A$ lattices are formed by a composite of three chains as is indicated by the black brackets. The black arrows illustrate the hopping from the $A$ lattice to the intermediate $B$ lattice, which leads to a zero-total hopping amplitude to the intermediate (green) site. 
  }
  \label{figurechainmodels}
\end{figure*}

\subsection{Two dimensions: Chern and quantum spin Hall insulators on kagome and related lattices} \label{sectionthreeA}

In this section, we study two-dimensional lattice models by stacking (periodic) chains containing two sites in the lattice unit cell (red and blue in Fig.~\ref{figurechainmodels}), which are connected via an intermediate chain with a single site per unit cell (green). The two types of chains are stacked in an alternating fashion in such a way that the lattice geometry is frustrated and a local hopping constraint is naturally realized, such that conditions (i)-(iii) are fulfilled. The four different stacking possibilities are shown in Fig.~\ref{figurechainmodels}, where the lattices in Figs.~\ref{figurechainmodelsa} and \ref{figurechainmodelsd} are related to the lattices in Figs.~\ref{figurechainmodelsc} and \ref{figurechainmodelsb}, respectively, via sublattice exchange in every other two-site chain.

The $A$ lattice for the models in Figs.~\ref{figurechainmodelsa} and \ref{figurechainmodelsd} are the two-site periodic chains in red and blue, and the intermediate $B$ lattice is the one-site periodic chain in green. For the lattices in Figs.~\ref{figurechainmodelsb} and \ref{figurechainmodelsc}, however, 
one has to consider the composite of three chains as the $A$ lattice indicated by the black brackets in the figures. The intermediate lattice is the one-site green chain in between. We treat all four models in detail and find a subset of solutions to the Schr\"odinger equation for each of them. First, we will show that when condition (iv) is not fulfilled, as is the case for the lattices in Figs.~\ref{figurechainmodelsa}, \ref{figurechainmodelsb} and \ref{figurechainmodelsc}, the system remains topologically trivial, signaled by an absence of edge states and vanishing topological invariants, regardless of the microscopic Hamiltonian (as long as it is local in the sense of condition (ii)). Then, by considering models living on the lattice displayed in Fig.~\ref{figurechainmodelsd}, which does fulfill condition (iv), we illuminate the relevance of condition (v). Strikingly, we find that whenever the system has a bulk band characterized by a unit Chern number our exact solutions describe the chiral edge states of the model.

Let us now proceed to demonstrate what is advertised in the preceding paragraph by considering a generic description of translation invariant tight-binding models on the aforementioned lattices. The Hamiltonian for each chain with two sublattices in red and blue is $H(k_x) = \Phi^\dagger(k_x) \mathcal{H}^{\rm ch}_{k_x} \Phi(k_x)$, where $\Phi$ is the annihilation operator of an electron in the $A$ lattice and
\begin{equation}
\mathcal{H}^{\rm ch}_{k_x} = {\bf d}(k_x) \cdot \boldsymbol\sigma + d_0(k_x) \sigma_0 . \label{standardformulahamiltonian}
\end{equation}
The corresponding energy eigenvalues are given by
\begin{equation}
E_\pm(k_x) = \pm|{\bf d}(k_x)| + d_0(k_x), \label{standardformulaenergy}
\end{equation}
and $\ket{\Phi_\pm(k_x)}$ are the eigenstates. In all four cases, the Hamiltonian for the full models is given in Eqs.~(\ref{generalmultilayerhamiltonian}) and (\ref{generalmultilayerhamiltonianperp}), and we set $h_{k_x} = 0$ and $t_{\perp, \alpha, s} = t_\perp \in \mathbb{R} \, \forall \, \alpha, \, s$. The latter can be interpreted as a gauge choice, and does not impede our general approach: one can always choose the perpendicular hopping parameter $t_\perp$ to be real by suitably redefining $H(k_x)$ to account for the ``flux" through each closed path of the lattice [\onlinecite{redderuhrig}]. 

 \begin{figure*}[t]
   \adjustbox{trim={0.\width} {.0\height} {0.1\width} {.0\height},clip}
  {\includegraphics[width=\textwidth]{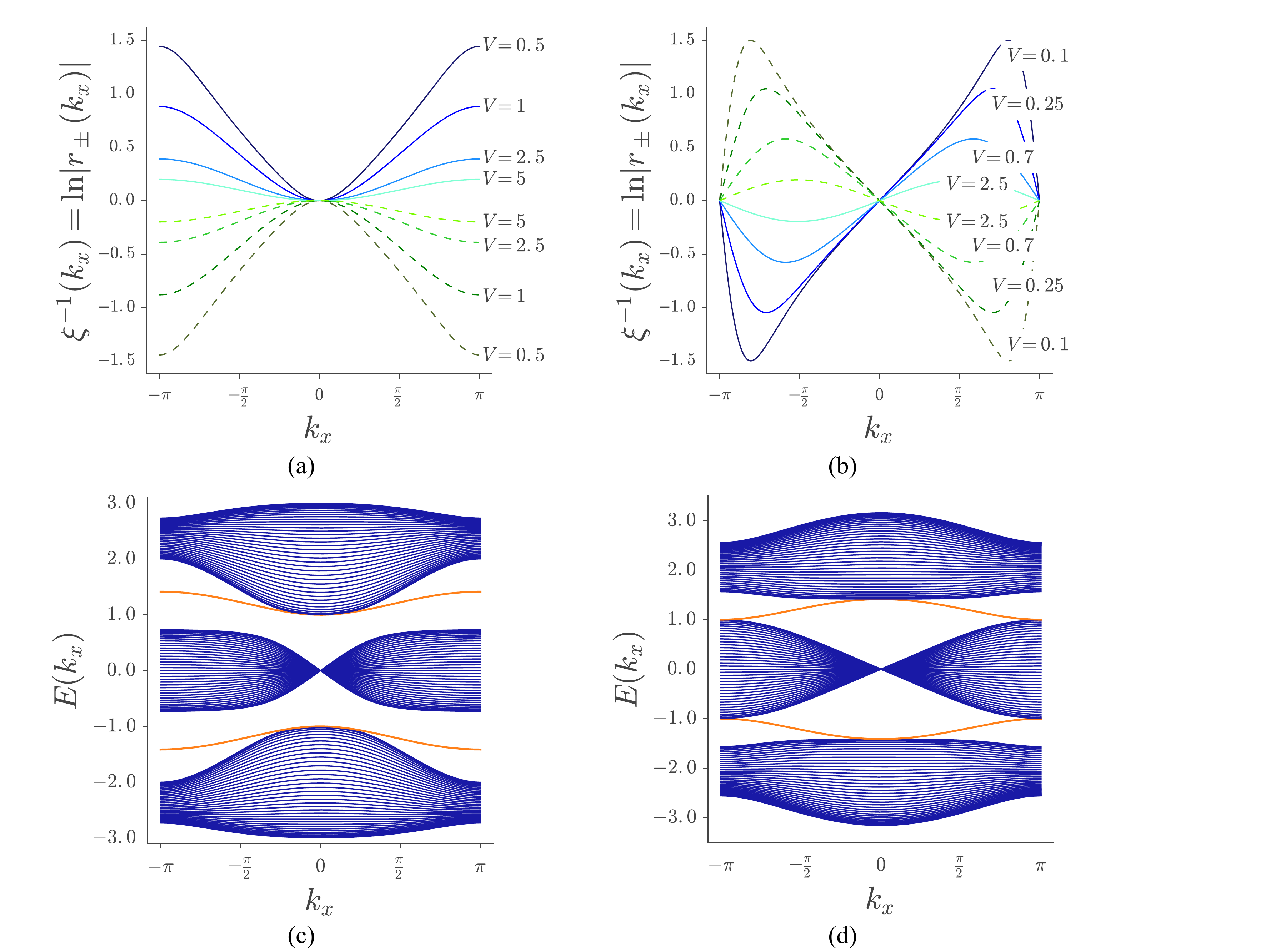}}
  \caption{Plots for the kagome model in Fig.~\ref{figurechainmodelsd} with the chain Hamiltonian given in Eqs.~(\ref{standardformulahamiltonian}) and (\ref{generalhamiltonianchains}). (a) and (b) show the inverse localization length ${\rm ln}|r_\pm (k_x)|$ as given in Eq.~(\ref{equationrofkxforthelayerlatticetrivialplusmodel}). (c) and (d) show the energy spectrum with $N = 40$ and the energy is plotted in units of $t'_1$ and $t'$, respectively. The blue and green lines in (a) and (b) correspond to the $+$ and $-$ solution of the wave function, respectively, and top and bottom orange lines in (c) and (d) depict the $+$ and $-$ energy solutions, respectively. (a) and (c) are plotted for $t/t'_1= t_1/t'_1 = t'/t'_1 = 0$, $t_\perp/t'_1 = 1$, and varying values of $V$ in units of $t'_1$ in (a), and $V/t'_1 = 1$ in (c). (b) and (d) are plotted for $t/t' = t_1/t' = t'_1/t' = 0$, $t_\perp/t' = 1$, and varying values of $V$ in units of $t'$ in (b), and $V/t' = 1$ in (d). (a) and (c) correspond to a topologically trivial system whereas (b) and (d) reveal that the system is a Chern insulator, where we find that the right movers are localized on the edge $m = 1$ and the left movers on the edge $m=N$. (a) and (b) also correspond to ${\rm ln}|s_\pm (k_x)|$ for the $A$ lattice of the model in Fig.~\ref{figurechainmodelsc} with $h_{k_x}^{\rm ch} = 0$ and the other parameters as mentioned before.}
  \label{figurekagomemodelforbrokensublatticesymm}
\end{figure*}

We first focus on the model in Fig.~\ref{figurechainmodelsa}. The $A$ lattice Hamiltonian $\mathcal{H}_{k_x}$ is given by Eq.~(\ref{standardformulahamiltonian}), and the pertinent phases are $f_{A,1}(k_x) = f_{B,1}(k_x) = {\rm exp}(-i k_x/4)$ and $f_{A,2}(k_x) = f_{B,2}(k_x) = {\rm exp}(i k_x/4)$ such that by using Eq.~(\ref{equationrofkintermsofbasishamiltonianeigenstatesandphase}) we immediately find that $r_\pm(k_x)= -1$ and condition (iv) is thus not fulfilled. The weight of the wave function on each individual chain given in Eq.~(\ref{equationweightofwavefctoneachlayer}) is thus equal for each chain $m$, which labels the $A$ lattice, meaning that the state is fully delocalized. We thus expect to find a topologically trivial system. Indeed, regardless of any details of ${\bf d}(k_x)$ for the $A$ lattice Hamiltonian, the Chern number remains zero.

Next, we look at the system in Fig.~\ref{figurechainmodelsb}. The suitably redefined $A$ lattice Hamiltonian now accounts for the five sites in the unit cell, hence we can find exact expressions for five edge state bands. The $A$ lattice Hamiltonian $\mathcal{H}_{k_x}$ is given by
\begin{equation*}
\mathcal{H}_{k_x} = \left(\begin{array}{ccccc}
\multicolumn{2}{c}{\multirow{2}{*}{$\mathcal{H}_{k_x}^{\rm ch}$}} & t_{\perp} {\rm e}^{-i \frac{k_x}{4}} & \multicolumn{2}{c}{\multirow{2}{*}{0}} \\
& & t_{\perp} {\rm e}^{i \frac{k_x}{4}} & & \\
t_{\perp} {\rm e}^{i \frac{k_x}{4}} & t_{\perp} {\rm e}^{-i \frac{k_x}{4}} & h^{\rm ch}_{k_x} & t_{\perp} {\rm e}^{i \frac{k_x}{4}} & t_{\perp} {\rm e}^{-i \frac{k_x}{4}} \\
\multicolumn{2}{c}{\multirow{2}{*}{0}} & t_{\perp} {\rm e}^{-i \frac{k_x}{4}}  & \multicolumn{2}{c}{\multirow{2}{*}{$\mathcal{H}_{k_x}^{\rm ch}$}} \\
& & t_{\perp} {\rm e}^{i \frac{k_x}{4}} & &  \end{array}\right),
\end{equation*}
and the concomitant phase factors are $f_{A,4}(k_x) = f_{B,1}(k_x) = {\rm exp}(i k_x/4)$, $f_{A, 5}(k_x) = f_{B,2}(k_x) = {\rm exp}(-i k_x/4)$ and $f_{A,\alpha}(k_x) = 0, \, \alpha = 1, 2, 3$ and $f_{B,\alpha'}(k_x) = 0, \, \alpha'= 3, 4, 5$. Before analyzing the five solutions to the Schr\"odinger equation for the full system, we first take a closer look at the solution for the $A$ lattice Hamiltonian $\mathcal{H}_{k_x}$. We observe that we can interpret the $A$ lattice as existing out of two sub-$A$ lattices, the two-site chains in red and blue, and a sub-intermediate $B$ lattice, the green sites. We thus find $n_{\rm sub} = 2$ solutions of the following form, which look similar to the solution in Eq.~(\ref{exactsolutionedgestate}):
\begin{equation}
\ket{\Phi_\pm (k_x)} \doteq \ \tilde{\mathcal{N}}_\pm(k_x) \begin{pmatrix}
\phi_{\pm,1} (k_x) \\
\phi_{\pm,2} (k_x) \\
0 \\
s_\pm (k_x)\phi_{\pm,1} (k_x) \\
s_\pm (k_x) \phi_{\pm,2} (k_x) \\
\end{pmatrix}, \label{equationeigenfunctionlayerlatticetrivialplusmodel}
\end{equation}
where $\tilde{\mathcal{N}}_\pm(k_x)$ is the normalization factor, $\ket{\Phi_{\pm} (k_x)}$ are the eigenstates of the two-site chain, and $s_\pm (k_x) = -1$, which can be straightforwardly derived. The corresponding eigenvalues are given in Eq.~(\ref{standardformulaenergy}). Plugging this and the phases into Eq.~(\ref{equationrofkintermsofbasishamiltonianeigenstatesandphase}) yields $r_\pm (k_x) = -1$ such that we retrieve a system in which the wave functions are completely delocalized. For the remaining three solutions, we find $\Phi_i ({\bf k}) = \bigoplus_{s=1}^5 \phi_{i,s} (\bfk)$ with $i = 1,2,3$ with the energy $E_i$ and again $r_i (k_x) = -1, \, \forall \, i$, independent of the Hamiltonian used for the chain. Again, regardless of the hopping terms included in the Hamiltonian, the system stays in a topologically trivial phase, as expected by the absence of edge states stemming from the breaking of condition (iv).

\begin{figure*}[t]
  \centering
  \subfigure[]{
  {\includegraphics[width=0.45\textwidth]{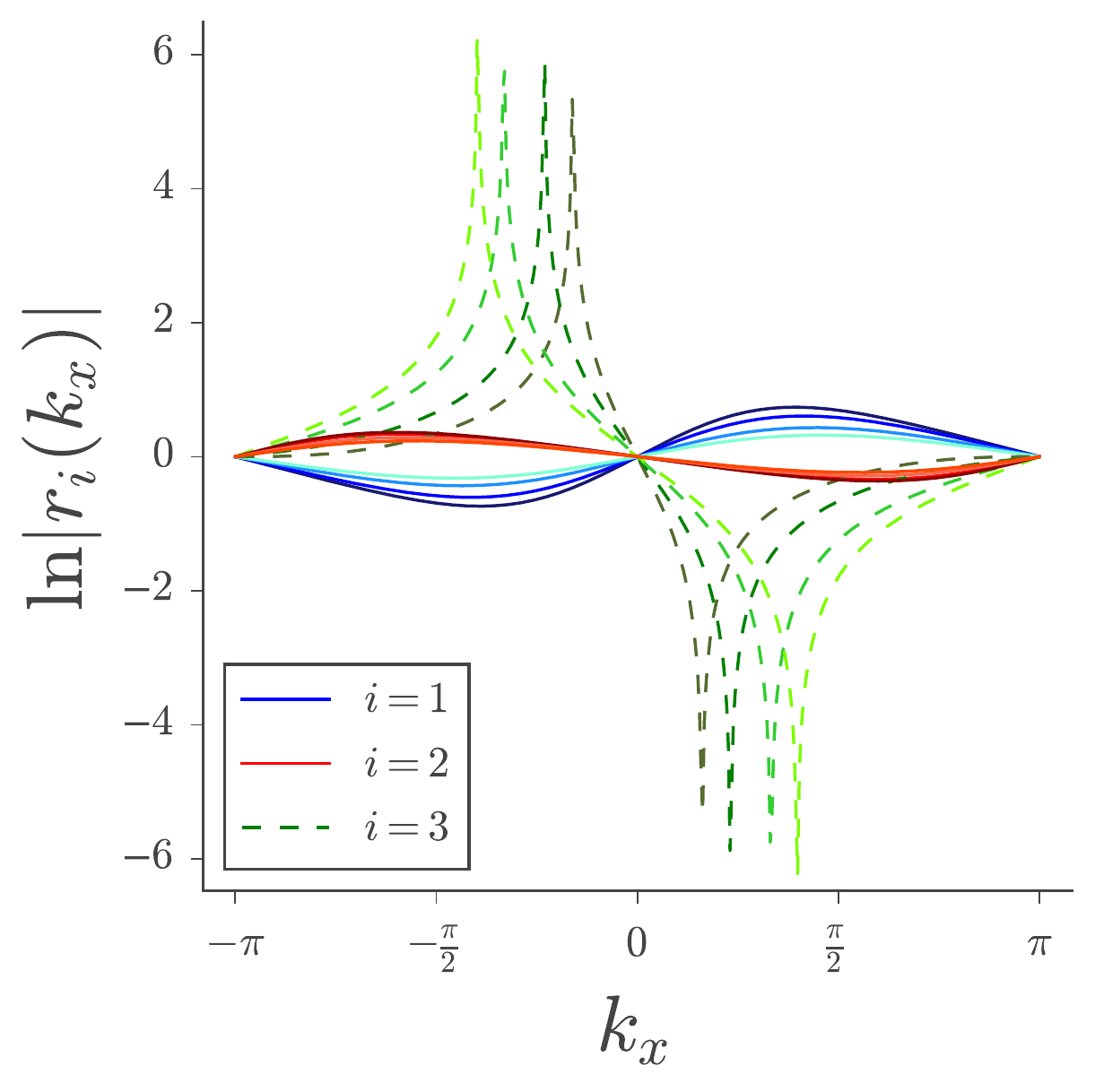}}
  \label{figurechainmodelnotsotrivialotherthreesolutionsa}}\quad
  \subfigure[]{
  {\includegraphics[width=0.45\textwidth]{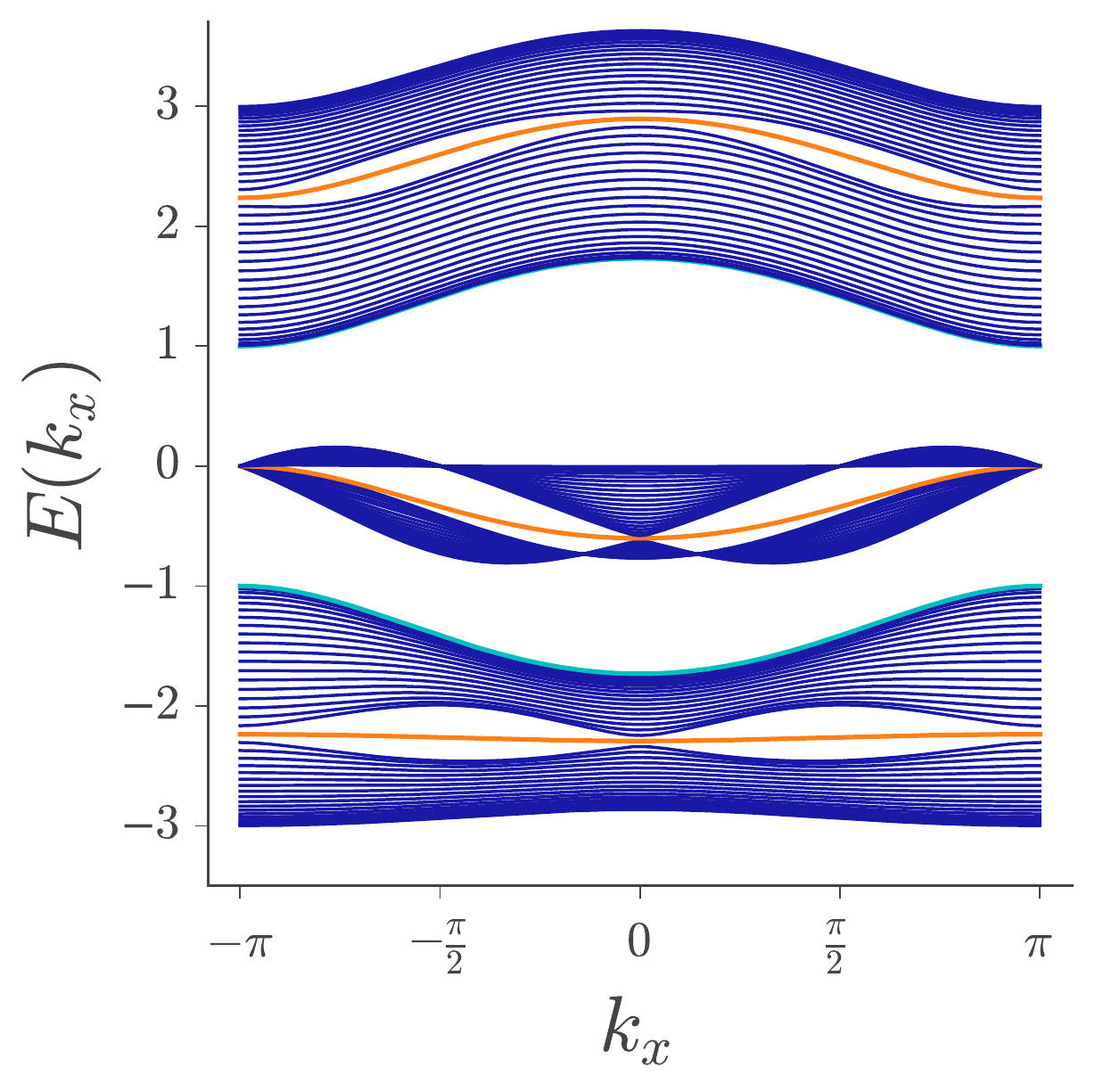}}
  \label{figurechainmodelnotsotrivialotherthreesolutionsb}}
  \caption{Plots for the model in Fig.~\ref{figurechainmodelsc} with the chain Hamiltonian given in Eqs.~(\ref{standardformulahamiltonian}) with $d_0(k_x) = 0$ and ${\bf d}(k_x) = \left( t \, {\rm cos}\frac{k_x}{2}, \,  t' \, {\rm cos}\frac{k_x}{2},  \, V\right)$ where $t$ and $t'$ are nearest-neighbor hopping parameters and $V$ is a staggering potential, and $h_{k_x}^{\rm ch}=0$. (a) Plot of the inverse localization length ${\rm ln}|r_i(k_x)|$ for $t'/t = t_\perp/t = 1$ and different values of $V/t = 0, 1, 2, 3$ corresponding from dark to light, and where blue corresponds to $r_1 (k_x)$, red to $r_2 (k_x)$ and green to $r_3 (k_x)$. (b) Energy spectrum with $N= 40$ shown in units of $t$ with $V/t = 1$, where the orange bands indicate the energies $E_i(k_x)$ with the lowest one corresponding to $E_1(k_x)$, the middle to $E_2(k_x)$ and the top to $E_3(k_x)$. The two bands in cyan correspond to the energy solutions $E_- (k_x)$ and $E_+(k_x)$, respectively, corresponding to the wave function solution in Eqs.~(\ref{equationeigenfunctionlayerlatticetrivialplusmodel}) and (\ref{equationrofkxforthelayerlatticetrivialplusmodel}). Even though, the inverse localization length shown in (a) indicates that the three solutions $\ket{\Psi_i (k_x)}$ localize, the band spectrum in (b) reveals that they are not topologically protected.}
  \label{figurechainmodelnotsotrivialotherthreesolutions}
\end{figure*}

\begin{figure}[t]
\centering
  \includegraphics[width=0.45\textwidth]{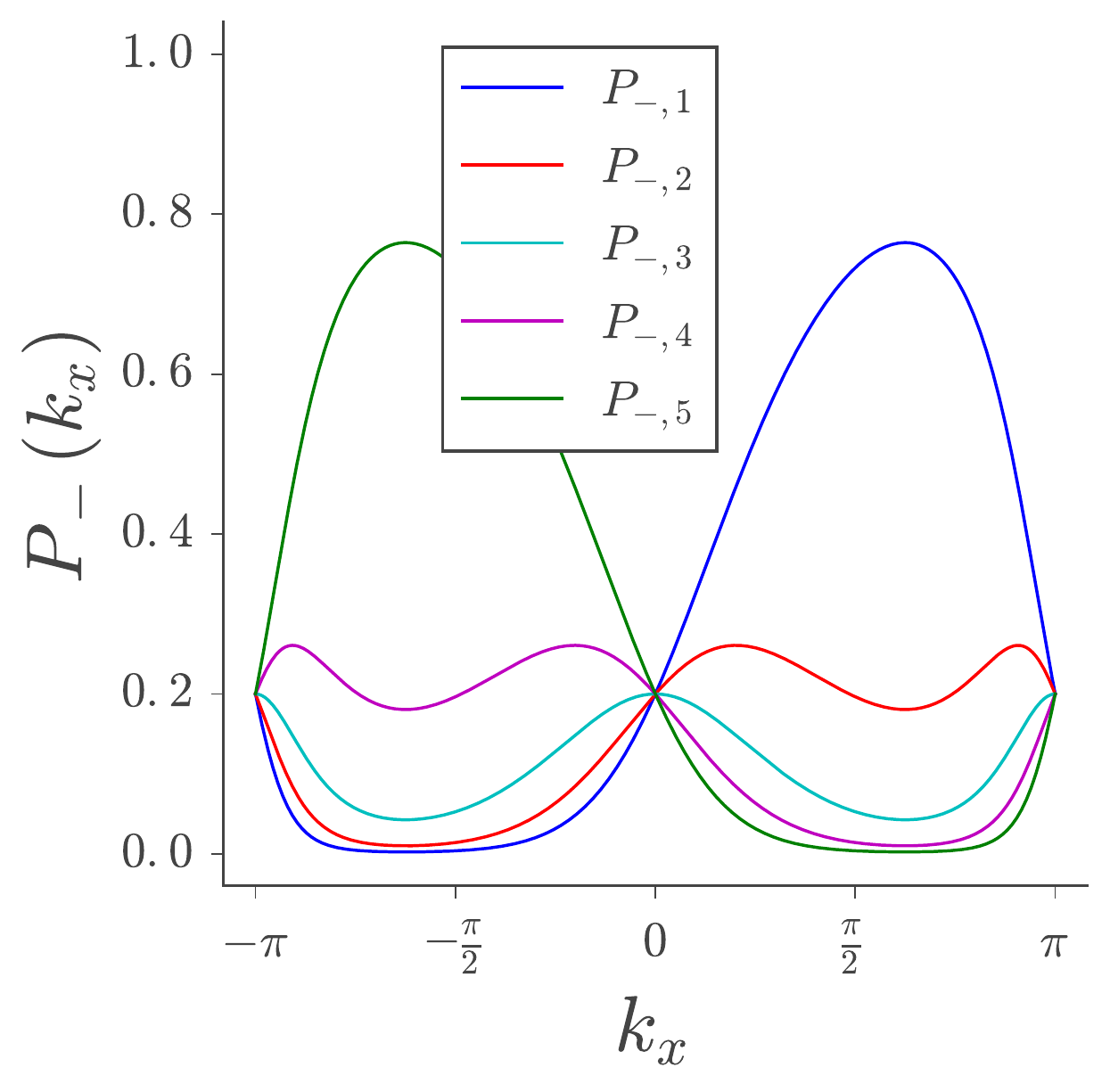}
\caption{Plot of the weight of the wave function $\ket{\Psi_-(k_x)}$ on each layer $m$ shown in Eq.~(\ref{equationweightofwavefctoneachlayer}) for $N = 5$ for the model in Fig.~\ref{figurechainmodelsd} with the $A$-lattice Hamiltonian given in Eq.~(\ref{generalhamiltonianchains}) with $t_\perp/t' = 1$ and $|t_\perp/V| = 2$ and all other parameters equal to zero. The wave functions are completely localized to layer $m$ if the weight equals $P_{-, m}(k_x) = 1$. Blue is $P_{-, 1}(k_x)$, red $P_{-, 2}(k_x)$, cyan $P_{-, 3}(k_x)$, magenta $P_{-, 4}(k_x)$, and green $P_{-, 5}(k_x)$. This plot is in perfect agreement with what is shown in Fig.~\ref{figurekagomemodelforbrokensublatticesymm}(b).}
\label{figurechainmodelweightofwavefctsfivebands}
\end{figure}

Now, we turn to the lattice model in Fig.~\ref{figurechainmodelsc}, which shows slightly more complex behavior. The $A$ lattice Hamiltonian again includes five sites in the unit cell and reads
\begin{equation*}
\mathcal{H}_{k_x} = \left(\begin{array}{ccccc}
\multicolumn{2}{c}{\multirow{2}{*}{$\mathcal{H}_{k_x}^{\rm ch}$}} & t_{\perp} {\rm e}^{-i \frac{k_x}{4}} & \multicolumn{2}{c}{\multirow{2}{*}{0}} \\
& & t_{\perp} {\rm e}^{i \frac{k_x}{4}} & & \\
t_{\perp} {\rm e}^{i \frac{k_x}{4}} & t_{\perp} {\rm e}^{-i \frac{k_x}{4}} & h^{\rm ch}_{k_x} & t_{\perp} {\rm e}^{-i \frac{k_x}{4}} & t_{\perp} {\rm e}^{i \frac{k_x}{4}} \\
\multicolumn{2}{c}{\multirow{2}{*}{0}} & t_{\perp} {\rm e}^{i \frac{k_x}{4}}  & \multicolumn{2}{c}{\multirow{2}{*}{$\mathcal{H}_{k_x}^{\rm ch}$}} \\
& & t_{\perp} {\rm e}^{-i \frac{k_x}{4}} & &  \end{array}\right).
\end{equation*}
The phases for the full model are $f_{A,4}(k_x) = f_{B,2}(k_x) = {\rm exp}(i k_x/4)$, $f_{A, 5}(k_x) = f_{B,1}(k_x) = {\rm exp}(-i k_x/4)$ and $f_{A,\alpha}(k_x) = 0, \, \alpha = 1, 2, 3$ and $f_{B,\alpha'}(k_x) = 0, \, \alpha'= 3, 4, 5$. Similar to the model corresponding to Fig.~\ref{figurechainmodelsb}, we find $n_{\rm sub} = 2$ wave function solutions as given in Eq.~(\ref{equationeigenfunctionlayerlatticetrivialplusmodel}) with
\begin{equation}
s_\pm(k_x) = -\frac{{\rm e}^{i \frac{k_x}{4}} \phi_{\pm,1} (k_x) + {\rm e}^{-i \frac{k_x}{4}} \phi_{\pm,2} (k_x)}{{\rm e}^{-i \frac{k_x}{4}} \phi_{\pm,1} (k_x) + {\rm e}^{i \frac{k_x}{4}} \phi_{\pm,2} (k_x)},\label{equationrofkxforthelayerlatticetrivialplusmodel}
\end{equation}
such that
\begin{align*}
r_\pm(k_x) & = -\frac{{\rm e}^{-i \frac{k_x}{4}} s_\pm(k_x) \phi_{\pm,1} (k_x) + {\rm e}^{i \frac{k_x}{4}} s_\pm(k_x) \phi_{\pm,2} (k_x)}{{\rm e}^{i \frac{k_x}{4}} \phi_{\pm,1} (k_x) + {\rm e}^{-i \frac{k_x}{4}} \phi_{\pm,2} (k_x)} \\
& = -s_\pm(k_x) \, s_\pm^{-1}(k_x) = -1.
\end{align*}
These two wave functions are thus equally localized to each $A$ lattice present in the full model. However, we notice that there may occur some localization inside the $A$ lattice as $|s_\pm (k_x)|$ is not trivially equal to $1$, which is shown in Figs.~\ref{figurekagomemodelforbrokensublatticesymm}(a) and \ref{figurekagomemodelforbrokensublatticesymm}(b).

For the remaining three solutions to the wave function, which read $\Phi_i ({\bf k}) = \bigoplus_{s=1}^5 \phi_{i,s} (\bfk)$ with $i = 1,2,3$, we find a nontrivial $|r_i (k_x)|$
as shown in Fig.~\ref{figurechainmodelnotsotrivialotherthreesolutionsa}. However, one can see in the corresponding energy spectrum in Fig.~\ref{figurechainmodelnotsotrivialotherthreesolutionsb} that the localized states are not topologically protected, which is supported by the retrieval of a zero-Chern number for these parameters.

Finally, we turn to the kagome lattice in Fig.~\ref{figurechainmodelsd}, which shows significantly richer and more complex behavior. This lattice is also of special interest as it occurs naturally in many materials and can also be engineered in cold atom systems [\onlinecite{joguzmanthomashosur, zhangchenmazhou}]. The $A$ lattice Hamiltonian is given by Eq.~(\ref{standardformulahamiltonian}), and the phases are $f_{A,1}(k_x) = f_{B,2}(k_x) = {\rm exp}(-i k_x/4)$ and $f_{A,2}(k_x) = f_{B,1}(k_x) = {\rm exp}(i k_x/4)$, such that $r_\pm (k_x)$ corresponds to the expression in Eq.~(\ref{equationrofkxforthelayerlatticetrivialplusmodel}), and constraint (iv) is fulfilled. This allows us to review condition (v). To specify the $A$ lattice Hamiltonian in Eq.~(\ref{standardformulahamiltonian}), we use the following:
\begin{equation}
{\bf d}(k_x) = \left( t \, {\rm cos}\frac{k_x}{2} + t_1 \, {\rm sin}\frac{k_x}{2}, \,  t' \, {\rm cos}\frac{k_x}{2} + t'_1 \, {\rm sin}\frac{k_x}{2},  \, V\right), \label{generalhamiltonianchains}
\end{equation}
where $t$, $t'$, $t_1$ and $t'_1$ are nearest-neighbor hopping parameters, and $V$ is a staggering potential, and $d_0(k_x) = 0$. We emphasize that details of the lattice Hamiltonian are irrelevant as long as it is local and translation invariant, and we have introduced nearest-neighbor hopping terms only to be able to review the localization of the state in a transparent fashion. We find that $|r_\pm (k_x)|\neq1$ only when at least $t'_1 \neq 0$ and/or $t' \neq 0$ as is shown in Fig.~\ref{figurekagomemodelforbrokensublatticesymm}. We first review the situation where $t'_1 \neq 0$ and $t' = 0$, in which case both sublattice and inversion symmetry are broken but time-reversal symmetry is preserved. We thus expect to be unable to find a Chern-insulator phase, which is indeed what is revealed in Figs.~\ref{figurekagomemodelforbrokensublatticesymm}(a) and \ref{figurekagomemodelforbrokensublatticesymm}(c). The case in which $t'_1 = 0$ and $t' \neq 0$ is more interesting. Now inversion symmetry is preserved but sublattice and time-reversal symmetry are broken such that we are able to find a Chern insulator as is shown in Figs.~\ref{figurekagomemodelforbrokensublatticesymm}(b) and \ref{figurekagomemodelforbrokensublatticesymm}(d). We see that the energy spectrum has two bulk gaps with a band crossing each of these gaps. These two bands (in orange) correspond to our exact wave function solution in Eq.~(\ref{exactsolutionedgestate}) and their inverse localization length ${\rm ln}|r_\pm (k_x)|$ reveals that the right mover localizes to the chain $m = 1$ and the left-mover to the chain $m=N$, which is in accordance with the chirality of the edge states in a Chern insulator. This localization is corroborated by the weight of the wave function on each $A$ lattice $m$ given in Eq.~(\ref{equationweightofwavefctoneachlayer}) shown in Fig.~\ref{figurechainmodelweightofwavefctsfivebands} for a system of five layers. The Chern number for the fully periodic kagome lattice is governed by the values of $V/t_\perp$, and we find for half-filling that the Chern number $C = 1$ for $|t_\perp/V|>0.7$. This leads to another interesting observation, namely, that when the system is no longer a Chern insulator, the exact solution still localizes, which is reflected by ${\rm ln}|r_\pm (k_x)|$ in Fig.~\ref{figurekagomemodelforbrokensublatticesymm}(b) still having a structure for $V/t' = 2.5$, $t_\perp/t' = 1$. We have thus found a Chern insulator on the kagome lattice with chiral-edge states whose exact wave function solution is given in Eq.~(\ref{exactsolutionedgestate}) when both time-reversal and sublattice symmetry are broken simultaneously. The breaking of the first is a minimal requirement to find a Chern insulator, whereas the breaking of the second is inherent to our specific choice of lattice as preserving sublattice symmetry would yield $|\phi_{\pm,1} (k_x)| = |\phi_{\pm,2} (k_x)|$ such that one trivially finds $|r_\pm (k_x)|=1$. Note that a quantum spin Hall insulator can simply be retrieved by taking two time-reversed copies of the Chern insulator, which introduces spin degree of freedom. In that case, the system supports four helical-edge states, whose wave function is described by our exact solution.

\subsection{Three dimensions, first example: Dirac and Weyl semimetals from stacked checkerboard models} \label{sectionfour}

\begin{figure}[ht]
  \centering
  \adjustbox{trim={0.25\width} {.275\height} {0.3\width} {.3\height},clip}
  {\includegraphics[width=0.7\textwidth]{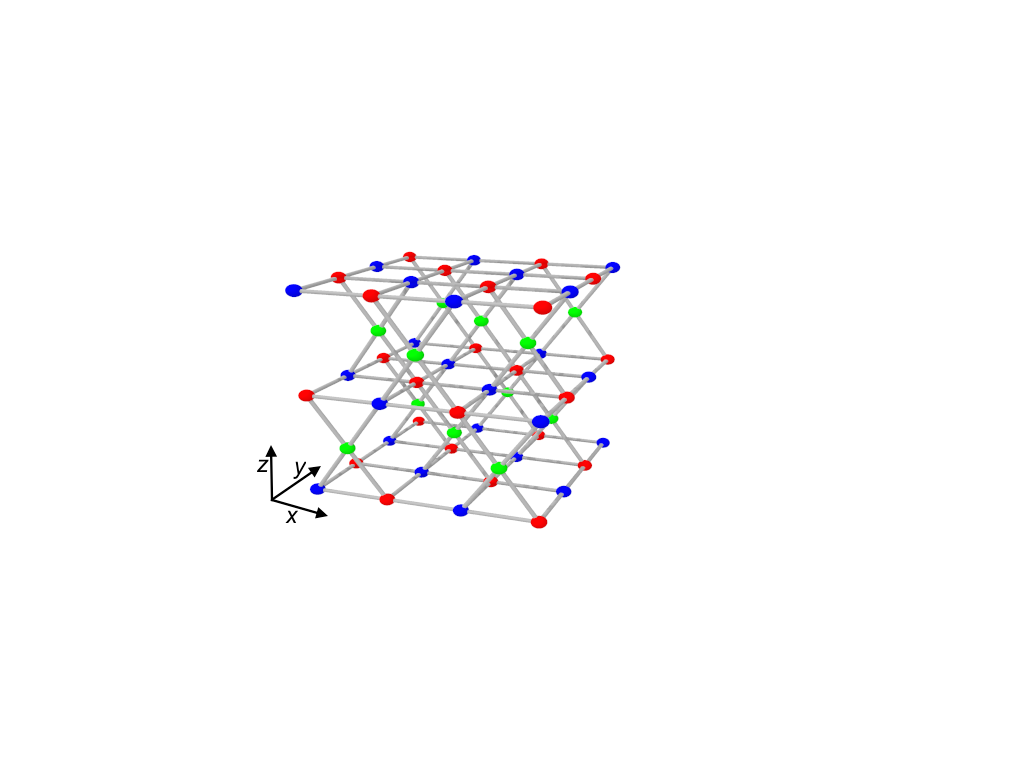}}
\caption{The checkerboard model featuring checkerboard lattices (red and blue sites) stacked on top of each other with a square lattice (green) in between. Notice the relative displacement between checkerboard lattices.}
\label{figurecheckerboardmodel}
\end{figure}

\begin{figure*}[t]
  \centering
  \subfigure[]{
  {\includegraphics[width=0.3\textwidth]{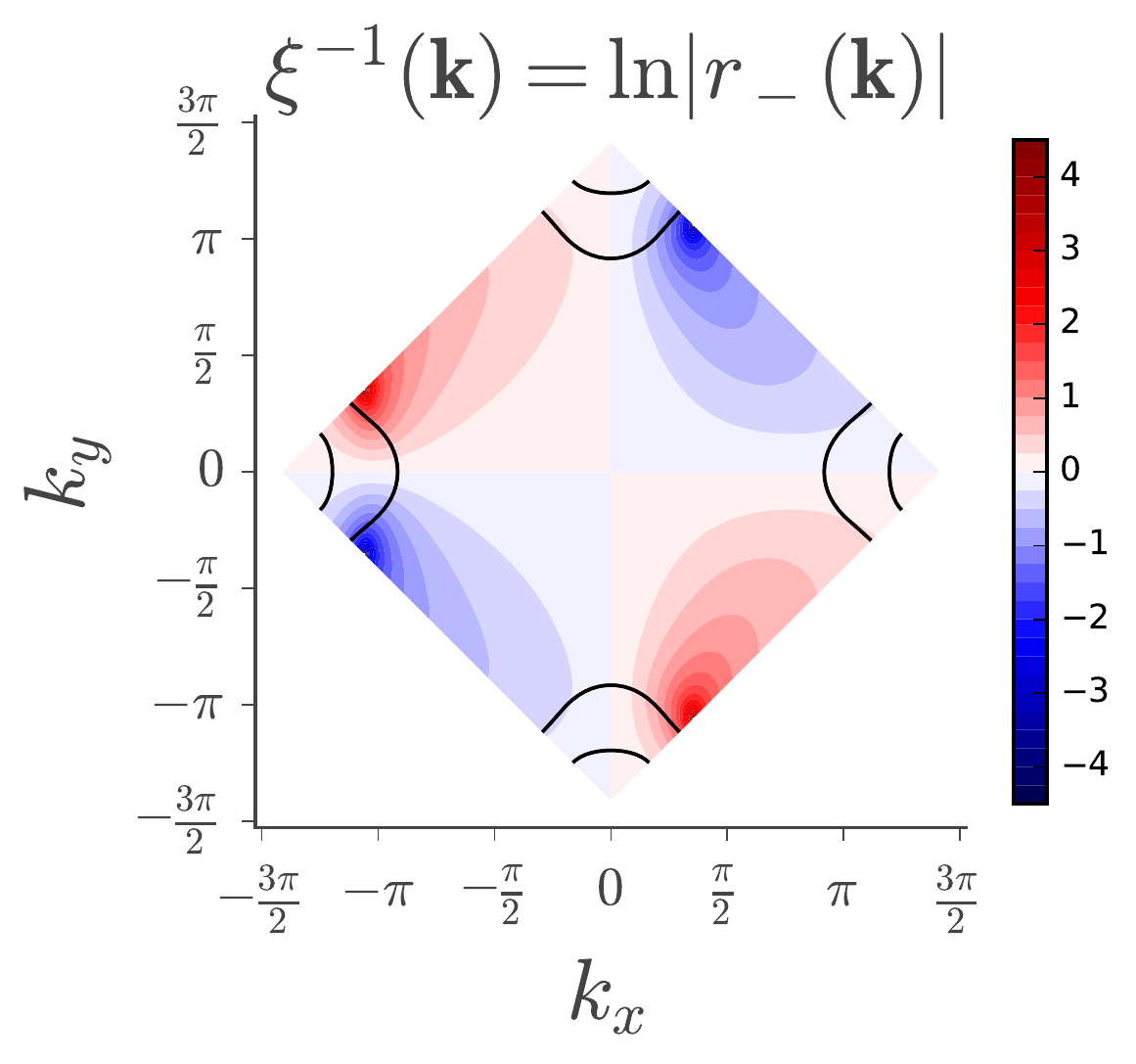}}
  \label{figurecheckerboardmodelcompilationtperp1mu1b}}\quad
    \subfigure[]{
  {\includegraphics[width=0.3\textwidth]{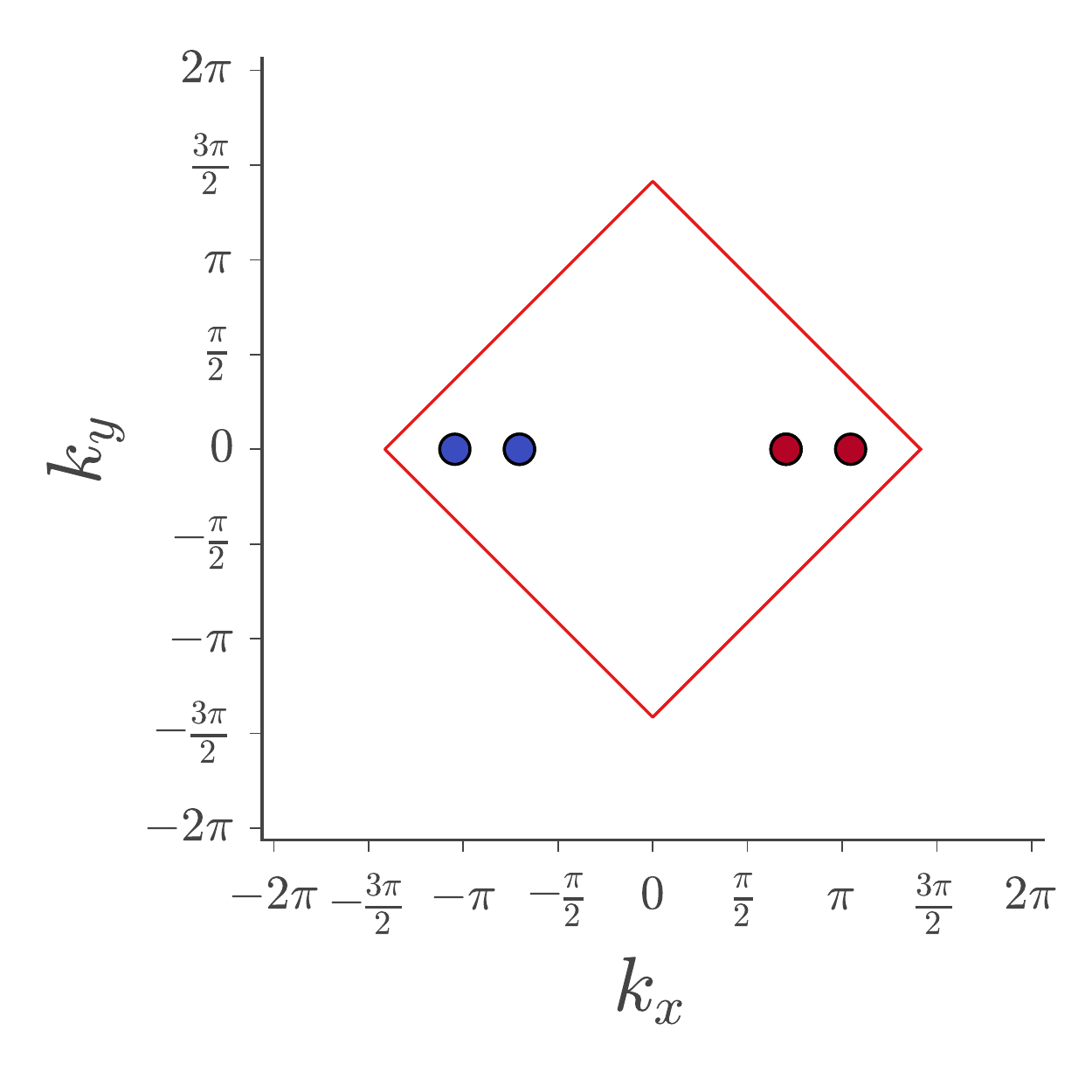}}
  \label{figurecheckerboardmodelcompilationtperp1mu1c}}\quad
  \subfigure[]{
  {\includegraphics[width=0.3\textwidth]{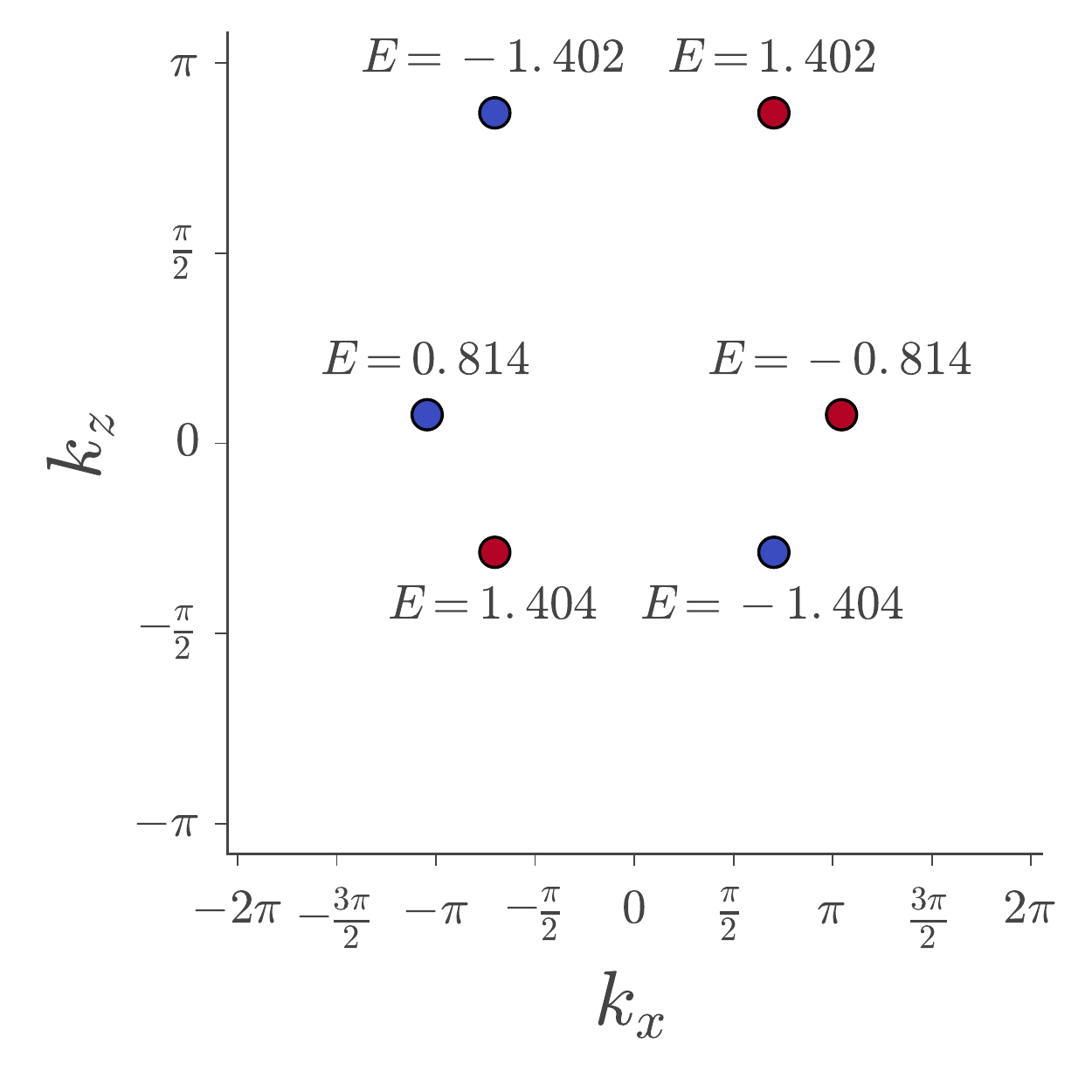}}
  \label{figurecheckerboardmodelcompilationtperp1mu1d}}\quad
  \subfigure[]{
  {\includegraphics[width=0.47\textwidth]{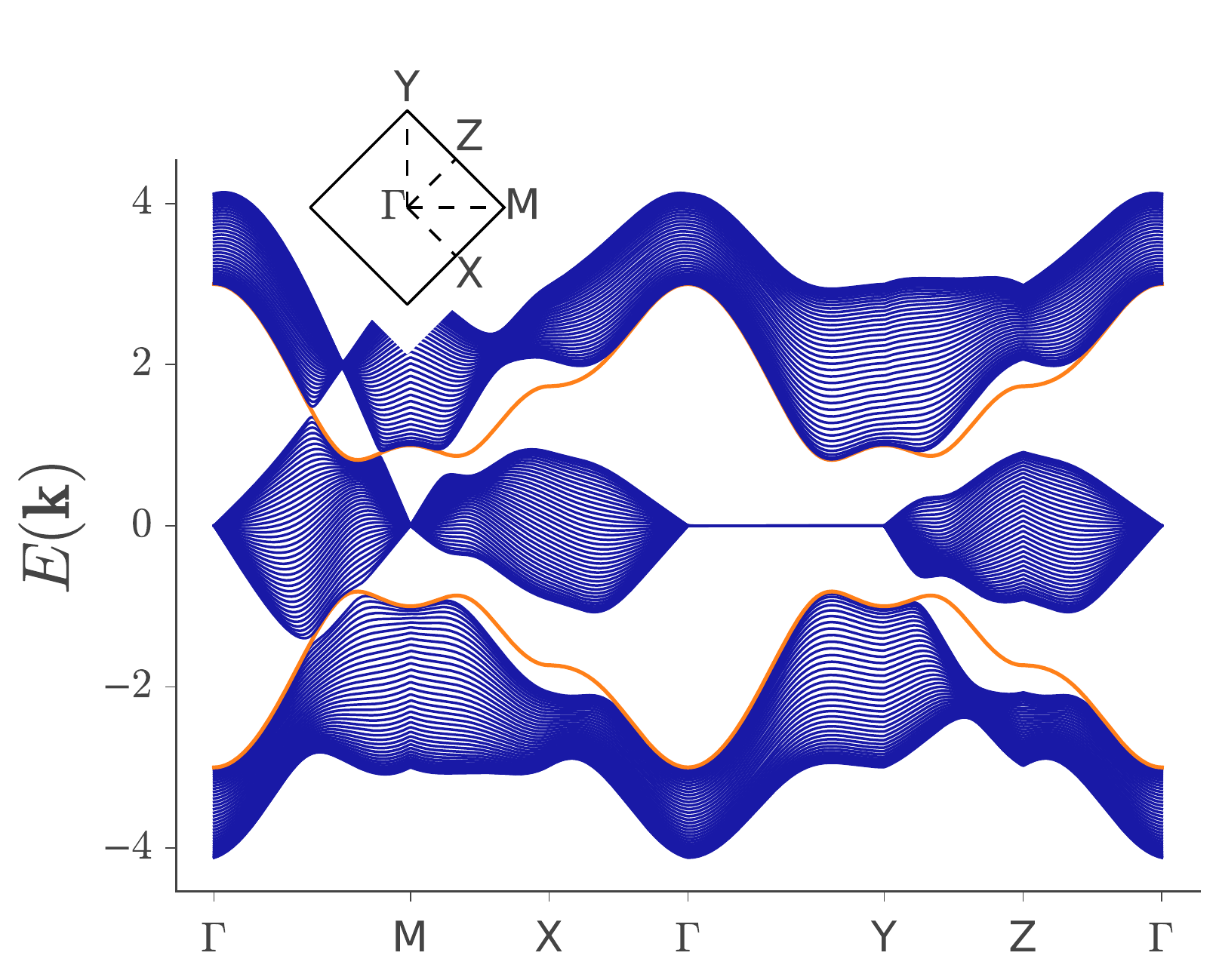}}
  \label{figurecheckerboardmodelcompilationtperp1mu1e}}\quad
  \subfigure[]{
  {\includegraphics[width=0.47\textwidth]{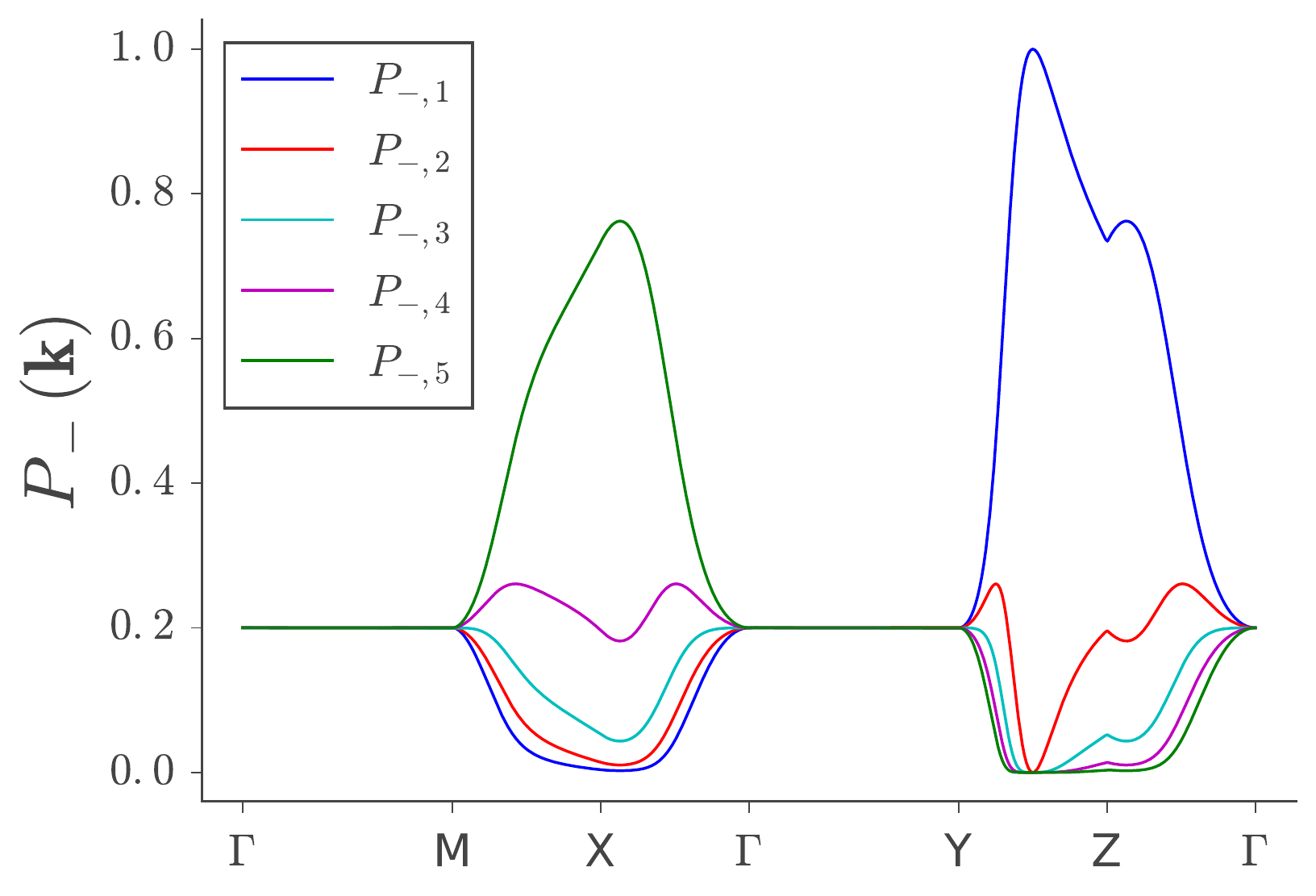}}
  \label{figurecheckerboardmodelcompilationtperp1mu1f}}
  \caption{Solutions to the checkerboard model with nonzero Chern number with $t'/t = t_2/t = t_\perp/t = V/t = 1$. In (a), we show the inverse localization length $\xi^{-1} (\bfk) = {\rm ln}|r_-(\bfk)|$ given in Eq.~(\ref{equationcheckerboardmodelrofk}). The black line is the energy contour of $E_-(\bfk)$ at the chemical potential $\mu/t = -0.9$. We see that the boundary state switches surfaces four times and is thus a Fermi arc. In (b) and (c), the location of the Weyl nodes in the three-dimensional Brillouin zone is shown from the top view in the $k_z$ direction in (b) and a two-dimensional cut at $k_y =0$ in (c). Red and blue correspond to positive and negative Berry charge, respectively. The energies corresponding to the Weyl nodes are included as insets in (c). The energy spectrum along a path in the Brillouin zone, indicated in the inset, is shown in (d) with $N = 40$ and the energy given in units of $t$. The orange bands correspond to the Fermi arc solution and one observes Weyl points at the locations and energies indicated in (b) and (c). The weight of the wave function associated with the lower orange band on each $A$ lattice is shown in (e) and is in perfect agreement with (a) and (d).}
  \label{figurecheckerboardmodelcompilationtperp1mu1}
\end{figure*}

We now turn to three-dimensional models built from stacking two-dimensional layers in a frustrated fashion. In the first example, the $A$ lattice is a checkerboard lattice, which has two sites in the unit cell, and the intermediary $B$ lattice is a square lattice with one site in the unit cell. By stacking them we obtain the lattice shown in Fig.~\ref{figurecheckerboardmodel}, which is a three-dimensional cousin of the kagome model in Fig.~\ref{figurechainmodelsd}. The two-dimensional surface Brillouin zone is shown as an inset in Fig.~\ref{figurecheckerboardmodelcompilationtperp1mu1e}. Conditions (i) - (iv) are fulfilled and we have a Weyl phase when condition (v) is fulfilled in which case we expect that the exact solution corresponds to (a family of) Fermi arcs. In particular, this model is well suited for exploring the connection between the topology of the two-dimensional layers and the band switching properties of the surface bands.

\begin{figure}[h]
    \includegraphics[width=0.5\textwidth]{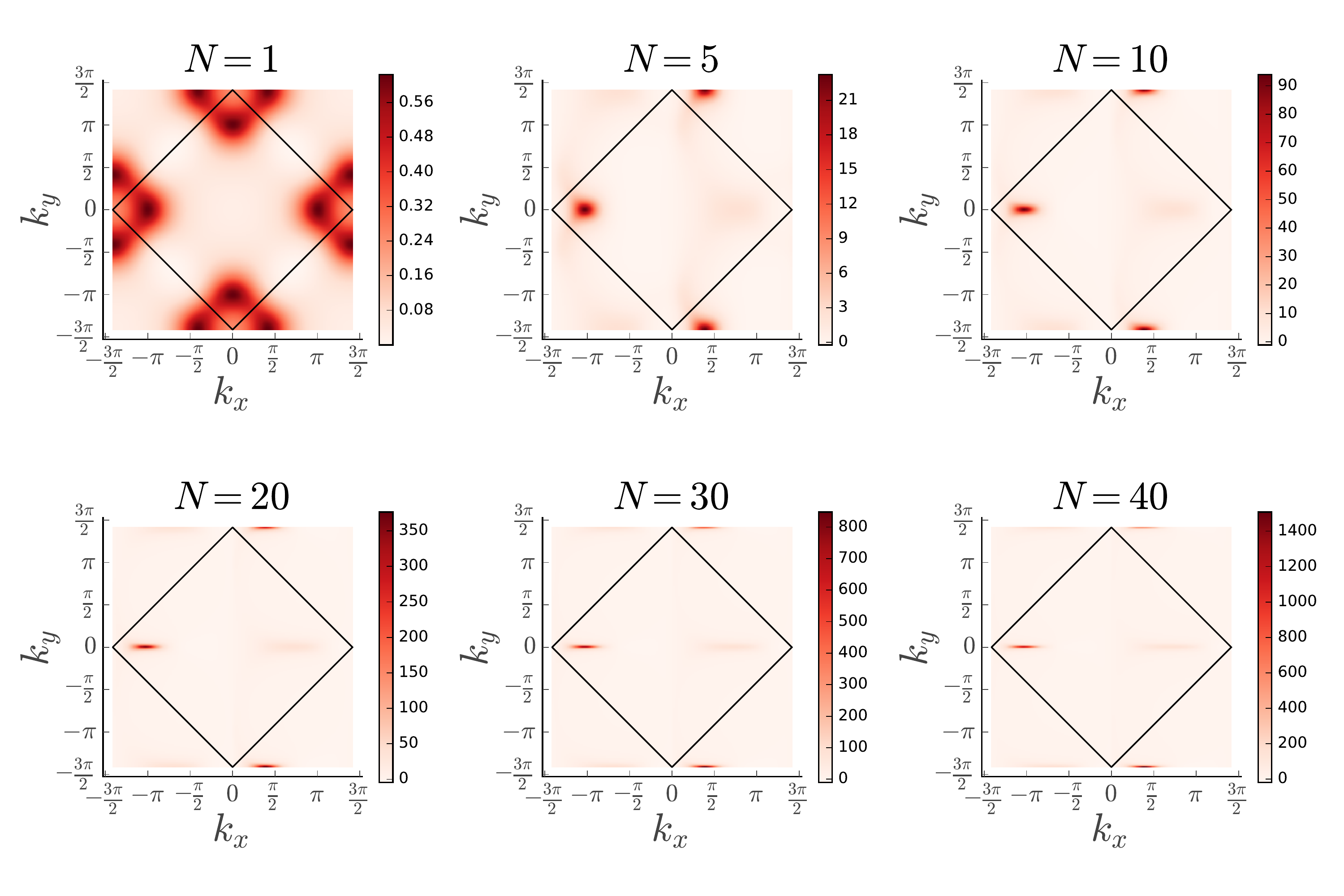}
\caption{(Color online.) Berry curvature associated with $\Psi_- ({\bf k})$ for the checkerboard model with $t'/t = t_2/t = t_\perp/t = V/t = 1$ for a different number of layers $N = 1, 5, 10, 20, 30, 40$. The black line indicates the Brillouin zone. We see that the Berry curvature has maxima at those points in the Brillouin zone corresponding to the location of the Weyl points.}
\label{figurecheckerboardmodelberrycurvature}
\end{figure}

The Hamiltonian for the individual checkerboard lattice is given by $H(\bfk) = \Phi^\dagger(\bfk) \mathcal{H}_{\bfk} \Phi (\bfk)$, where $\Phi (\bfk)$ is the annihilation operator of an electron in the checkerboard layer and
\begin{equation}
\mathcal{H}_{\bfk} = {\bf d} (\bfk) \cdot \boldsymbol\sigma + d_0(\bfk) \sigma_0.\label{equationgeneralhamiltonianwithboldmomentum}
\end{equation}
The eigenvalues read
\begin{equation}
E_\pm (\bfk) = \pm|{\bf d}(\bfk)| + d_0(\bf k). \label{equationgeneralenergywithboldmomentum}
\end{equation}
The Hamiltonian for the full model is given in Eqs.~(\ref{generalmultilayerhamiltonian}) and (\ref{generalmultilayerhamiltonianperp}) with $h_{\bfk} = 0$, $t_{\perp, s, \alpha} = t_\perp \in \mathbb{R} \, \forall \, s, \, \alpha$, and $f_{A,1}(k_x) = f_{B,2}(k_x) = {\rm exp}[-i k_x/2\sqrt{2}]$ and $f_{A,2}(k_x) = f_{B,1}(k_x) = {\rm exp}[i k_x/2\sqrt{2}]$. We find two exact solutions to the wave function, which are given in Eq.~(\ref{exactsolutionedgestate}) with
\begin{equation}
r_\pm (\bfk) = -\frac{{\rm e}^{i \frac{k_x}{2\sqrt{2}}} \phi_{\pm,1} (\bfk) + {\rm e}^{-i \frac{k_x}{2\sqrt{2}}} \phi_{\pm,2} (\bfk)}{{\rm e}^{-i \frac{k_x}{2\sqrt{2}}} \phi_{\pm,1} (\bfk) + {\rm e}^{i \frac{k_x}{2\sqrt{2}}} \phi_{\pm,2} (\bfk)}. \label{equationcheckerboardmodelrofk}
\end{equation}
It is trivial to see that $|r_\pm (\bfk)| = 1$ when $k_x = 0$. In Sec.~\ref{sectiontwof}, we discussed that $r_i(\bfk)$ must have zeros and infinities when the $A$ lattice is a Chern insulator. For this specific model, this means that we need to solve
\begin{equation}
{\rm e}^{i s \frac{k_x}{2\sqrt{2}}} \phi_{\pm,1} (\bfk) + {\rm e}^{-i s \frac{k_x}{2\sqrt{2}}} \phi_{\pm,2} (\bfk) = 0,\label{arccond}
\end{equation}
with $s = 1$ ($s = -1$) which corresponds to setting the numerator (denominator) in Eq.~(\ref{equationcheckerboardmodelrofk}) to zero. Using the standard representation of the Pauli matrices, we have 
\begin{equation}
\ket{\Phi_\pm(\bfk)} \doteq \tilde{\mathcal{N}}_\pm(\bfk) \begin{pmatrix}
\pm |{\bf d}(\bfk)| + d_z(\bfk) \\
d_x(\bfk) + i \, d_y(\bfk)
\end{pmatrix},
\end{equation}
where $\tilde{\mathcal{N}}_\pm(\bfk)$ is the normalization factor. The above requirement (\ref{arccond}) thus decomposes into
\begin{align}
& \pm |{\bf d}(\bfk)| + d_z ({\bf k}) + {\rm cos}\left( \frac{k_x}{\sqrt{2}}\right) d_x (\bfk) + s \, {\rm sin}\left( \frac{k_x}{\sqrt{2}}\right) d_y (\bf k) \nonumber \\
& +i \left[{\rm cos}\left( \frac{k_x}{\sqrt{2}}\right) d_y(\bfk) - s \, {\rm sin}\left( \frac{k_x}{\sqrt{2}}\right) d_x(\bfk) \right] = 0, \label{equationconditionsurfaceswitchingcheckerboard}
\end{align}
such that the real and imaginary part of this equation have to be zero simultaneously. When the Hamiltonian in Eq.~(\ref{equationgeneralhamiltonianwithboldmomentum}) describes a Chern insulator, we know that ${\bf d}(\bfk)$ points in every direction in the Brillouin zone, which means we can always satisfy Eq.~(\ref{equationconditionsurfaceswitchingcheckerboard}) at, except in pathological cases, {\it different} $\bfk$, thus implying Fermi arc like surface bands. To be more specific, the Fermi arcs are formed as one-dimensional constant energy contours in the two-dimensional surface bands. In these two-dimensional bands, the argument above shows that there are momenta where the state is entirely localized at the topmost layer [diverging $r(\bfk)$] and other points where it is identically localized to the bottom layer [$r(\bfk)=0$]. It is also important to note that this is independent of the presence of Weyl nodes in the bulk as those only appear at sufficiently strong inter-layer tunneling. 

\begin{figure*}[t]
  \centering
  \subfigure[]{
  {\includegraphics[width=0.3\textwidth]{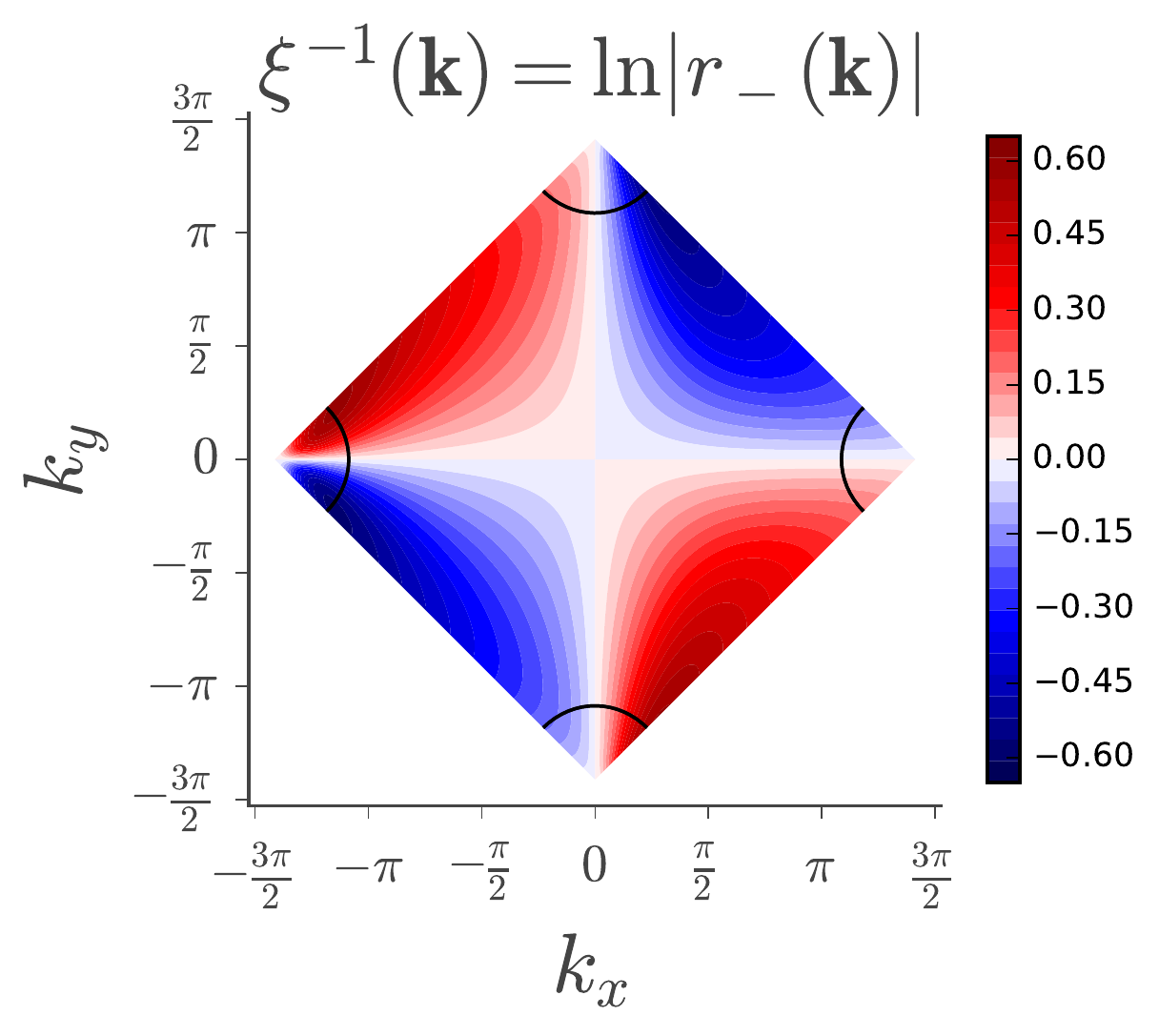}}
  \label{figurecheckerboardmodelcompilationtperp1mu2halfa}}\quad
    \subfigure[]{
  {\includegraphics[width=0.3\textwidth]{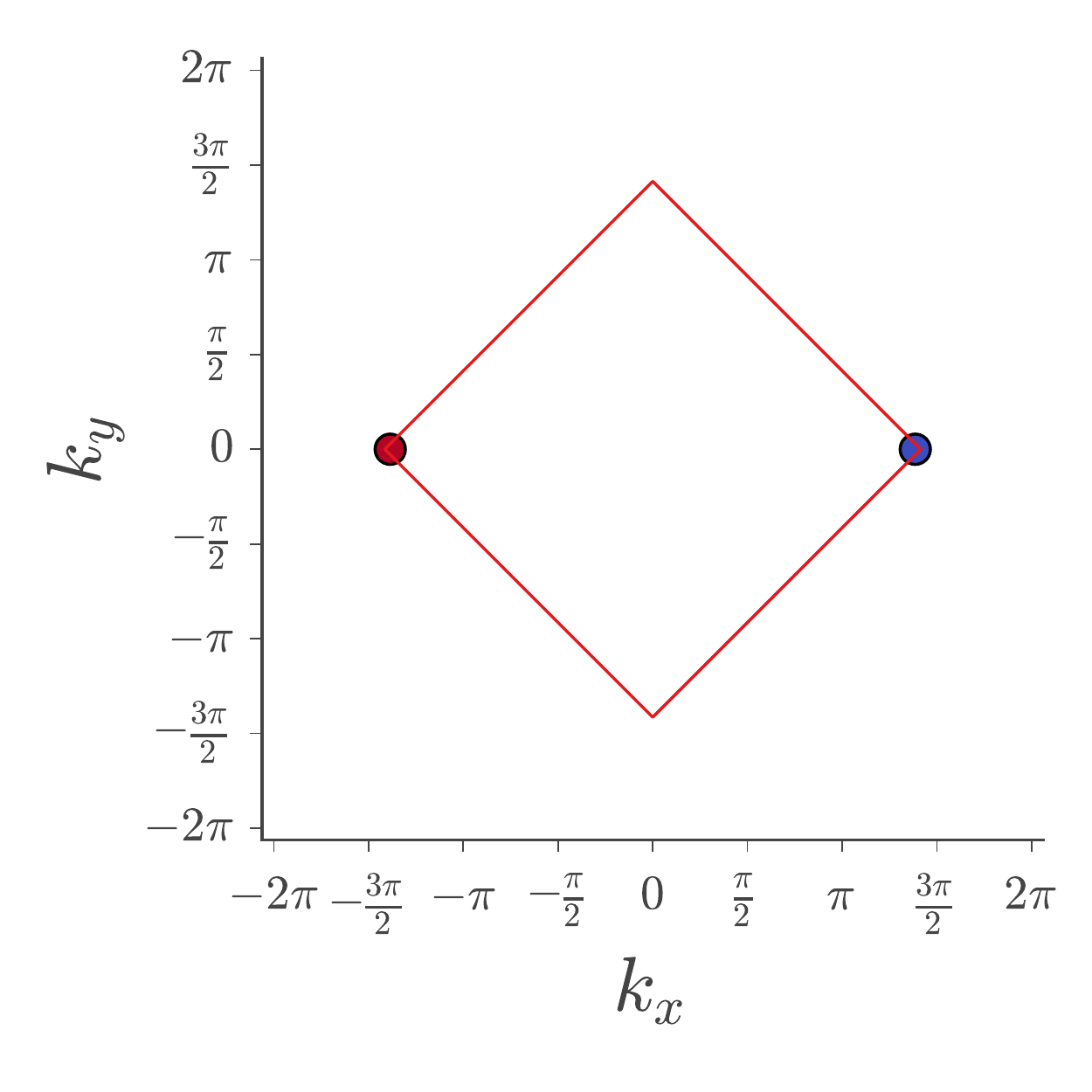}}
  \label{figurecheckerboardmodelcompilationtperp1mu2halfb}}\quad
  \subfigure[]{
  {\includegraphics[width=0.3\textwidth]{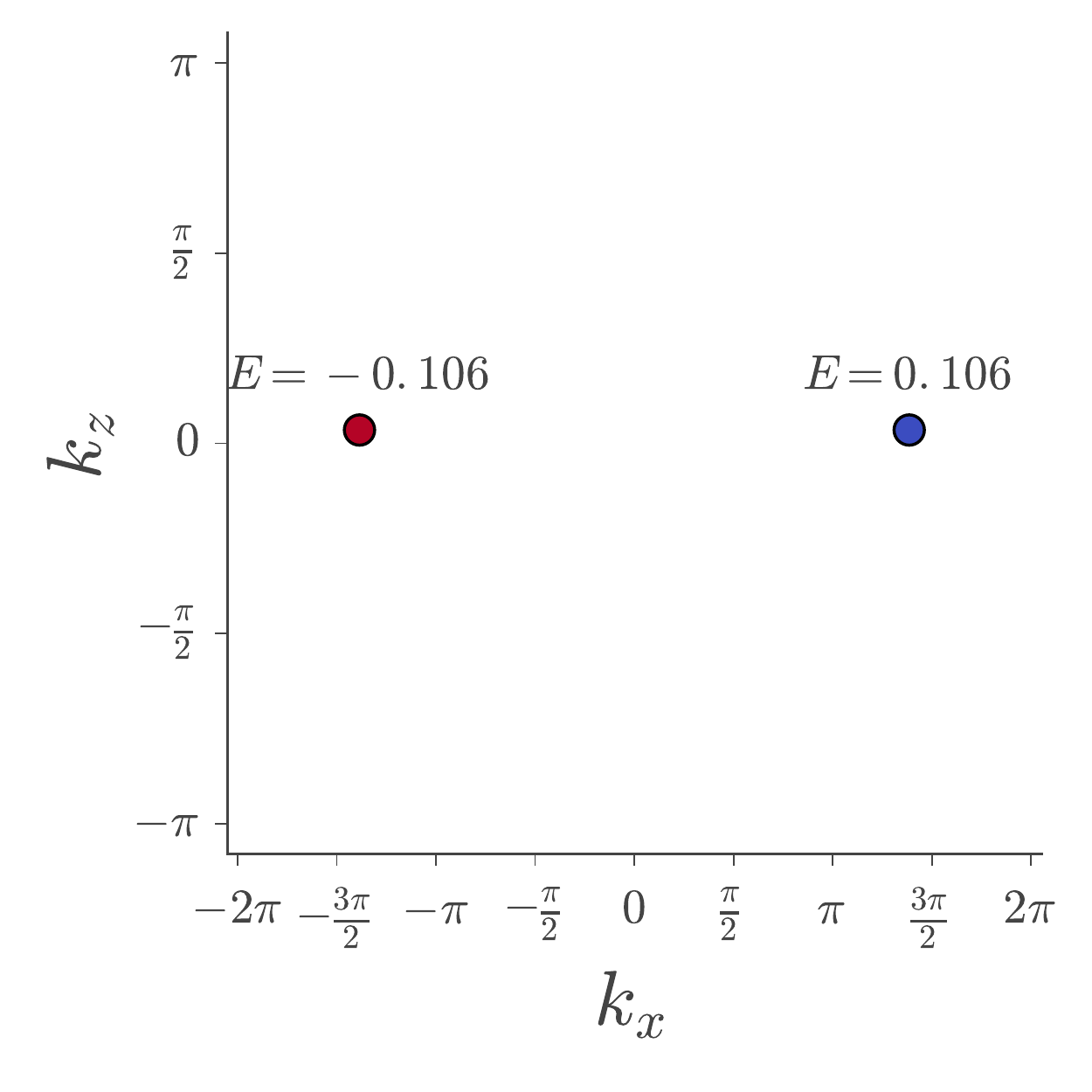}}
  \label{figurecheckerboardmodelcompilationtperp1mu2halfc}}\quad
  \subfigure[]{
  {\includegraphics[width=0.47\textwidth]{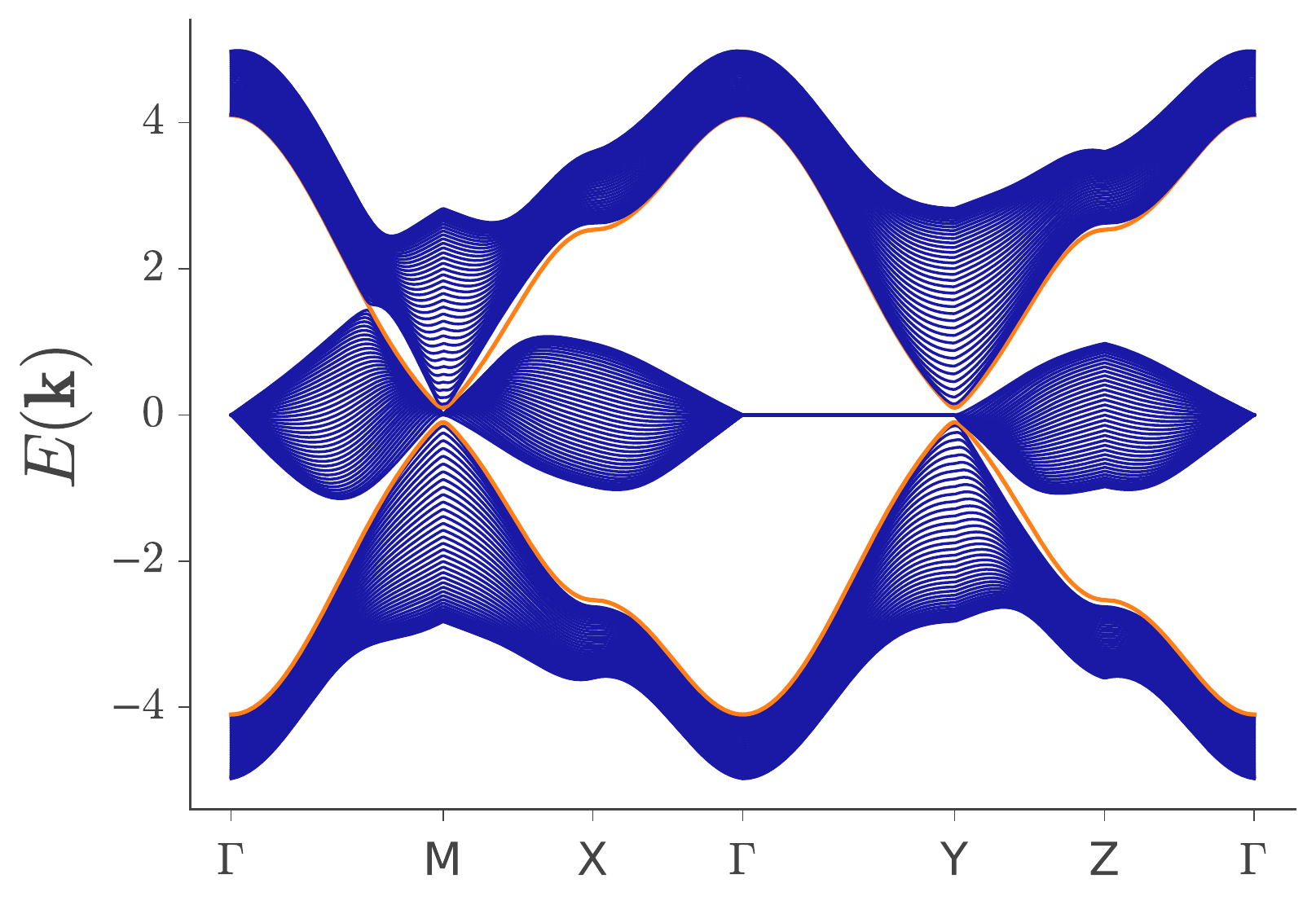}}
  \label{figurecheckerboardmodelcompilationtperp1mu2halfd}}\quad
  \subfigure[]{
  {\includegraphics[width=0.47\textwidth]{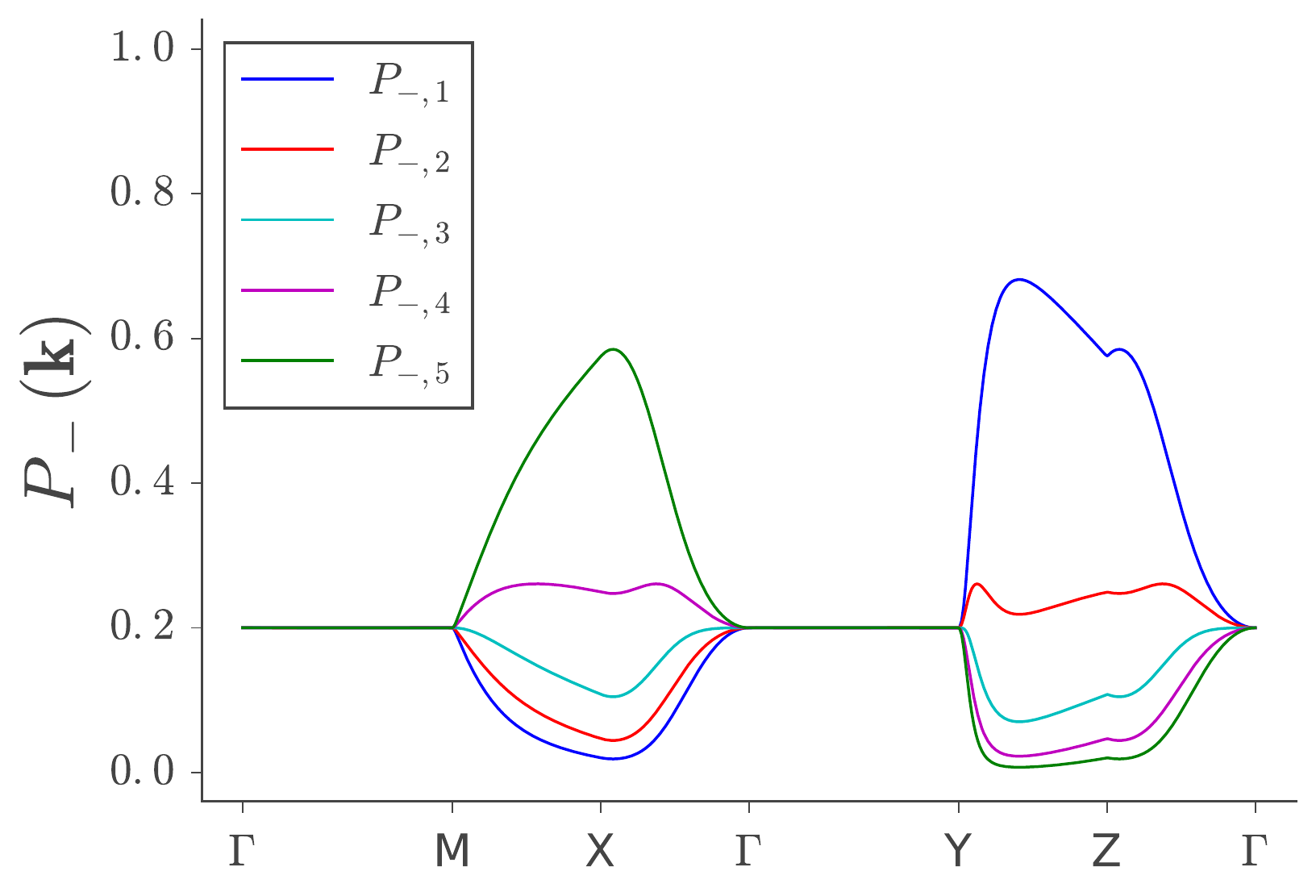}}
  \label{figurecheckerboardmodelcompilationtperp1mu2halff}}
  \caption{Solutions to the checkerboard model with zero-Chern number with $t'/t = t_2/t = t_\perp/t = 1$ and $V/t = 2.1$. In (a), we show the inverse localization length $\xi^{-1} (\bfk) = {\rm ln}|r_-(\bfk)|$, where the black line is the energy contour of $E_-(\bfk)$ at the chemical potential $\mu/t = -0.9$. The location of the Weyl nodes is shown in (b) and (c) with red and blue corresponding to positive and negative Berry charge, respectively. The energy spectrum along a path in the Brillouin zone, shown as an inset in Fig.~\ref{figurecheckerboardmodelcompilationtperp1mu1}(d), is shown in (d) with $N = 40$ and the energy given in units of $t$. The weight of the wave function associated with the lower orange band on each $A$ lattice is shown in (e).}
  \label{figurecheckerboardmodelcompilationtperp1mu2half}
\end{figure*}

\begin{figure*}[t]
  \centering
  \subfigure[]{
  {\includegraphics[width=0.5\textwidth]{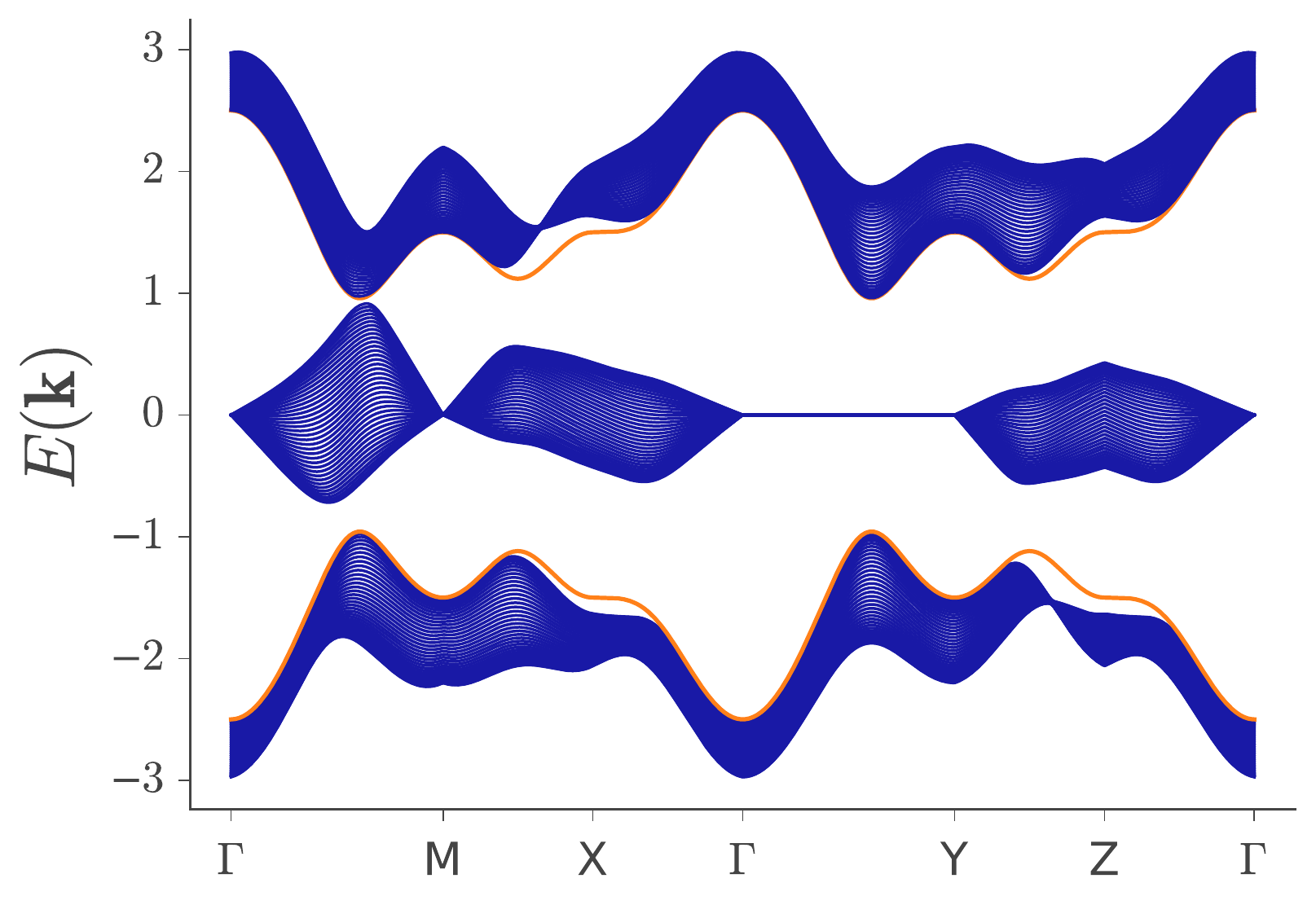}}
  \label{figurecheckerboardmodelspecialcasesa}}\quad
    \subfigure[]{
  {\includegraphics[width=0.35\textwidth]{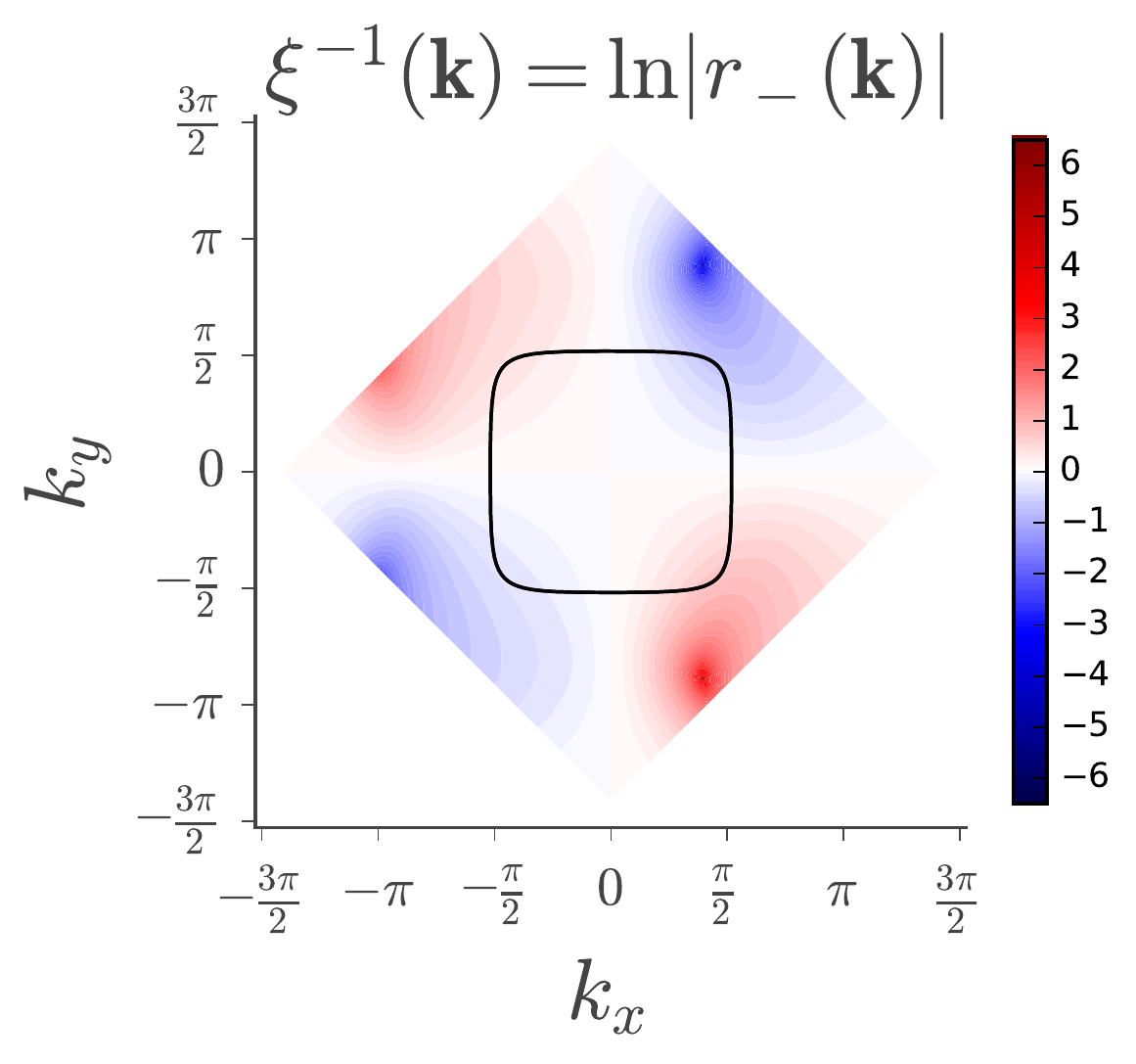}}
  \label{figurecheckerboardmodelspecialcasesb}}\quad
  \subfigure[]{
  {\includegraphics[width=0.5\textwidth]{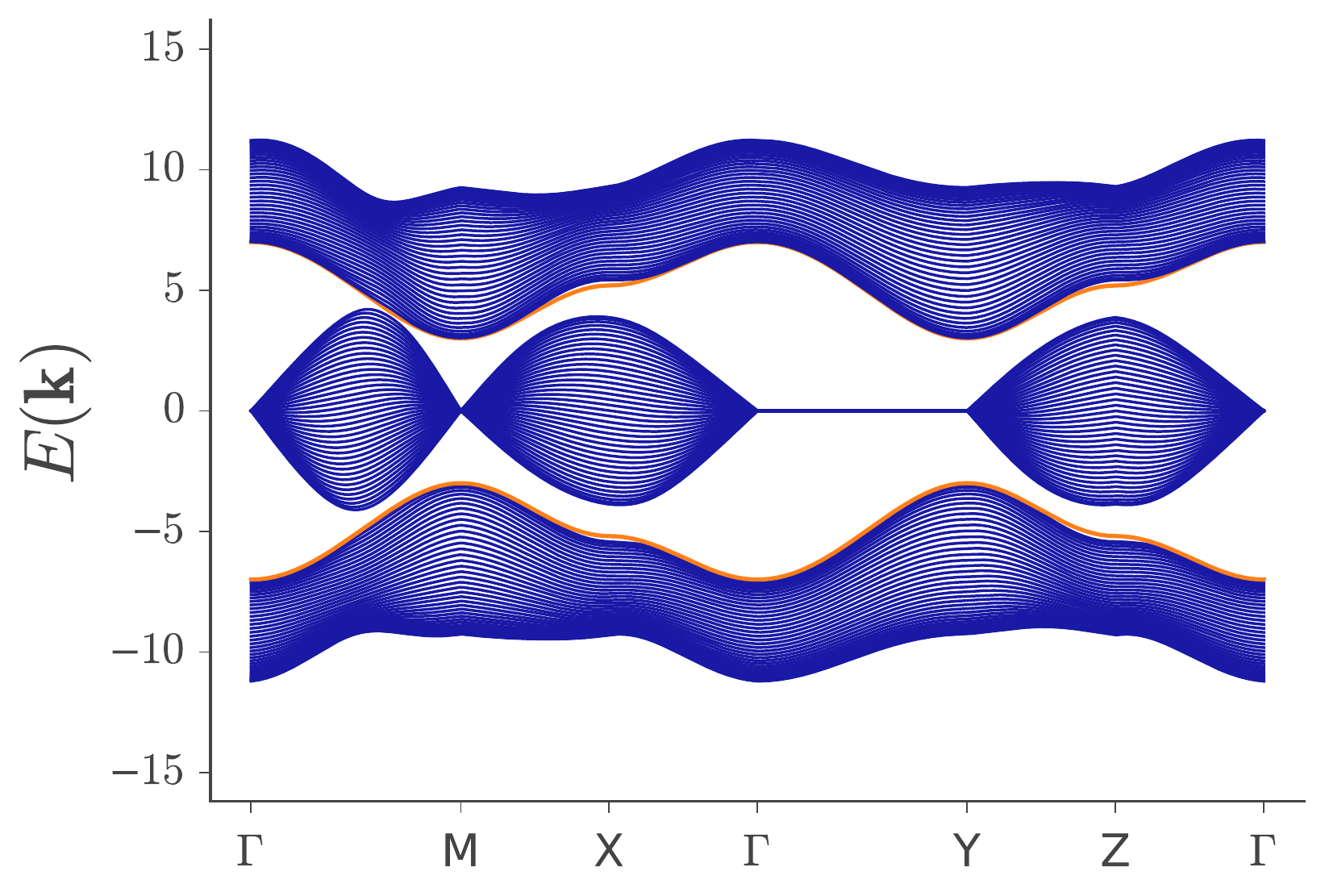}}
  \label{figurecheckerboardmodelspecialcasesc}}\quad
  \subfigure[]{
  {\includegraphics[width=0.35\textwidth]{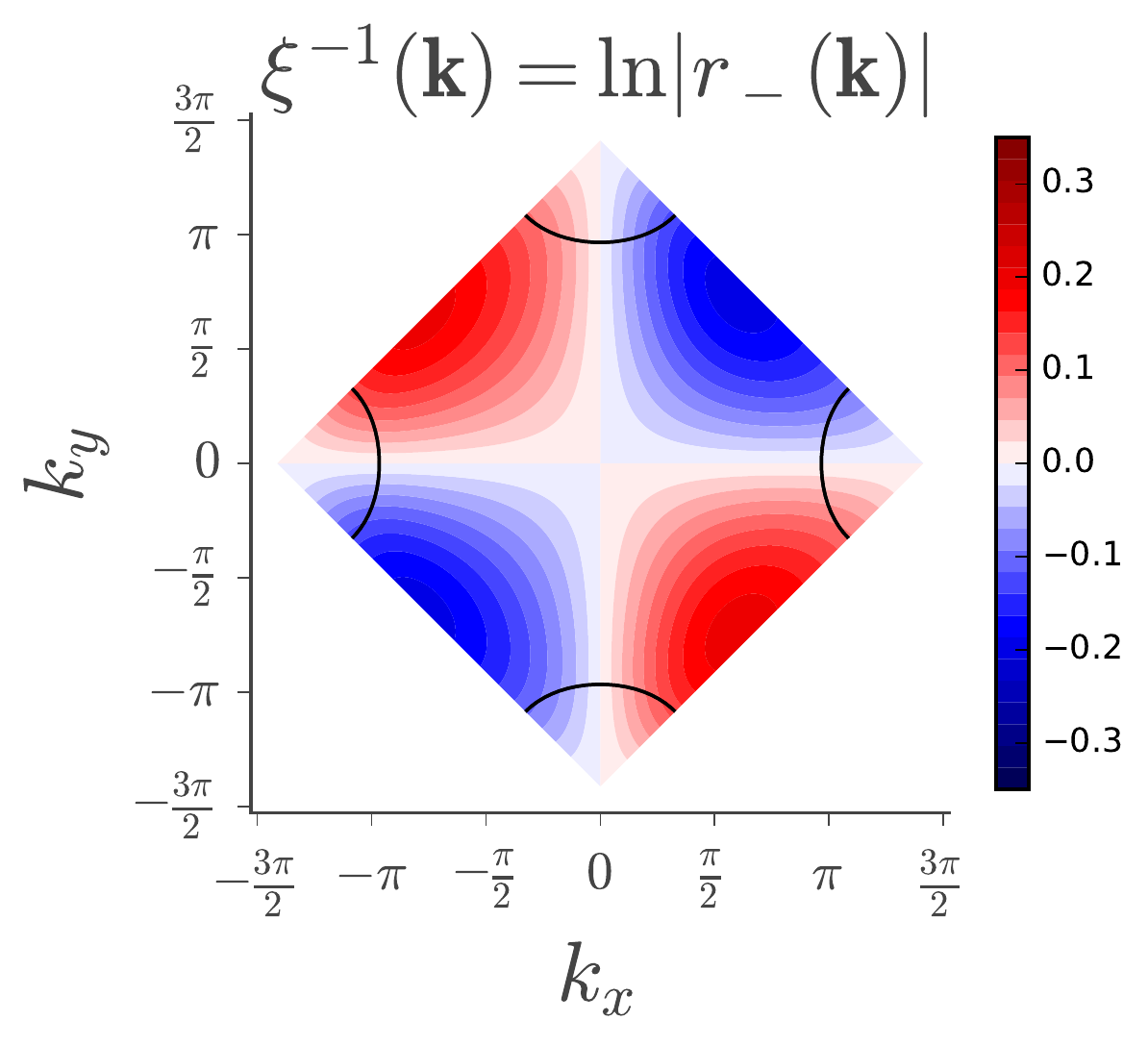}}
  \label{figurecheckerboardmodelspecialcasesd}}
  \caption{(a) and (b) Solutions to the checkerboard model with nonzero-Chern number with $t'/t = t_2/t = 1$ and $V/t = 0.5$ and $t_\perp/t = 0.57$. In (a), we show the energy spectrum with $N = 40$ and the energy given in units of $t$ along a certain path in the Brillouin zone. Even though, the $A$ lattice carries a Chern number, there are no Weyl points in the system. In (b), we show the inverse localization length $\xi^{-1} (\bfk) = {\rm ln}|r_-(\bfk)|$, where the black line is the solution to $E_-(\bfk)$ at the chemical potential $\mu/t = -1.6$. (c) and (d) Solutions with zero-Chern number with $t'/t = t_2/t = 1$ and $V/t = 5$ and $t_\perp/t = 3.1$. (c) The energy spectrum with $N = 40$ and the energy given in units of $t$. There are no Weyl points. (d) The inverse localization length with an equal-energy line at $\mu/t = -4$.}
  \label{figurecheckerboardmodelspecialcases}
\end{figure*}

We will now proceed by considering a specific layered checkerboard model which supports a Weyl semimetal phase at strong inter-layer coupling. The Hamiltonian of a single $A$ layer, as specified in Eq.~(\ref{equationgeneralhamiltonianwithboldmomentum}), is known to describe a Chern insulator when it has the following components
\begin{align}
d_0(\bfk) & = 0, \quad d_x({\bf k}) = t \, {\rm sin}\left(\frac{k_x}{\sqrt{2}} \right), \quad d_y({\bf k}) = t' \, {\rm sin}\left(\frac{k_y}{\sqrt{2}} \right), \nonumber \\
d_z ({\bf k}) & = V + t_2 \left[{\rm cos}\left(\frac{k_x + k_y}{\sqrt{2}}\right) + {\rm cos} \left(\frac{k_x - k_y}{\sqrt{2}}\right)\right], \nonumber
\end{align}
where $t$ and $t'$ are the nearest-neighbor hopping amplitudes, $t_2$ is the next-nearest-neighbor hopping amplitude, and $V$ is the staggering potential. The single-layer system has $C = 1$ at half-filling for $t'/t = t_2/t = 1$ and $|V/t| \leq 2$ and $C=0$ for $|V/t| > 2$. The solution of the inverse localization length $\xi^{-1} (\bfk) = {\rm log}|r_-(\bfk)|$ is shown in Fig.~\ref{figurecheckerboardmodelcompilationtperp1mu1b} for
a set of parameters in which the checkerboard lattice features a Chern insulator, and they agree with the solutions to Eq.~(\ref{equationconditionsurfaceswitchingcheckerboard}) explicitly mentioned in a footnote below.\footnote{We set $t'/t = t_2/t = V/t = 1$ such that they agree with the parameters used to plot the inverse localization length $\xi^{-1} (\bfk) = {\rm ln}|r_-(\bfk)|$ in Fig.~\ref{figurecheckerboardmodelcompilationtperp1mu1b}. The solutions for setting the numerator and denominator of $r_i(\bfk)$ zero are the same for $i = +$ and $i = -$. The numerator, i.e. $s = 1$, is zero when $\bfk = \left(0, \pm \sqrt{2} \pi \right)$, $\left(\pm \sqrt{2} \pi, 0 \right)$, $\left(- \frac{\pi}{2 \sqrt{2}}, \frac{3 \pi}{2 \sqrt{2}} \right)$, $\left( \frac{3 \pi}{2 \sqrt{2}}, - \frac{\pi}{2 \sqrt{2}} \right)$, $\left( \pm \frac{\sqrt{2}}{{\rm cos}(\mp \sqrt{5/2})}, \pm \frac{\sqrt{2}}{{\rm cos}(\pm 1/\sqrt{10})} \right)$, and for the denominator, $s = -1$, we find $\bfk = \left(0, \pm \sqrt{2} \pi \right)$, $\left(\pm \sqrt{2} \pi, 0 \right)$, $\left(- \frac{\pi}{2 \sqrt{2}}, -\frac{3 \pi}{2 \sqrt{2}} \right)$, $\left( \frac{3 \pi}{2 \sqrt{2}}, \frac{\pi}{2 \sqrt{2}} \right)$, $\left( \pm \frac{\sqrt{2}}{{\rm cos}(\mp \sqrt{5/2})}, \pm \frac{\sqrt{2}}{{\rm cos}(\mp 1/\sqrt{10})} \right)$. Note that these solutions can easily be generalized for different values of the parameters. While some of the zeros are common for the nominator and denominator, we generically find that they also have distinct zeros as is also the case in the example given.} We notice that $\xi^{-1} (\bfk)$ respects the symmetry of the lattice model, which has a mirror symmetry along the line $y = 0$. The black line in the plot corresponds to the energy contour of the surface state $E_-(\bfk)$ at the chemical potential $\mu/t = -0.9$. $\xi^{-1} (\bfk)$ changes sign four times when we trace the equal-energy line of the surface state meaning that this state switches surfaces an equal number of times. The solution to the boundary state in Eq.~(\ref{exactsolutionedgestate}) with $r_\pm(\bfk)$ given in Eq.~(\ref{equationcheckerboardmodelrofk}) thus describes a family of Fermi arcs, while a constant energy contour in the manifold of these solutions, as the one just described, describes a Fermi arc. 

Moreover, we find numerically that the two-dimensional surface band carries a Chern number whose value corresponds to the total number of checkerboard layers as given in Eq.~(\ref{equationchernnumberscaling}). The surface switching along the line $\Gamma - M$ in the Brillouin zone pins the presence of Weyl nodes to occur along these lines as shown in Figs.~\ref{figurecheckerboardmodelcompilationtperp1mu1c}-\ref{figurecheckerboardmodelcompilationtperp1mu1e}. Figures~\ref{figurecheckerboardmodelcompilationtperp1mu1c} and \ref{figurecheckerboardmodelcompilationtperp1mu1d} show the location of the Weyl nodes in the three-dimensional Brillouin zone from a top view in the $k_z$ direction and a two-dimensional cut at $k_y =0$, respectively, where red dots correspond to positive and blue dots to negative Berry charge. These nodes were found numerically by scanning the energy spectrum of the fully periodic lattice model over the entire Brillouin zone for energy gaps smaller than $E/t < 0.1$ and a subsequent computation of the Berry charge of a small sphere around these points. The numerical search for the Weyl points is greatly simplified by the fact that they are only allowed to occur at energies corresponding to the single-layer eigenvalues (the Fermi arcs traverse the Weyl points), and that they can occur only along the lines of diverging $r(\bfk)$  as shown in Fig.~\ref{figurecheckerboardmodelcompilationtperp1mu1e}. We see that the orange bands, which correspond to $E_\pm(\bfk)$ and are thus associated with the Fermi arcs, cross the bulk gaps exactly at the same momentum and energy values as found for the three-dimensional periodic model in Figs.~\ref{figurecheckerboardmodelcompilationtperp1mu1c} and \ref{figurecheckerboardmodelcompilationtperp1mu1d}. In Fig.~\ref{figurecheckerboardmodelberrycurvature}, we show plots for the Berry curvature for different $N$, which show that the value of the Berry curvature diverges along the line in the surface Brillouin zone where the Weyl nodes sit. The surface switching of the boundary state along the line $\Gamma - Y$ in the Brillouin zone takes place also in the absence of Weyl nodes and comes about due to the complete attachment of the boundary-state energy bands to the bulk yielding its corresponding wave function completely delocalized by which surface switching is facilitated. The energy band at $E/t = 0$ along $\Gamma - Y$ is highly degenerate and corresponds to the energy solutions of the intermediate $B$ (green) sites. That the states belonging to these sites have zero energy along the line $k_x = 0$ can be understood by realizing that hopping from the $A$ lattices to the intermediate $B$ lattices only take place in the $x$ direction. The boundary state switching surfaces four times is also confirmed by the plot of the weight of the wave function of this state on each individual checkerboard layer shown in Fig.~\ref{figurecheckerboardmodelcompilationtperp1mu1f}. We see that the surface state localizes to the top and bottom layer in complete accordance with Fig.~\ref{figurecheckerboardmodelcompilationtperp1mu1b}. Upon varying the value of the staggering potential with upper bound $|V/t|\leq 2$ such that the checkerboard lattices remain in the nonzero Chern number regime, we find that the number, location in the Brillouin zone and energy value of the Weyl nodes change. However, they always remain on the $k_y = 0$ line. As we increase the inter-layer coupling, we find numerically that only two Weyl nodes remain, which are pinned in the three-dimensional Brillouin zone slightly below $k_z = 0$ at the points $M$ and $-M$ in the $k_x k_y$-plane with energies going to $E/t \approx 1$ and $-1$, respectively. 

It is tempting to assume that the Weyl nodes and Fermi arcs disappear when the Chern number carried by the individual checkerboard lattices goes to zero. However, if the Hamiltonian of the checkerboard lattice in the trivial phase is fine-tuned to remain close to the topological phase transition, we find that a Weyl-semimetal phase is still supported by the three-dimensional lattice as shown in Fig.~\ref{figurecheckerboardmodelcompilationtperp1mu2half} for $V/t = 2.1$. Figures~\ref{figurecheckerboardmodelcompilationtperp1mu2halfa} and \ref{figurecheckerboardmodelcompilationtperp1mu2halff} shows that the surface state still switches surfaces four times, albeit with a weaker localization to the surfaces as before. Again, the surface switching along the line $\Gamma - M$ is facilitated by the presence of Weyl nodes, whose location and energy values are shown in Figs.~\ref{figurecheckerboardmodelcompilationtperp1mu2halfb} and \ref{figurecheckerboardmodelcompilationtperp1mu2halfc} and they correspond to what we observe in the energy spectrum shown in Fig.~\ref{figurecheckerboardmodelcompilationtperp1mu2halfd}. Similarly, the surface switching along the line $\Gamma - Y$ is due to the attachment of the surface state energy to the bulk bands. Upon increasing $V/t$, such that the Hamiltonian on the checkerboard lattice moves further away from the topological phase transition, we find that the Weyl nodes disappear completely when $V/t \gtrsim 2.55$.

Lastly, we briefly mention two more cases. In the first, the $A$ lattice is in the Chern-insulator phase but the energy spectrum of the three-dimensional model does not support a Weyl phase as shown in Fig.~\ref{figurecheckerboardmodelspecialcasesa}, which is due to the weak inter-planar coupling. In Fig.~\ref{figurecheckerboardmodelspecialcasesb}, we have plotted the inverse localization length and one can indeed see that there is a Fermi arc, which carries a Chern number. In the second case, we present an example where the $A$ lattice is in the trivial topological phase, there are no Weyl points in the three-dimensional lattice as shown in Fig.~\ref{figurecheckerboardmodelspecialcasesc}, but there are still Fermi arcs as shown in Fig.~\ref{figurecheckerboardmodelspecialcasesd}. These Fermi arcs carry a zero Chern number and are thus not topologically protected.

\subsection{Three dimensions, second example: Weyl semimetals and layered Chern insulators in pyrochlore iridate slabs} \label{sectionsix}

\begin{figure}[t]
\centering
\adjustbox{trim={0\width} {0.4\height} {0\width} {0\height},clip}
  {\includegraphics[scale=0.25]{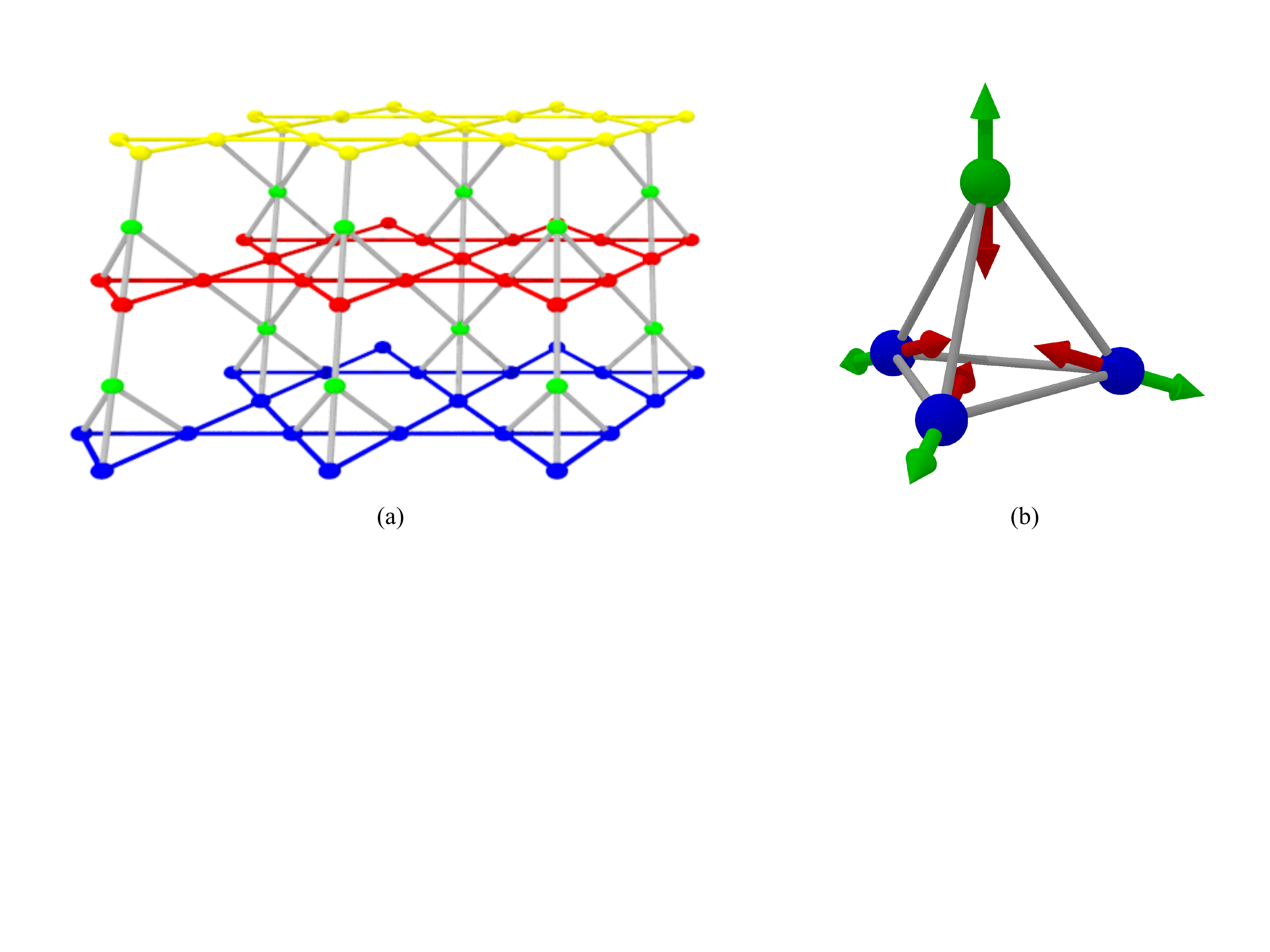}}
\caption{(a) The pyrochlore model featuring kagome lattices stacked on top of each other with a triangular lattice (green) in between. The kagome layers are shifted relative to each other and repeat every third layer. (b) The preferred spin directions pointing towards (red) or away from (green) the center.}
\label{figurepyrochloremodel}
\end{figure}

\begin{figure*}[t]
\centering
  \subfigure[]{
    {\includegraphics[height=0.31\textwidth]{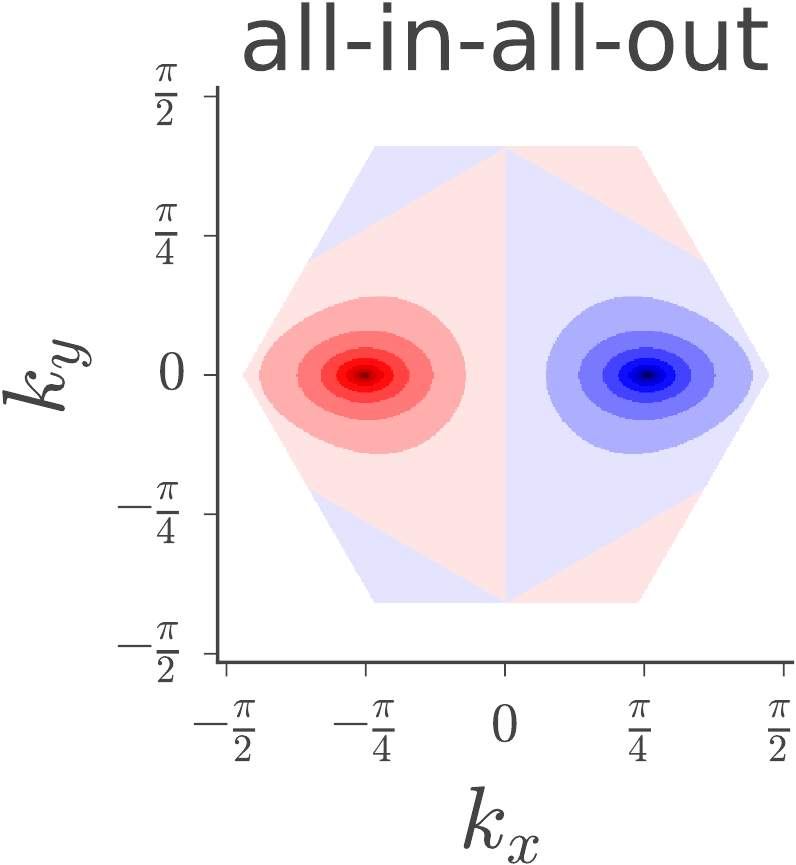}}
    \label{figurerofka}}\quad
  \subfigure[]{
    {\includegraphics[height=0.31\textwidth]{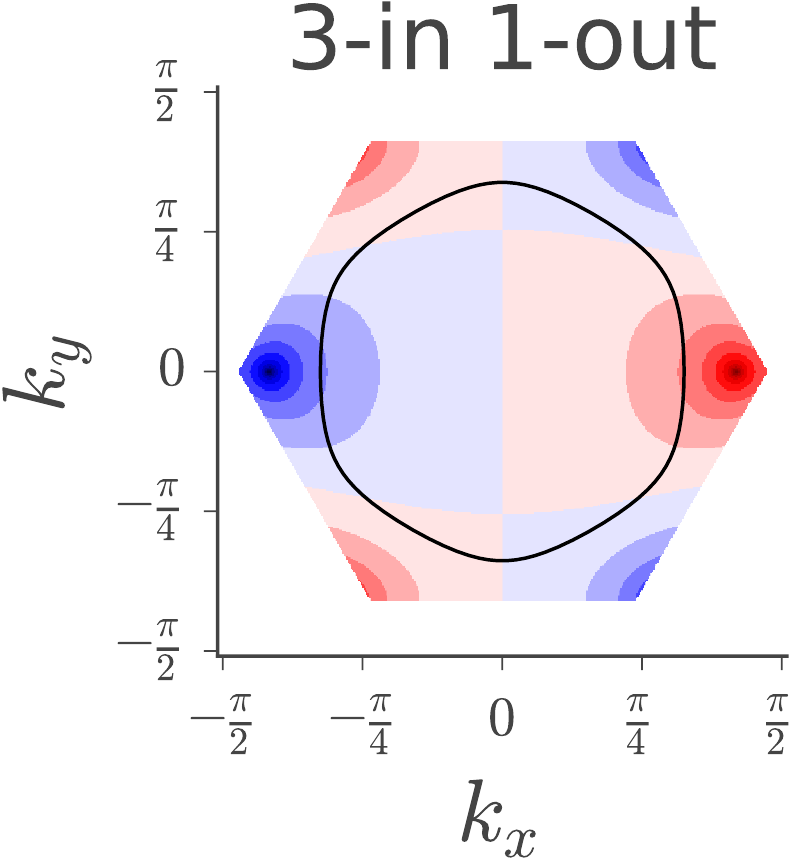}}
    \label{figurerofkb}}\quad
  \subfigure[]{
    {\includegraphics[height=0.31\textwidth]{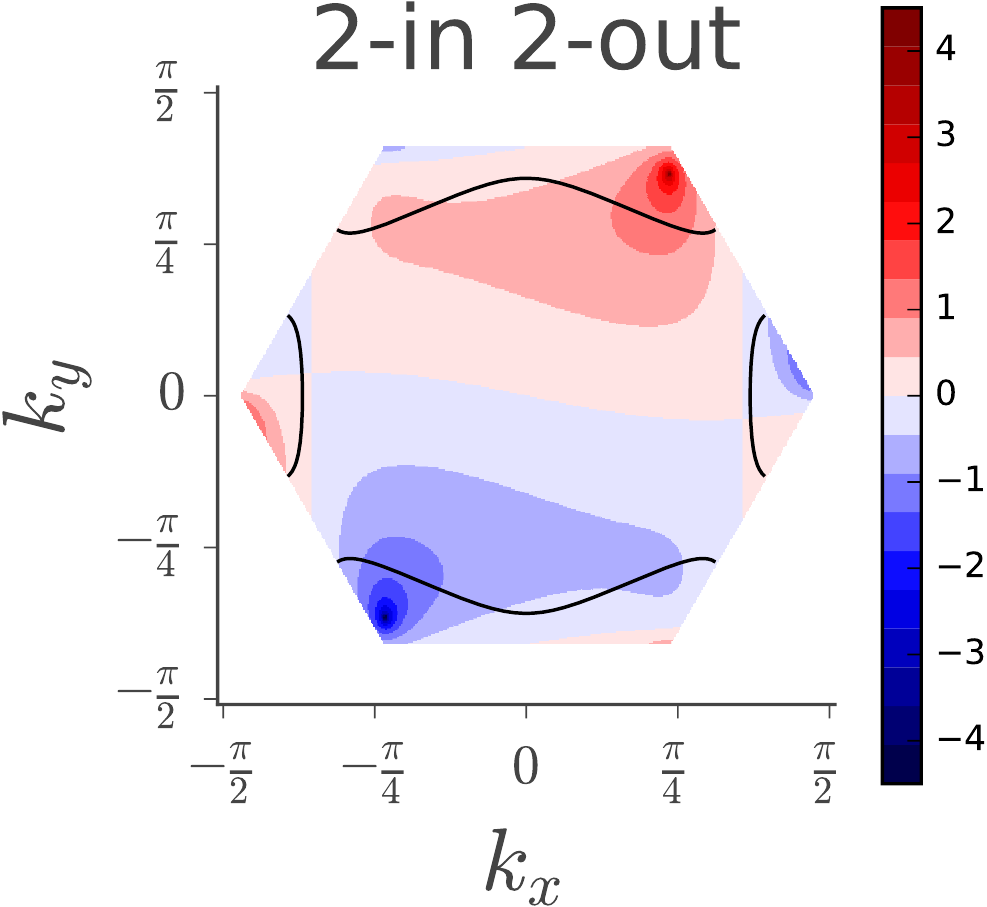}}
    \label{figurerofkc}}\quad
\subfigure[]{
  {\includegraphics[width=0.3\textwidth]{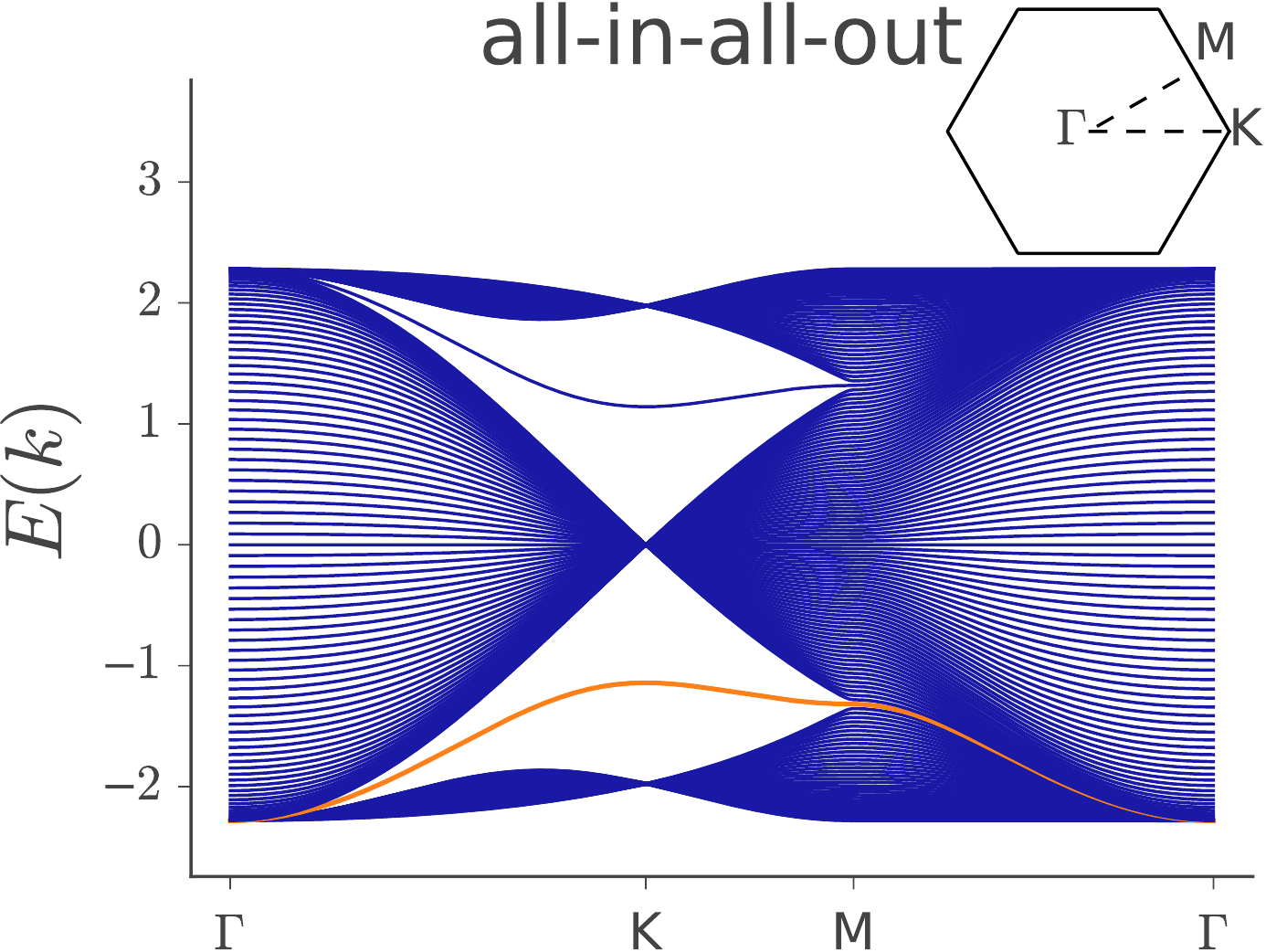}}
  \label{figurerofkd}}\quad
    \subfigure[]{
  {\includegraphics[width=0.3\textwidth]{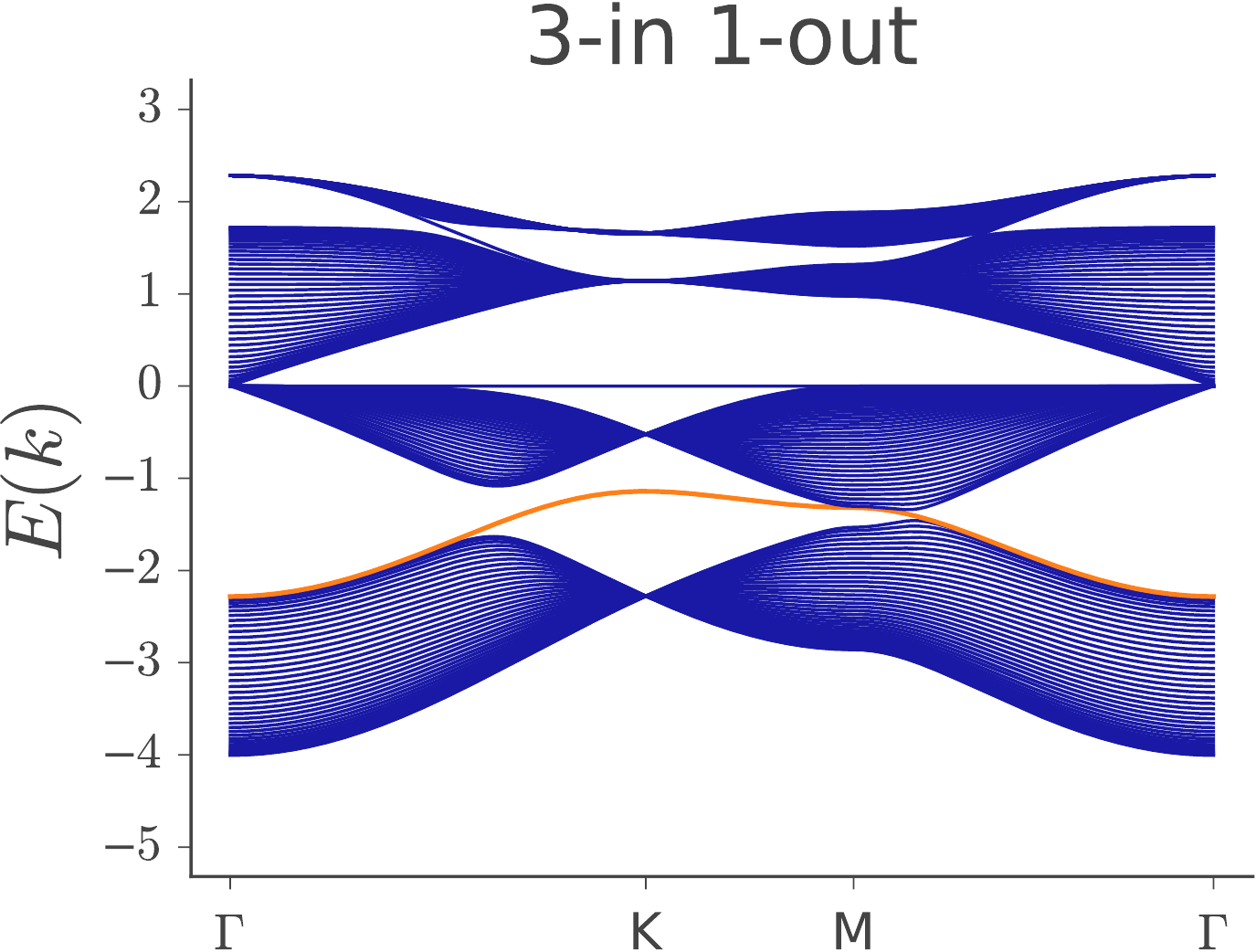}}
  \label{figurerofke}}\quad
  \subfigure[]{
  {\includegraphics[width=0.3\textwidth]{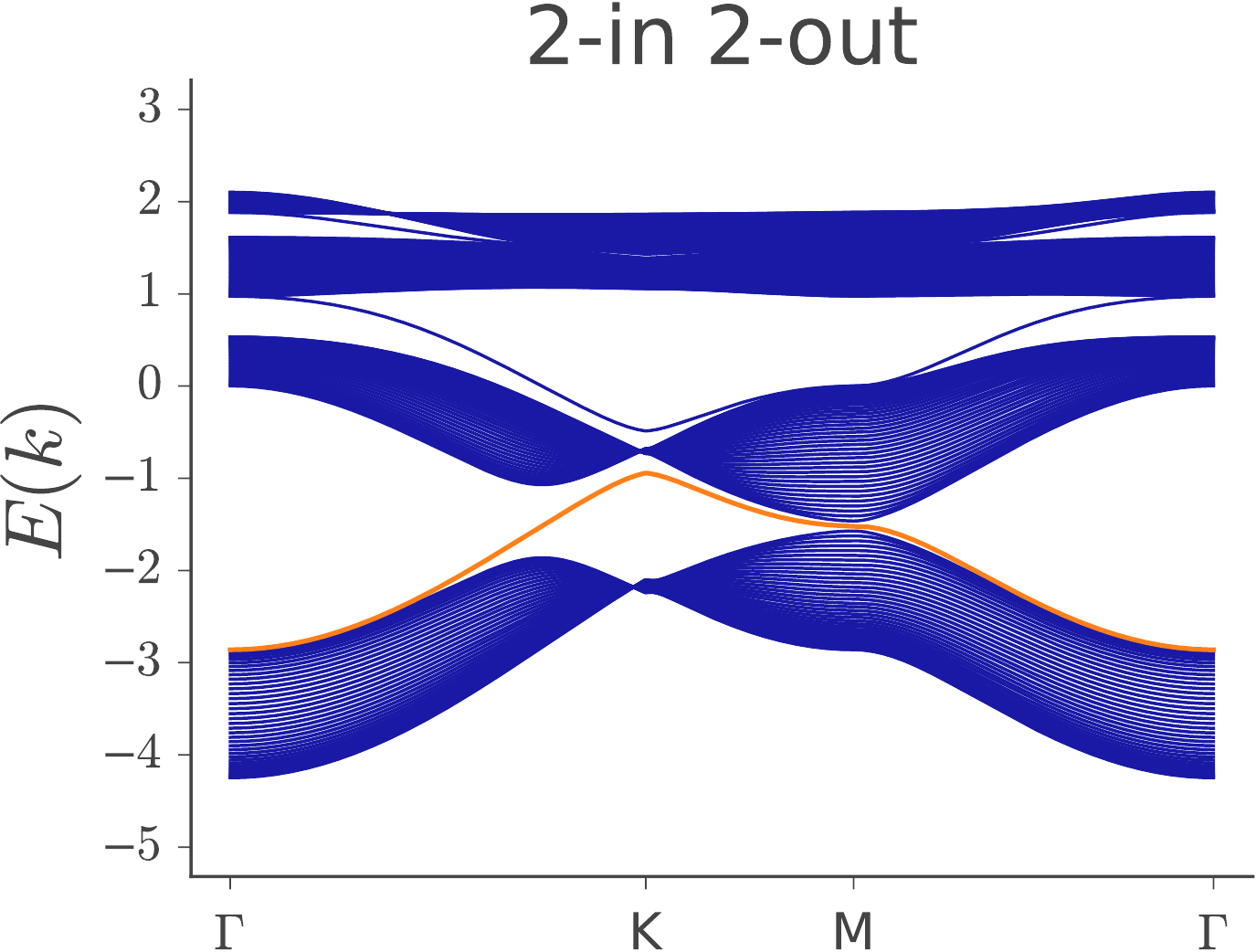}}
  \label{figurerofkf}}
\caption{(a) - (c) The inverse localization length $\xi^{-1} (\bfk) = {\rm ln}|r_1(\bfk)|$ for the lowest surface states in the pyrochlore model shown for different spin configurations indicated in (d)-(f), respectively, with hopping parameters $t_{\mathrm{ SOC}}/t =0.1$. The black line in (b) and (c) is the energy contour at $\mu/t = -1.4$. (d) - (f) The energy spectra for different spin configurations with $N = 40$ and  $t_{\mathrm{ SOC}}/t =0.1$. Energy is given in units of $t$.}
\label{figurerofk}
\end{figure*}

A pyrochlore lattice is built from alternating kagome ($A$) and triangular ($B$) layers, where successive kagome lattices are shifted relative to each other as shown in Fig.~\ref{figurepyrochloremodel}(a). The lattice structure implies that conditions (i)-(iv) are fulfilled hence our general expression for the surface states applies. 

In this section, we use the formalism introduced in section~\ref{sectiontwo} to show that pyrochlore slabs very generically support Fermi-arc-like surface states, regardless of the presence of Weyl points in the bulk. The novel scenario of Fermi-arc-like surface states in the absence of Weyl points is particularly pertinent for recent experiments.

A special case of the pyrochlore model where the spin-orbit coupling was restricted to only the kagome layers has been discussed extensively by two of the authors in Refs.~\onlinecite{trescherbergholtz, bergholtzliutreschermoessnerudugawa}. 
Here we consider more realistic models of pyrochlore lattices making direct contact to real materials and experiments, namely  Eu$_2$Ir$_2$O$_7$ \cite{fujita2015} and Nd$_2$Ir$_2$O$_7$ \cite{gallagher2016} for which pertinent slabs have recently been grown and for potential future experiments on yet to be grown slabs of Pr$_2$Ir$_2$O$_7$. 
The model includes spin degree of freedom and has two parameters: the nearest-neighbor-hopping strength $t$ and the spin-orbit-coupling parameter $t_{\rm SOC}$.
It is directly based on the model in Ref.~\onlinecite{yamaji}, up to a rotation of the [111] direction in the $z$ direction to describe the layered structure as in the models presented earlier.
In contrast to the models discussed earlier, we do not assume a spin polarization in a globally fixed direction but we make use of the internal spin ordering of pyrochlore materials. 
This procedure is justified by the observation that different materials featuring the pyrochlore lattice structure realize phases with fixed but different spin configurations at low temperatures \cite{onose2004, bramwell2001}.
In the bulk, the pyrochlore lattice can be seen as a collection of corner sharing tetrahedra, where the lattice sites are located at the corners. 
In all these spin-ordered phases, the preferred orientation of a spin in the pyrochlore lattice is to point either towards or away from the center of the tetrahedron. A spin configuration with $n$ spins pointing inwards is called $n$-in-(four-$n$)-out, which is equivalent to the (four-$n$)-in-$n$-out configuration. The configuration four-in-zero-out is often called the ``all-in-all-out'' configuration as for any tetrahedron either all spins are pointing outwards or all spins are pointing inwards [\onlinecite{fujita2015, arima2013}]. In order to obtain a tight-binding model with a fixed spin configuration, we apply a different spin rotation on each site to transform the Hamiltonian to the all-in-all-out spin basis as shown in Fig.~\ref{figurepyrochloremodel}(b),
and finally we project the Hamiltonian to the subspace of the spin configuration of interest. 
The Hamiltonians for the different spin configurations are listed in the Appendix.

Based on geometric considerations and intuition developed in the models before, we expect localized states for any spin configuration,
as the local constraint involves three unit cells in the upper layer. Furthermore, the three-in-one-out and two-in-two-out configurations have topological nontrivial bands in the single-layer case, and hence fulfill condition (v). Applying Eq.~(\ref{equationrofkintermsofbasishamiltonianeigenstatesandphase}) to the Hamiltonians given in Eq.~(\ref{eq:pyrochlore}) reveals that all of these models indeed have nontrivial surface states.
The results for $r(\bf{k})$ are shown in Figs.~\ref{figurerofka}-\ref{figurerofkc}.
While in the all-in-all-out configuration the surface state might be ``hidden'' by the bulk bands, in the two other configurations we find Fermi arcs when choosing the Fermi level in the bulk gap. The two-in-two-out configuration possesses Fermi arcs even without having Weyl points in the bulk.

\begin{figure}[t]
\centering
    \includegraphics[width=0.47\textwidth]{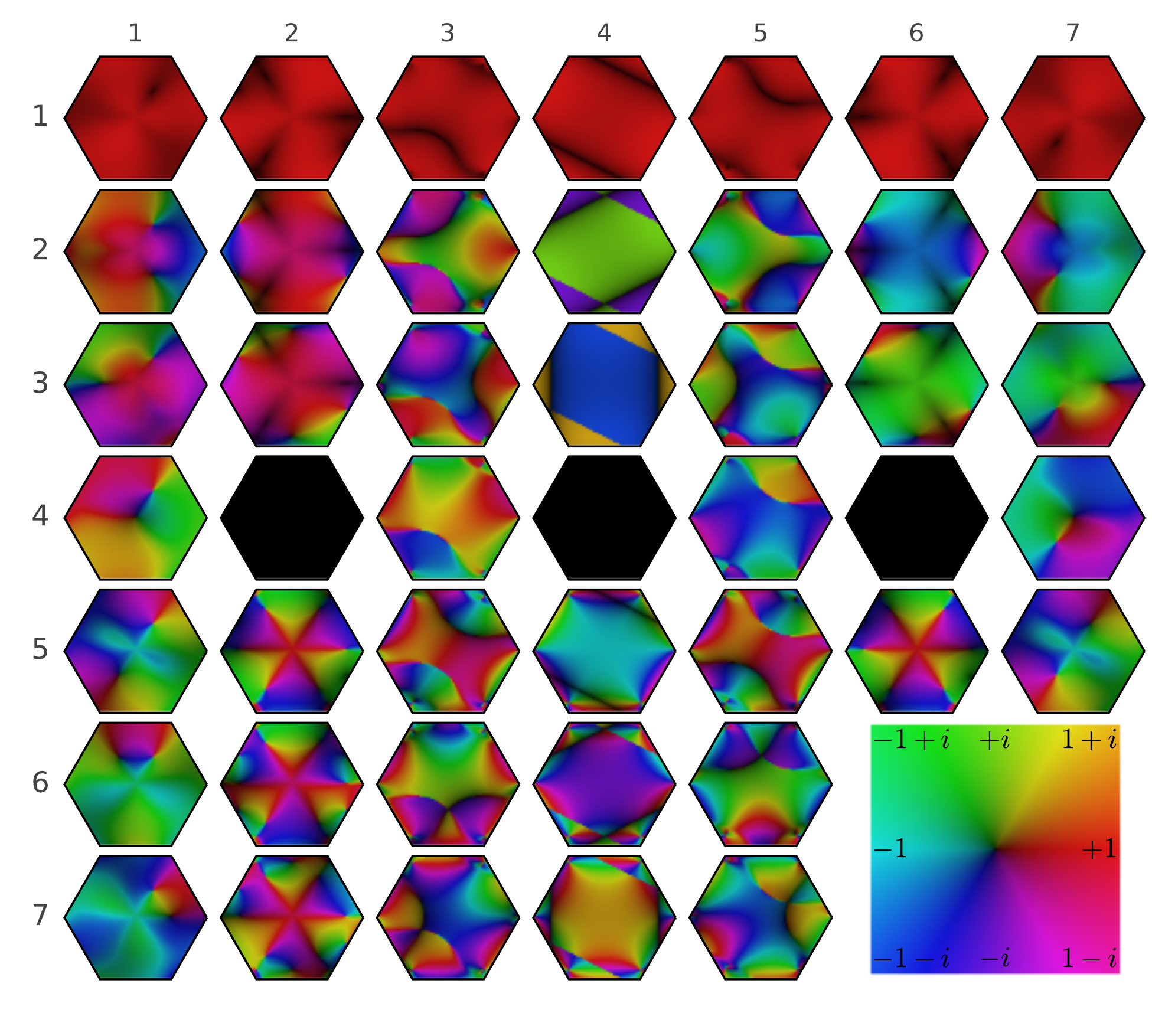}
\caption{The wave function components for a two layer slab of the all-in-all-out model. The columns correspond to the different eigenstates and the rows correspond to the sites, the wave function values are shown throughout the BZ using the domain coloring method shown in the right bottom corner.}
\label{pyrochlore-wavefunctions}
\end{figure}

In the all-in-all-out configuration, the model's two parameters $t$ and $t_{\rm SOC}$ only occur in the linear combination $t + \sqrt{2}t_{\rm SOC}$ after the spin transformation, hence we are left with a single effective parameter in this case.
By diagonalizing the $(3\times3)$ $A$-lattice Hamiltonian $\mathcal{H}_{\bf k, 4-0}$, we find a completely flat band, which is topologically trivial. Note, however, that this spin configuration cannot be physically realized in a single layer, because it would lead to a large magnetization, and should be viewed only in combination with the fourth spin on the intermediate $B$ layers.
In Fig.~\ref{pyrochlore-wavefunctions}, a graphical representation of the full wave functions of a slab with two $A$ lattices and one intermediate $B$ lattice in the all-in-all-out configurations is shown. One easily recognizes the three bands that are given by our construction as they have zero weight (shown in black) on the intermediate $B$ site with index 4.

\begin{figure}[t]
\centering
    \includegraphics[width=0.5\textwidth]{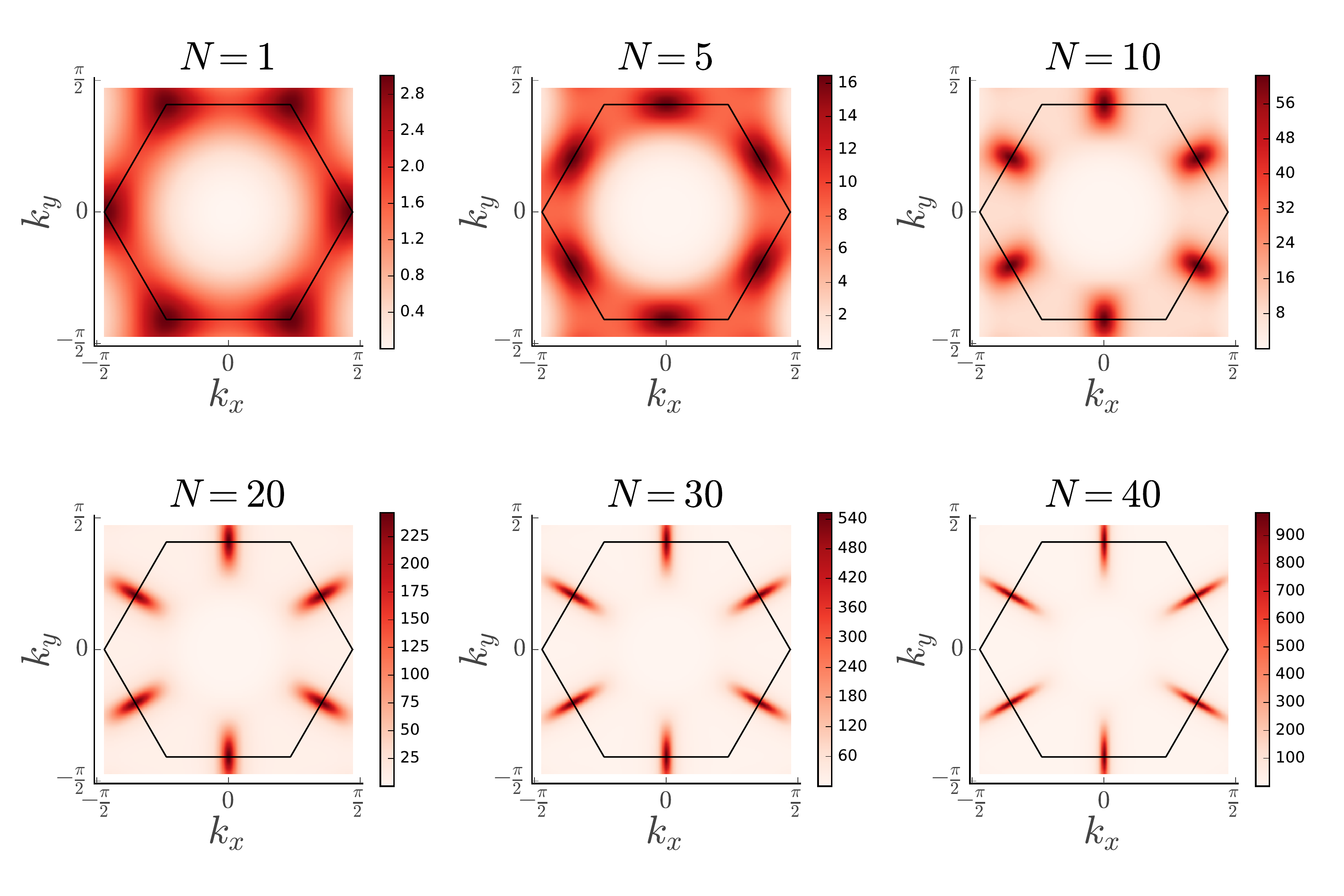}
\caption{Berry curvature associated with $\Psi_1 ({\bf k})$ for the pyrochlore model with hopping parameters $t_{\mathrm{ SOC}}/t =0.1$ in the three-in-one-out configuration.}
\label{equationpyrochloreberrycurvature}
\end{figure}

The three-in-one-out configuration is not uniquely defined in a finite slab, as the surface normals represent a special direction in the model.
For the sake of brevity, we restricted the discussion to the most natural choice of the three-in-one-out configurations, where we take the single ``out"-spin to be on the intermediate site [green in Fig.~\ref{figurepyrochloremodel}(a)]. Hence the kagome layers of the all-in-all-out and the three-in-one-out configurations have the same $A$-lattice Hamiltonian $\mathcal{H}_{\bf k}$. 
In contrast to the existence of the surface states in all spin configurations, only the three-in-one-out model harbors Weyl points in the bulk, and does so only for small values of $t_{\rm SOC}/t$ in accordance with Ref.~\onlinecite{goswami2016}, as is shown in Figs.~\ref{figurerofkd}-\ref{figurerofkf}. 
In Fig.~\ref{equationpyrochloreberrycurvature}, the Berry curvature is shown for this model for a different number of $A$ lattices. One indeed sees a divergence of the Berry curvature along those lines in the Brillouin zone where the Weyl points appear.

\subsection{Three dimensions, final example: trivial states on layered honeycomb lattices} \label{sectionfive}

\begin{figure}[b]
    \adjustbox{trim={0.1\width} {.45\height} {0\width} {0.1\height},clip}
    {\includegraphics[width=0.55\textwidth]{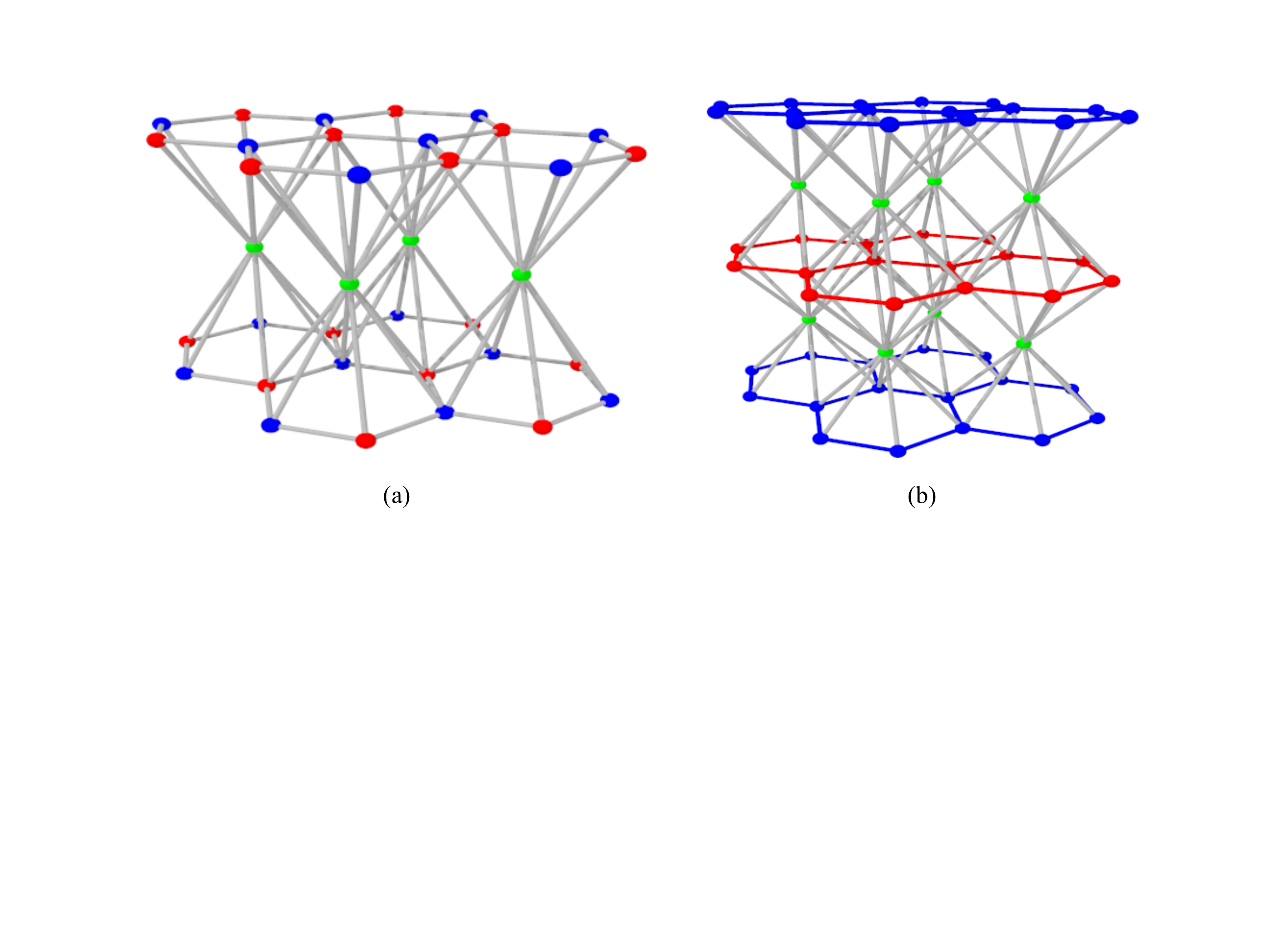}}
\caption{(a) The $A$ lattice of the honeycomb model featuring honeycomb lattices (red and blue sites) stacked on top of each other with a triangular lattice (green) in between. Note the sublattice exchange in the alternating honeycomb lattices. (b) The honeycomb model where one $A$ lattice is composed of one blue and red honeycomb lattice and the green triangular lattice in between.}
\label{figurehoneycombmodel}
\end{figure}

In this section, we study tight-binding models inspired by KHg$X$ ($X = $ As, Sb, Bi) whose underlying lattice structure is a three-dimensional analog of the two-dimensional models in Figs.~\ref{figurechainmodelsb} and \ref{figurechainmodelsc}. The lattice is formed by stacked honeycomb lattices, which are connected via triangular lattice layers. In Fig.~\ref{figurehoneycombmodel}(a), we show the $A$ lattice of this model, which has five sites in the unit cell. Figure~\ref{figurehoneycombmodel}(b) features the honeycomb model with one and a half $A$ lattices.  

The surface states of KHg$X$ are known to be intriguing featuring novel so-called hourglass fermions [\onlinecite{wangalexandracavabernevig}]. However, the surface that we study is not expected to support these states, which we will here corroborate by showing that our exact solution inevitably leads to trivial states delocalized over the full sample. While the actual KHg$X$ materials are time-reversal symmetric we here treat ``half" of the model as this turns out to be sufficient for understanding the absence of surface states. Our argument proceeds in a very similar fashion as for the analogous two-dimensional lattices [Figs.~\ref{figurechainmodelsb} and \ref{figurechainmodelsc}] that defy surface states. 

To describe the honeycomb layers, we use Haldane's model [\onlinecite{haldane}], such that the Hamiltonian $\mathcal{H}_\bfk^{\rm honey}$ and energy for the odd-numbered honeycomb layers are given in Eqs.~(\ref{equationgeneralhamiltonianwithboldmomentum}) and (\ref{equationgeneralenergywithboldmomentum}), respectively, with
\begin{align}
& d_0 (\bfk) = t_2 \, {\rm cos} \phi \, \sum_{i=1}^3 {\rm cos}({\bf k} \cdot \boldsymbol\delta'_i), \quad d_x({\bf k}) - i d_y({\bf k}) = t \, \gamma_\bfk , \label{equationhoneycombhamiltonian} \\
& d_z({\bf k}) = V - t_2 \, {\rm sin} \phi \, \sum_{i=1}^3 {\rm sin}({\bf k} \cdot \boldsymbol\delta'_i), \nonumber
\end{align}
where $t$ ($t_2$) is the (next-)nearest-neighbor hopping amplitude, $\phi$ is the phase picked up in a next-nearest-neighbor hopping, which we from now on set as $\phi = \pi/2$, $V$ is the staggering potential, $\gamma_\bfk \equiv \sum_{i=1}^3 {\rm exp}\left(i {\bf k} \cdot \boldsymbol\delta_i\right)$ and $\boldsymbol\delta_i$ ($\boldsymbol\delta'_i$) are the (next-)nearest-neighbor vectors specified in a footnote below.\footnote{The nearest-neighbor vectors read $\boldsymbol\delta_1 = (1, \, \sqrt{3})/2$, $\boldsymbol\delta_2 = (1, \, -\sqrt{3})/2$, and $\boldsymbol\delta_3 = -(1, \, 0)$. The next-nearest-neighbor vectors are $\boldsymbol\delta'_1 = (3, \, \sqrt{3})/2$, $\boldsymbol\delta'_2 = (-3, \, \sqrt{3})/2$ and $\boldsymbol\delta'_3 = (0, \, -\sqrt{3})$, where the complete set of next-nearest-neighbor vectors is given by $\{\pm\boldsymbol\delta'_i\}$.} The Chern number at half-filling in a single layer is $C = 1$ for $t/t_2 = 1$ and $|V/t_2| \lesssim 1.75$ and zero otherwise. The Hamiltonian for the even-numbered honeycomb layers is simply given by $\mathcal{H}_{-\bfk}^{\rm honey}$, such that the Hamiltonian for the $A$ lattice $\mathcal{H}_{\bfk}$ reads
\begin{equation*}
\mathcal{H}_{\bfk} = \left(\begin{array}{ccccc}
\multicolumn{2}{c}{\multirow{2}{*}{$\mathcal{H}_{\bfk}^{\rm honey}$}} & t_{\perp} \, \gamma_{-\bfk} & \multicolumn{2}{c}{\multirow{2}{*}{0}} \\
& & t_{\perp}  \, \gamma_{\bfk} & & \\
t_{\perp} \, \gamma_{\bfk} & t_{\perp} \, \gamma_{-\bfk} & 0 & t_{\perp} \, \gamma_{-\bfk} & t_{\perp}  \, \gamma_{\bfk} \\
\multicolumn{2}{c}{\multirow{2}{*}{0}} & t_\perp \, \gamma_{\bfk}  & \multicolumn{2}{c}{\multirow{2}{*}{$\mathcal{H}_{-\bfk}^{\rm honey}$}} \\
& & t_{\perp}  \, \gamma_{-\bfk} & &  \end{array}\right),
\end{equation*}
where we used that $\gamma_{\bfk}^* = \gamma_{-\bfk}$. The Hamiltonian for $N$ $A$ lattices is given in Eqs.~(\ref{generalmultilayerhamiltonian}) and (\ref{generalmultilayerhamiltonianperp}) with $h_{\bfk} = 0$, $t_{\perp, s, \alpha} = t_\perp \in \mathbb{R} \, \forall \, s, \,\alpha$, and $f_{A,4}(\bfk) = f_{B,2}(\bfk) = t_\perp \, \gamma_{\bfk}$ and $f_{A,5}(\bfk) = f_{B,1}(\bfk) = t_\perp \, \gamma_{-\bfk}$.

We first discuss the case where the staggering potential is zero, i.e. $V = 0$, such that $\mathcal{H}_{-\bfk}^{\rm honey} = \sigma_x \mathcal{H}_{\bfk}^{\rm honey} \sigma_x$. As in the analogous two-dimensional examples we notice that the $A$ lattice itself exists out of two sub-$A$ lattices and one sub-intermediate $B$ lattice. Solving the Schr\"odinger equation for these $n_{\rm sub} = 2$ solutions yields
\begin{equation*}
\ket{\Phi_\pm (\bfk)} \doteq \tilde{\mathcal{N}}_\pm(\bfk) \begin{pmatrix}
\phi_{\pm,1} (\bfk) \\
\phi_{\pm,2} (\bfk) \\
0 \\
s_\pm (\bfk)\phi_{\pm,2} (\bfk) \\
s_\pm (\bfk) \phi_{\pm,1} (\bfk) \\
\end{pmatrix},
\end{equation*}
where we have absorbed the sublattice exchange into the wave function, $\tilde{\mathcal{N}}_\pm(\bfk)$ is the normalization factor, and we see immediately that $s_\pm (\bfk) = -1$. This leads to
\begin{equation*}
r_\pm(\bfk) = - s_\pm (\bfk) \frac{\gamma_{-\bfk} \, \phi_{\pm,2} (\bfk) + \gamma_{\bfk} \, \phi_{\pm,1} (\bfk)}{\gamma_{\bfk} \, \phi_{\pm,1} (\bfk) + \gamma_{-\bfk} \, \phi_{\pm,2} (\bfk)} = 1.
\end{equation*}
The other three solutions have the solution $\Phi_i ({\bf k}) = \bigoplus_{s=1}^5 \phi_{i,s} (\bfk)$ with $i = 1,2,3$ and the energy $E_i$, and we again find $|r_i (k_x)| = 1, \, \forall i$.

When $V \neq 0$, we no longer find two solutions whose wave function has zero weight on the sub-intermediate $B$ site. This is simply the case because the Hamiltonian on the even-numbered honeycomb layers $\mathcal{H}_{-\bfk}^{\rm honey}$ has wave function solutions who can only be related to the wave function solutions of the Hamiltonian $\mathcal{H}_\bfk^{\rm honey}$ via $\bfk \rightarrow -\bfk$ such that a solution $\ket{\Phi_\pm(\bfk)}$ cannot be defined in terms of the wave function solutions of $\mathcal{H}_\bfk^{\rm honey}$ alone. However, we still find solutions of the form Eq.~(\ref{exactsolutionedgestate}) with $|r_i (k_x)| = 1, \, \forall i$ with $i = 1,2,3,4,5$ indicating that this system remains trivial.

\begin{figure}[t]
\centering
  \includegraphics[width=0.45\textwidth]{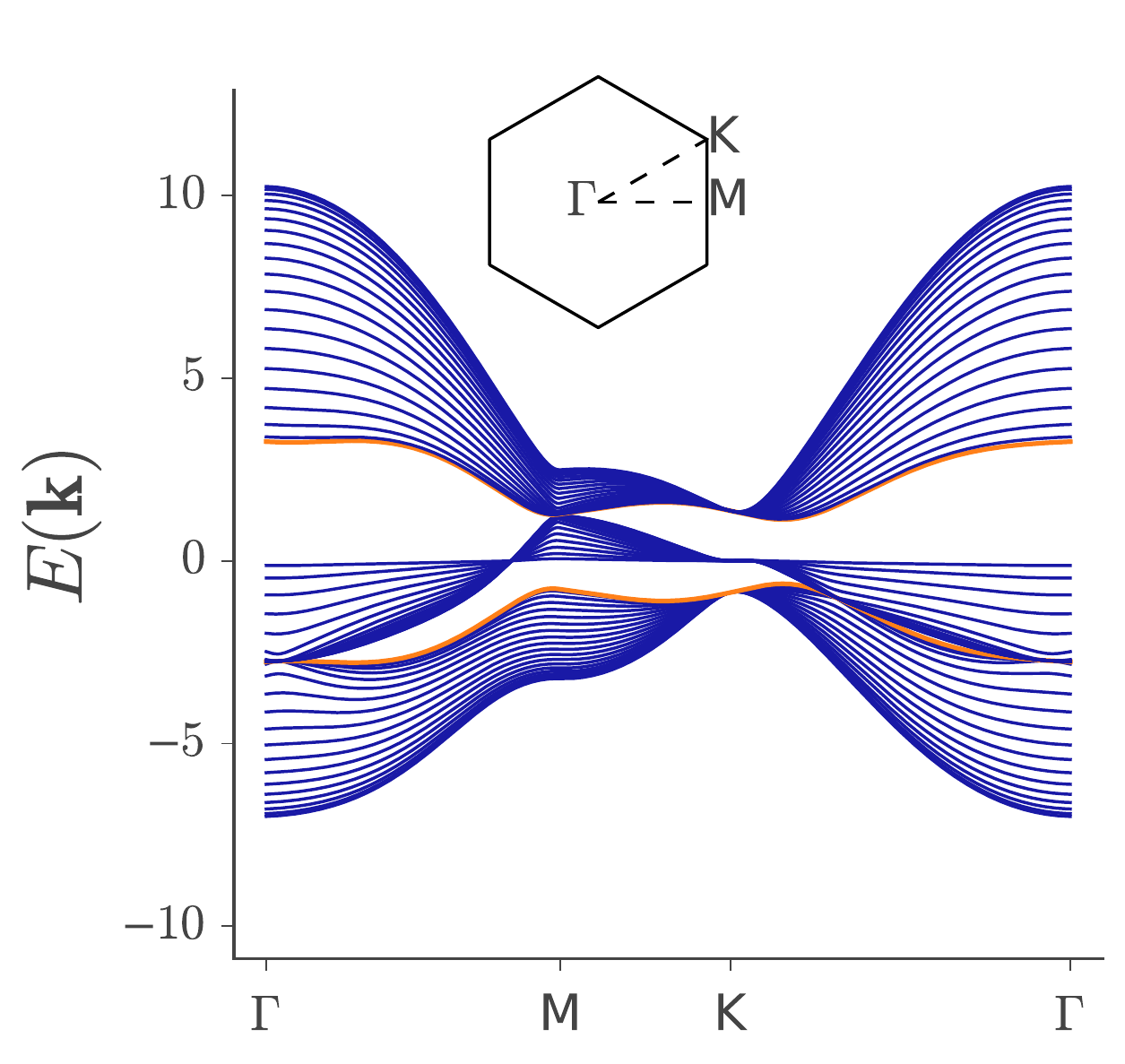}
\caption{Plot of the energy spectrum of the honeycomb model with $N = 20$ $A$ lattice layers and the energy given in units of $t$, with $t_2/t = t_\perp/t = 1$ and $V/t = 0.5$. The Brillouin zone is shown as an inset in the figure. The orange bands correspond to the exactly-solvable states. The total Chern number is one.}
\label{figurehoneycombmodelenergyspectrum}
\end{figure}

The energy spectrum for this Hamiltonian along a path in the two-dimensional Brillouin zone is shown in Fig.~\ref{figurehoneycombmodelenergyspectrum}. There are touching points, which are not Weyl points as there is no state crossing the bulk gap. When each individual honeycomb layer is a Chern insulator, the Chern number of the full system equals the Chern number of one of the honeycomb layers, i.e. there is no enhancement of the band Chern number as we generically observe in models where $r(\bfk)$ behaves nontrivially. Finally, we note that taking two time-reversed copies of our model yields a time-reversal symmetric model, which makes contact to the arguments put forward in Ref.~\onlinecite{wangalexandracavabernevig} where no specific lattice Hamiltonian is investigated but instead symmetry arguments are used to study possible boundary states on several surfaces. Our findings are corroborating the conclusion of Ref.~\onlinecite{wangalexandracavabernevig} that there are no nontrivial surface states on the $(001)$ surface of the KHg$X$ materials class.

\section{Discussion} \label{conclusion}

In this work we have presented exact solutions for the surface states of a variety of topological and nontopological phases on geometrically frustrated lattices. Rather being based on fine-tuning or numerical approximations, our solutions are remarkably general. All we need to assume is the appropriate frustrated lattice structure---which is frequently represented in real materials---in combination with local coupling between neighboring layers and translation symmetry. It should be emphasized that the exact solutions are stable; they do {\it not} represent fragile states on the verge of being topological but are rather representatives that can be obtained arbitrarily deep into the various topological phases we discuss. For instance, while the exact solvability depends on having a local hopping coupling the different $A$ layers only via hopping to the intermediate $B$ layers, the various (topological) phases are stable to the inclusion of such terms. Considering the pyrochlore iridates including Eu$_2$Ir$_2$O$_7$  and Nd$_2$Ir$_2$O$_7$  as particularly prominent examples, quantitatively more accurate descriptions have next-nearest-neighbor hopping terms of the order $t_2/t\approx 0.08-0.2$ \cite{wanturnerbishwanathsavrasov,Witczak-Krempa}. This is, albeit certainly large enough to be quantitatively important, likely small enough such that the intriguing phenomena, present in our large class of solvable models, such as Fermi arcs despite the absence of Weyl nodes, may be realized these materials.

Given the far reaching success of harnessing the properties of frustrated systems through the microscopic understanding of their ubiquitous flat bands in terms of localized modes, we hope that our work will spur a development in which this large body of knowledge will be exploited in order to deepen our understanding also of topological phases of matter. We have here initiated such a program by analytically calculating a number of properties that have previously been accesible by numerical simulation and in fine-tuned models. Building on this, aspects of quantum transport, stability to disorder and interactions provide obvious new avenues.

Our investigation has thus far highlighted several generic facts. This includes the role of lattice geometry for topological phases that are not intrinsically symmetry protected; although Chern insulators and Weyl semimetals rely on breaking symmetries, in contrast to phases being protected thereby, we have shown that they cannot be realized with strictly local (nearest neighbor) constraints on lattices that satisfy conditions (i)-(iii) but fail to satisfy condition (iv), while they are readily realized once also condition (iv) is satisfied. We provided explicit examples of lattices---including two-dimensional lattices akin to kagome lattices but with different stacking properties as well as a three-dimensional system corresponding to the $(100)$ grown surface of the KHgX materials class---whose structure prohibits the formation of topological phases. In glaring contrast, spin-orbit-coupled particles on the kagome and pyrochlore lattices are ideally suited for realizing topological phases and, even more generally, they support novel surface states. 

In fact, we have here corroborated the recent observation \cite{bergholtzliutreschermoessnerudugawa} that there is no one-to-one dichotomy between surface Fermi arcs and the existence of bulk Weyl nodes; while the Fermi arc-like surface bands are always partially attached to bulk bands, these surface bands exist also in the absence of Weyl nodes in the bulk. The Fermi arcs occurring without Weyl nodes are similar to those associated with type-II Weyl excitations in that they are generally blurred in parts of the surface Brillouin zone due the finite density of states in the bulk. While this is a feature that distinguishes them from the surface states corresponding to type-I Weyl systems, our exact solutions, which remain identical tuning from weak to strong inter-layer coupling, highlight that both the surface eigenstates and dispersion are in principle completely unrelated to the existence of bulk Weyl nodes. These observations have direct bearing for engineered as well as naturally existing materials. Notably, single-crystal slabs of the pyrochlore iridates Eu$_2$Ir$_2$O$_7$ \cite{fujita2015} and Nd$_2$Ir$_2$O$_7$ \cite{gallagher2016}, grown precisely in the $(111)$ direction that we consider have been recently synthesized. The possibilities of experimental probing are further extended by growing the pertinent pyrochlore slabs with variable chemical composition and doping \cite{variableslab}. Beyond being of great interest on their own, the study of the pyrochlores on the slab geometry may lead to a better understanding of the rich variety of phenomena associated with bulk pyrochlore iridates \cite{yamaji, yanamaeno,matswakehinataka,tomimatsuiwawata,uedafujiakasuzuishi,ishifarrellnaka,matsutokusakehinatakagi}. 

Our work also opens up questions of fundamental nature that call for further disquisition. Saliently, we have observed that the solvable eigenstates generically carry a Chern number being equal to the sum of the Chern numbers of the pertinent bands in all layers. This happens immediately for arbitrarily weak inter-layer coupling as the eigenstates of the different layers hybridize and amounts to a ``topological selection rule" whose implications and generality we have only touched upon here.  

Finally, we note that the anatomy of the surface states we have provided, dissecting them into tinker toy-like constituents, may serve as a guide when engineering artificial materials. In the context of designing devices and quantum simulators, either in optical, cold atom systems or in the solid state, the fact that the exact solution holds at any finite size is a particularly attractive feature.

\acknowledgments
We acknowledge discussions with Claudio Castelnovo, Nigel Cooper, Sian Dutton, Roderich Moessner, and Masafumi Udagawa. This work is supported by DFG\textquoteright{}s Emmy Noether program (BE 5233/1-1), the Helmholtz VI ``New States of Matter and Their Excitations'', the Swedish research council (VR) and the Wallenberg Academy Fellows program of the Knut and Alice Wallenberg Foundation.

\appendix
\begin{widetext}

\section{Pyrochlore Hamiltonians} \label{appb}
The following Hamiltonians for the pyrochlore model correspond to the different spin configurations, where the specific configurations are given as subscripts and the blocks of the Hamiltonian matrix are given as defined in Eq.~(\ref{generalmultilayerhamiltonian}):

\allowdisplaybreaks
\begin{align}
    k_1 &= \sqrt{2} k_x ,&
    k_2 &= \frac{k_x + \sqrt{3}k_y}{\sqrt{2}} ,&
    k_3 &= \frac{k_x - \sqrt{3}k_y}{\sqrt{2}} ,&
    k_a &= \frac{k_x}{\sqrt{2}}- \frac{k_y}{\sqrt{6}} ,&
    k_b &= \frac{k_x}{\sqrt{2}}+ \frac{k_y}{\sqrt{6}}, &
    k_c &= \frac{\sqrt{2}k_y}{\sqrt{3}},
\end{align}
\begin{align}
    \mathcal{H}_{\bf k, 4-0} = 
    \mathcal{H}_{\bf k, 3-1} &= 
    -\frac{\sqrt{2}t + 2 t_{\rm SOC}}{\sqrt{3}}
    \begin{pmatrix}
        0 & \sqrt{2} \cos(k_1) & (1+i) \cos(k_3) \\
        \sqrt{2} \cos(k_1) & 0 &  (1-i) \cos(k_2) \\
        (1-i) \cos(k_3) & (1+i) \cos(k_2) & 0 \\
     \end{pmatrix}
     + m, \nonumber\\
    \mathcal{H}_{\bf k, 2-2} &= 
    -\frac{\sqrt{2}t + 2 t_{\rm SOC}}{\sqrt{3}}
    \begin{pmatrix}
        0 & \sqrt{2} \cos(k_1) &  0 \\
        \sqrt{2} \cos(k_1) & 0 & 0 \\
        0 & 0 & 0 \\
     \end{pmatrix} \nonumber\\
    & \quad +\frac{1}{3} (\sqrt{2} t - t_{\rm SOC})
    \begin{pmatrix}
        0 & 0 & (-3i + \sqrt{3}) \cos(k_3) \\
        0 & 0 & (3 - i\sqrt{3}) \cos(k_2) \\
        (3i + \sqrt{3}) \cos(k_3) & (3 + i \sqrt{3}) \cos(k_2) & 0
     \end{pmatrix}
     + m,  \nonumber\\
     \mathcal{H}_{\perp,4-0}^{A}(\bf k) &= 
     -\frac{\sqrt{2}t + 2t_{\rm SOC}}{2 \sqrt{3} }
            \begin{pmatrix}
                (1-i) e^{-i k_a } 
                & (1+i) e^{+i k_b} 
                & \sqrt{2} e^{-i k_c}
            \end{pmatrix}^{T}, \nonumber\\
     \mathcal{H}_{\perp,3-1}^{A}(\bf k) &= 
     - \frac{\sqrt{2}t - t_{\rm SOC}}{6 }
            \begin{pmatrix}
                (3 - \sqrt{3}i) e^{-i k_a}
                & (\sqrt{3} - 3i) e^{+i k_b}
                & \sqrt{6} (1-i) \sqrt{2} e^{-i k_c}
            \end{pmatrix}^{T}, \nonumber\\
     \mathcal{H}_{\perp,2-2}^{A}(\bf k) &= 
     - \frac{\sqrt{2}t - t_{\rm SOC}}{6 }
            \begin{pmatrix}
                (3 - \sqrt{3}i) e^{-i k_a}
                & (\sqrt{3} - 3i) e^{+i k_b}
                & \frac{2 \sqrt{3} (t+2 t_{\mathrm{SOC}})}{\sqrt{2}t - t_{\mathrm{SOC}}} e^{-i k_c}
            \end{pmatrix}^{T}, \nonumber\\
            \mathcal{H}_{\perp}^{B}(\bf k) &= \mathcal{H}_{\perp}^{A}(\bf k \rightarrow -\bf{k}).
    \label{eq:pyrochlore}
\end{align}

\end{widetext}

\end{document}